# High-Mass X-ray Binaries in the Milky Way

## A closer look with *INTEGRAL*

**R. Walter · A. A. Lutovinov · E. Bozzo · S. S. Tsygankov**

**Abstract** High-mass X-ray binaries are fundamental in the study of stellar evolution, nucleosynthesis, structure and evolution of galaxies and accretion processes. Hard X-rays observations by *INTEGRAL* and *Swift* have broadened significantly our understanding in particular for the super-giant systems in the Milky Way, which number has increased by almost a factor of three. *INTEGRAL* played a crucial role in the discovery, study and understanding of heavily obscured systems and of fast X-ray transients. Most super-giant systems can now be classified in three categories: classical/obscured, eccentric and fast transient.

The classical systems feature low eccentricity and variability factor of $\sim 10^3$, mostly driven by hydrodynamic phenomena occurring on scales larger than the accretion radius. Among them, systems with short orbital periods and close to Roche-Lobe overflow or with slow winds, appear highly obscured. In eccentric systems, the variability amplitude can reach even higher factors, because of the contrast of the wind density along the orbit. Four super-giant systems, featuring fast outbursts, very short orbital periods and anomalously low accretion rates, are not yet understood.

Simulations of the accretion processes on relatively large scales have progressed and reproduce parts of the observations. The combined effects of wind

Roland Walter & Enrico Bozzo
ISDC, Geneva Observatory, University of Geneva, Ch. d'Ecogia 16, CH-1290 Versoix, Switzerland. E-mail: Roland.Walter@unige.ch, Enrico.Bozzo@unige.ch

Alexander A. Lutovinov
Space Research Institute of the Russian Academy of Science, Profsoyuznaya str. 84/32, RU-117997 Moscow, Russian Federation. E-mail: aal@iki.rssi.ru

Sergey S. Tsygankov
Tuorla Observatory, Department of Physics and Astronomy, University of Turku, Väisäläntie 20, FI-21500 Piikkiö, Finland. E-mail: stsygankov@gmail.com

clumps, magnetic fields, neutron star rotation and eccentricity ought to be included in future modelling work.

Observations with *INTEGRAL* in combination with other observatories were also important for detecting cyclotron resonant scattering features in spectra of X-ray pulsars, probing their variations and the geometry of the accretion column and emission regions. Finally, the unique characteristics of *INTEGRAL* and its long life time played a fundamental role for building a complete catalogue of HXMBs, to study the different populations of these systems in our Galaxy, and to constrain some of the time scales and processes driving their birth and evolution.

## 1 Introduction

Neutron stars and stellar mass black holes stand out as luminous X-ray sources in the Galaxy when they are accreting matter from nearby stars. When these companions have masses above $\sim 10$ $M_{\odot}$, the systems are known as High-Mass X-ray Binaries (HMXB). Such systems can be formed when one of the initial member stars looses a significant part of its mass, through stellar wind or mass transfer, before the first supernova explosion occurs (van den Heuvel and Heise, 1972). They are young (several dozen million years old), in contrast to the Low-Mass X-ray Binary systems (LMXBs) that are several billion years old.

In most HMXBs, the compact objects capture a very small fraction of the stellar wind of their companions and the resulting accretion rates are low (Bondi and Hoyle, 1944; Davidson and Ostriker, 1973; Lamers et al., 1976). High X-ray luminosities ($> 10^{35}$ erg/s) are observed in two situations. Strong and transient X-ray flares, reaching the Eddington luminosity, occur when the compact object crosses a dense component of the stellar wind, usually expelled by a fast rotating main sequence star (featuring emission lines in the optical and hence identified as “Be” systems). High accretion rates are also observed in close systems where the companion is practically filling its Roche lobe (giant and super-giant systems). These systems become very luminous (up to $10^{40}$ erg s$^{-1}$; Bachetti et al., 2014) when the donor is close to the Roche limit and the accretion becomes dominated by a tidal stream. Roche-lobe overflow is rarely observed as the compact object is quickly enshrouded, unless the radial expansion of the companion is slow.

The very large majority of the HMXB systems harbour accreting pulsars (Liu et al., 2006; Lutovinov and Tsygankov, 2009). In such systems, the plasma approaching the neutron star is stopped by the pressure of the dipolar magnetic field and forced to move along the field lines toward the magnetic poles, where the captured matter releases its gravitational energy in the form of X-rays. The X-ray continuum of accreting pulsars is characterized by a powerlaw of photon index 0.3-2 with a high-energy exponential cutoff (7-30 keV, White et al., 1983; Filippova et al., 2005), sometimes modified by absorption and emission lines in the soft X-rays and by Cyclotron Resonance Scattering Features (CRSF) at higher energies (Coburn et al., 2002; Filippova et al., 2005; Caballero and Wilms, 2012). The plasma falls in the accretion column at almost the speed of light and heats to $10^8$ K close to the neutron star surface (see, e.g. Basko and Sunyaev, 1976; Nagel, 1981; Meszaros and Nagel, 1985; Araya-Góchez and Harding, 2000; Nishimura, 2008; Mushtukov et al., 2015). Bulk and thermal Comptonization plays a key role in the formation of the non thermal X-ray emission (Becker and Wolff, 2007).

CRSFs are caused by the scattering of hard X-ray photons on electrons whose energy is quantized by the magnetic field according to the Landau levels (Gnedin and Sunyaev, 1974; Truemper et al., 1978; Araya-Góchez and Harding, 2000). This electron energy can be measured from the source spectra and hence the magnetic field strength in the scattering region. Variability of the CRSF energy with luminosity on long and spin period time scales indicate that the accretion flow is not uniform nor stationary (Mihara et al., 1998; Mowlavi et al., 2006; Staubert et al., 2007; Tsygankov et al., 2006, 2010; Klochkov et al., 2011).

Emission lines and absorption observed in the soft X-ray band are the imprints of the companion stellar wind. Photo-ionisation and other effects of the pulsar on the wind structure, as well as inhomogeneities of the wind, either genuine or induced by the compact object, lead to additional variability.

The HMXBs of the Milky Way include three microquasars and black-hole candidates and three gamma-ray loud binaries. Because of their peculiarities, these six sources will not be discussed in this review. Their high energy emission and variability patterns are very different from those described above and dominated by inverse Compton scattering of electron accelerated close to the black-hole or in the interaction regions between the companion stellar winds and pulsar winds or microquasar jets (Dubus, 2013).

The INTernational Gamma-Ray Astrophysics Laboratory (*INTEGRAL*), a medium size mission from the European Space Agency (Winkler et al., 2003), observes the Universe in the hard X-ray and soft gamma-ray band. The wide field of view ($\sim 30°$) of its main instruments, its unique energy coverage and its frequent scans of the galactic plane allowed *INTEGRAL* to observe the Galaxy in a parameter space not well studied before and to discover strongly absorbed and transient HMXBs with low duty cycles.

110 HMXB systems were known in the Milky Way before the launch of *INTEGRAL* (Liu et al., 2000): 13 super-giant, 52 Be and 45 systems of unclear or other types. The serendipitous discovery by *INTEGRAL* of many new HMXB systems, in particular 23 likely of super-giant type, came as a surprise. The mere fact that these new systems had not been identified in the past indicates that the HMXB phenomenology is more diverse and rich than anticipated. This review concentrates on these new aspects.

HMXBs are generally concentrated towards the Galactic plane, close to their birthplace (Fig. 1; see also e.g. Grimm et al., 2002). The X-ray luminosity of normal star forming galaxies, dominated by HMXBs and by the hot ionized inter-stellar gas, correlates well with the star formation rate (Grimm et al., 2003; Ranalli et al., 2003; Lehmer et al., 2010; Mineo et al., 2012a,b; Lutovinov et al., 2013b). The discovery by *INTEGRAL* of many new HMXBs close to the tangent directions to the inner galactic arms also allowed to understand better their distribution in the Milky Way and their relation with star forming regions (Lutovinov et al., 2005a; Bodaghee et al., 2012c; Coleiro and Chaty, 2013). Finally the small fraction of black-hole HMXB systems, probably originating from very high mass stars, and their higher masses when compared to neutron stars, can be related to the physics of supernova explosions (Belczynski et al., 2012).

Section 2 and 3 review the new observations, source discoveries and catalogue and the properties of the various classes of HMXBs in the light of the new observations. In section 4 and 5, we discuss several new aspects of the phenomenology of wind accretion revealed by the individual objects and the global properties of their population at the scale of the Galaxy. Finally a summary of the new results is presented in section 6.

## 2 Observations and source catalogue

### 2.1 Hard X-ray sources and their identification

The large field of view, hard X-ray coded mask imagers on board *INTEGRAL* and *Swift* are observing the full sky regularly; *INTEGRAL* focussing more on the galactic plane. The observations consist of numerous short pointings of $(1-5)\times10^3$ sec, enhancing the sensitivity to flaring activities on such time scales. Many new sources and flares were detected and about a thousand Astronomer's Telegrams were issued.

The value of any sky survey to study the properties of a population of sources (in particular HMXBs) depends on the survey completeness and on the identification of the nature of the detected sources. Surveys performed with *INTEGRAL* and *Swift* have a very high identification completeness, reaching 92% in the Galactic plane (Krivonos et al., 2012). Such a high identification completeness results from follow-up observations performed by several research groups in the soft X-rays ($<10$ keV), optical, infrared and radio wavelengths (see, e.g. Walter et al., 2003; Bikmaev et al., 2006, 2008; Masetti et al., 2006b,

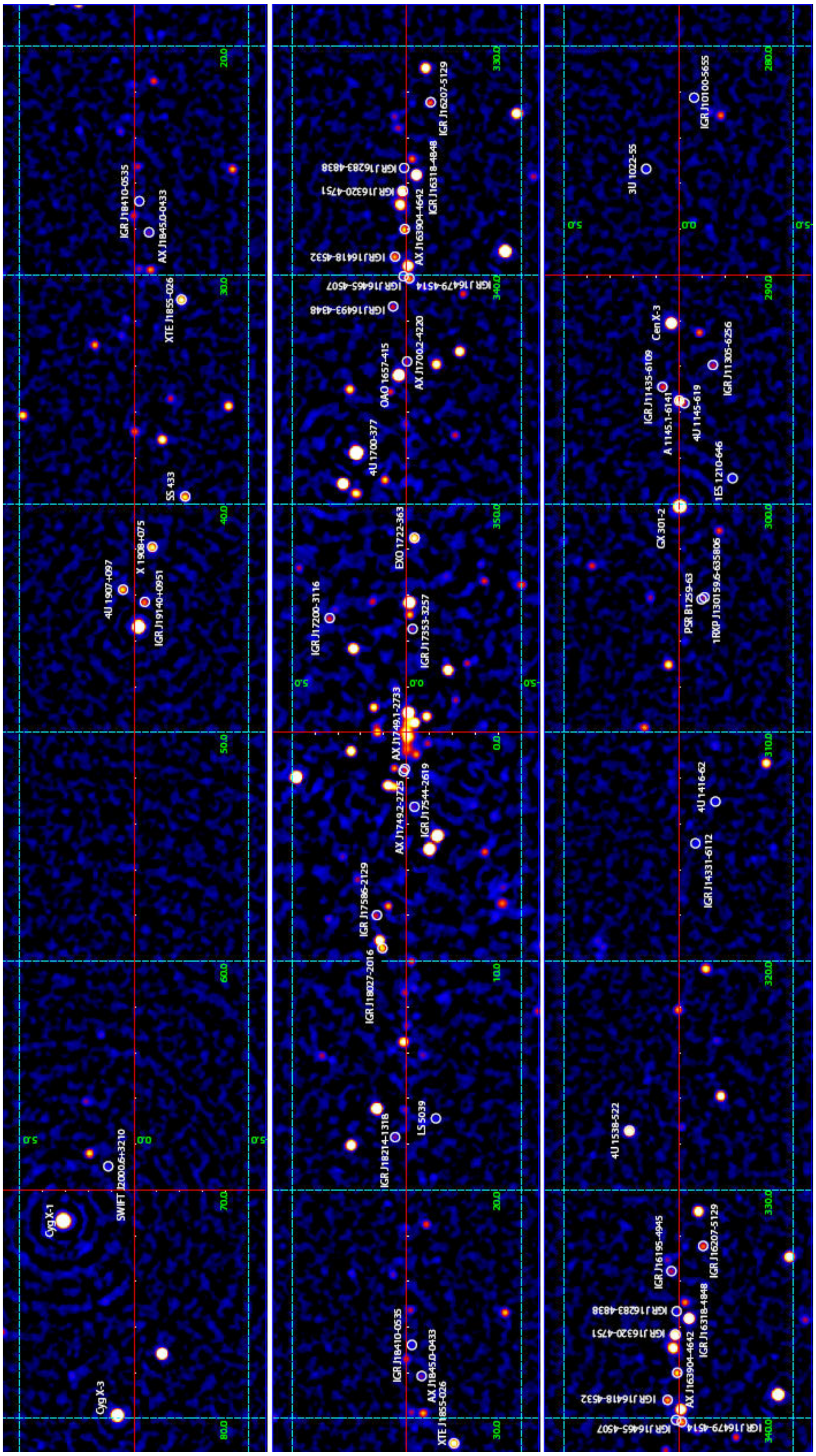

**Fig. 1** Image of the inner part of the Galactic plane, obtained with INTEGRAL/IBIS in the 17-60 keV energy band. Persistent HMXBs are identified with circles (Lutovinov et al., 2013b). The horizontal red line is the Galactic equator.

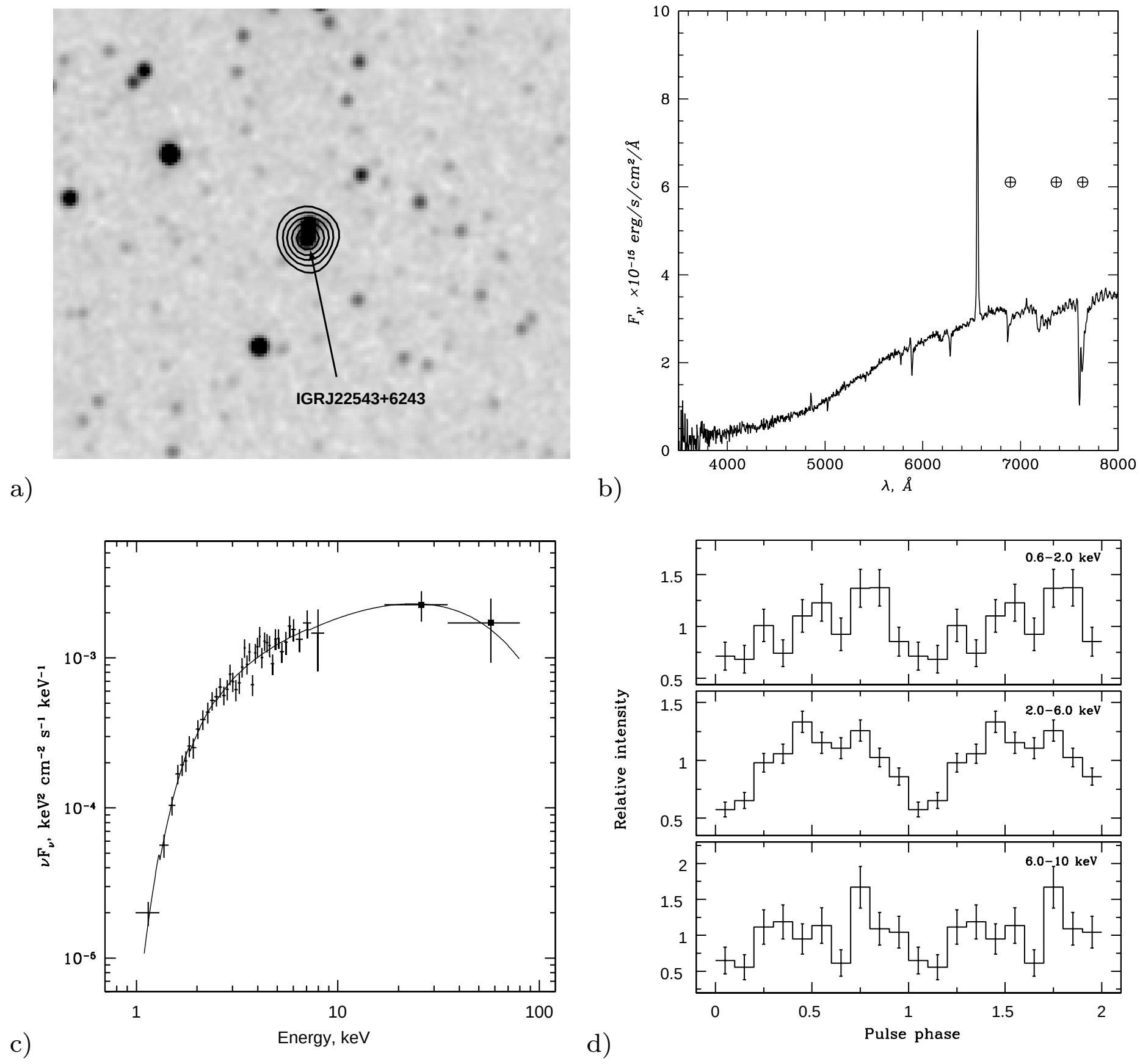

**Fig. 2** (a) Sky field around the source IGR J22534+6243 in the *J*-band (*2MASS* survey). Contours indicate levels of the source intensity in the X-rays, obtained by *Swift/XRT*. The infrared counterpart is indicated by the arrow. (b) Optical spectrum of the source. (c) Broadband energy spectrum of IGR J22534+6243. The best-fit model is indicated by the solid line. (d) Pulse profile in three energy bands, folded with the period of 46.675 sec. See Lutovinov et al. (2013a) for details.

2009, 2012b; Tomsick et al., 2006a, 2008, 2009a; Rahoui et al., 2008a; Burenin et al., 2008; Chaty et al., 2008; Lutovinov et al., 2012b; Karasev et al., 2012).

As the source localization accuracy provided by the imagers on board *INTEGRAL* and *Swift* (about $2-5$ arcminutes depending on the source significance) is not enough for an unambiguous optical identification, a significant improvement of the localization accuracy is required as a first step. This is achieved by follow-up observations (or archival studies) carried out with focussing X-ray telescopes such as *Swift/XRT, XMM-Newton* or *Chandra.* In densely populated regions, such as the inner part of the Galaxy, sub arcsecond resolution is required and only follow-up observations with *Chandra* can help to identify a hard X-tray source.

In the case of HMXBs, an accurate X-ray position is usually good enough to identify the likely counterpart in optical and infrared surveys or catalogues (such as *DSS, USNO-A2(B1), 2MASS, UKIDSS,* or *VVV*). The photometry obtained from these surveys together with the high-energy spectra and lightcurves allow us to make first assumptions on the nature of the sources. In particular, the presence of X-ray absorption together with a counterpart well detected in the infrared and much weaker in the optical is a good indication for the massive nature of the binary system. The detection of X-ray pulsations unambiguously points at a rotating neutron star with a strong magnetic field.

A final confirmation of the nature of the sources can only be obtained from infrared/optical spectroscopic observations with low to medium resolution ($\lambda/\Delta\lambda \simeq 500 - 3000$). Several classification parameters are used: the reddening, different absorption and emission lines typical for different object classes, line flux ratios, lines width and their redshift.

The identification process is illustrated in Fig. 2 for IGR J22534+6243, a hard X-ray source discovered by *INTEGRAL*. An infrared image in the $J$-band around the position obtained by a follow-up observation with *Swift/XRT* is shown in Fig. 2a. Two close (4.4 arcsecond separation) relatively bright ($m_J \simeq 11.64$ and $m_J \simeq 11.78$) objects are detected in the X-ray error circle. The optical spectrum of the most central object obtained with the Russian-Turkish Telescope *RTT-150* is typical for an early-type star (Fig. 2b). The broadened $H\alpha$ emission line, together with the $H\beta$ and $HeI$ emission lines are often observed for Be stars, which have a fast-rotating equatorial disc. The broadband X-ray spectrum of IGR J22534+6243 obtained with *Chandra* and *INTEGRAL* is typical for an accreting neutron star with a cutoff power law model and photo-absorption at low energies (Fig. 2c). Finally, X-ray pulsations with a period of $P_s \simeq 46.67$ s were detected from this source (Fig. 2d). These observations allow to classify IGR J22534+6243 as a new X-ray pulsar in a Be high-mass X-ray binary system (Lutovinov et al., 2013a). Other examples of the optical and infrared spectra of high-mass X-ray binaries, discovered by *INTEGRAL* are shown in Fig. 3.

### 2.2 Source catalogue

Our catalogue of HMXBs in the Milky Way includes a total of 87 sources listed in Table 1, organized per source category as commonly known in the literature. For each source we list coordinates, spin and orbital periods, spectral type, distance, system type (cl: classical; abs: obscured; SFXT: transients; ?: unclear type; e: eccentric orbit; P: pulsar; BH: black-hole) and the average 17-60 keV flux in units of $10^{-11}$ erg s$^{-1}$ cm$^{-2}$ (taken from Krivonos et al., 2007, 2012). If the source is missing in these catalogues, then its flux was taken from other papers (appropriate references and energy bands are mentioned).

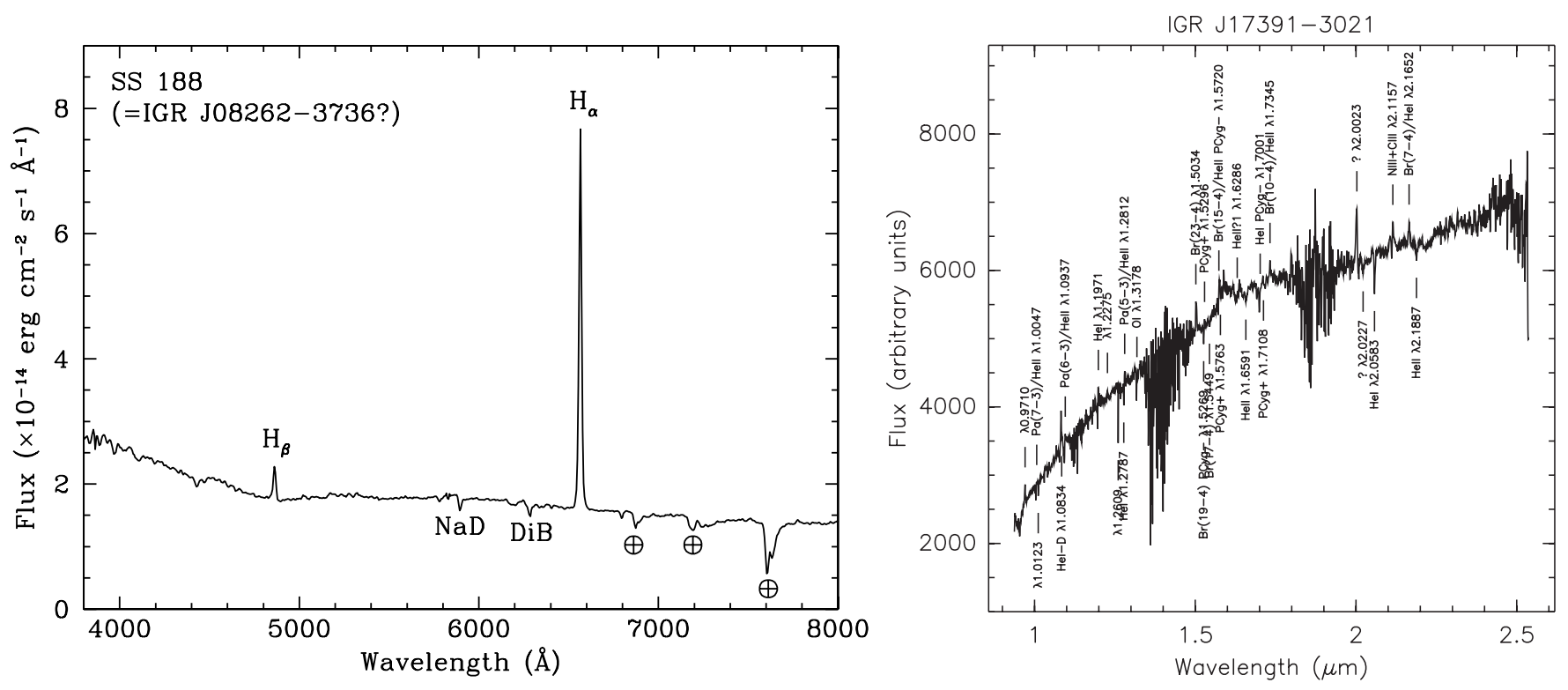

**Fig. 3** An example of optical and infrared spectra of two high mass X-ray binary systems IGR J08262-3736 (left) and IGR J17391-3021 (right), discovered by *INTEGRAL*. The spectra are from Masetti et al. (2010b) and Chaty et al. (2008).

The system type is based on our analysis of the available data, presented in this review, and can be different from the previously published ones.

The Milky-Way HMXBs can be categorized as follow:

- 24 systems have super-giant companions and are persistent at hard X-rays. These are the classical systems. Six of them are characterised by high obscuration. Seven of them are known in the literature as super-giant fast X-ray transients but can be understood as classical systems.
- 10 systems are super-giant fast X-ray transients detected above 10 mCrab only for short periods and with a low ($\lesssim$10%) duty cycle. They feature likely super-giant companions and show impressive variability factors in the range $10^{2-5}$. Most of them have been discovered by INTEGRAL (some had been discovered previously but not identified as super-giant systems).
- Cen X-3, the only Roche Lobe Overflow giant system identified in the Galaxy.
- 57 systems have likely a Be stellar companion (32 detected by *INTEGRAL*).
- 3 gamma-ray loud binaries (of Be type as well).
- 3 black-hole systems (2 are super-giant systems).
- 4 giant and main sequence systems (two of them discovered by *INTEGRAL*).
- 12 systems of unclear type, 4 among them have likely a main sequence or giant companion, and their identification is therefore more difficult. IGR J10101-5654 is a sgB[e] system which was detected in outburst for two months in 2004 and has been faint otherwise.

The galactic plane observations of *INTEGRAL* had an important impact on our knowledge of super-giant systems. They tripled the the number of these systems identified in the Galaxy (Fig. 4) and new types of behaviour were discovered, in particular systems featuring strong and persistent obscuration or high variability and low duty cycles (respectively 6 and 13 sources). Even while pulsations have not yet been detected in 12 of these systems, their hard X-ray spectra are typical of accreting pulsars. Not a single new high-mass black-hole system has been discovered.

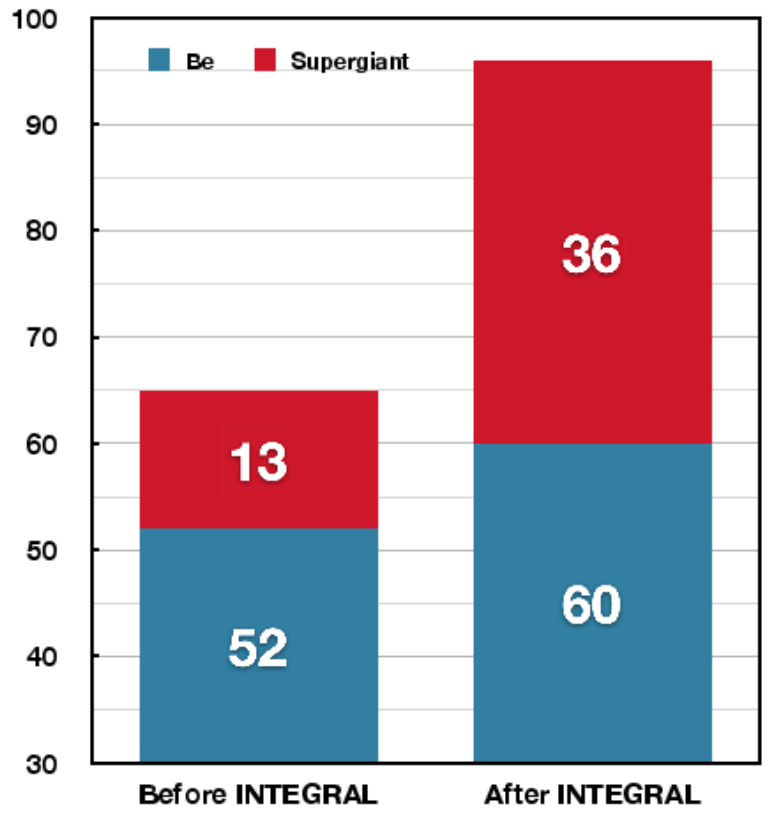

**Fig. 4** Number of HMXBs identified as Be or super-giant systems in the Galaxy, before and after the discoveries triggered by the *INTEGRAL* mission.

There are about 20’000 O stars in the galaxy and 33% of them are double systems evolving through envelope stripping (Sana et al., 2012). Assuming that half of these systems will survive the supernova kicks, about one HMXB forms every 1500 years. The larger number of super-giant HMXBs discovered by *INTEGRAL* points to a lifetime of $\sim 10^5$ years for the HXMB phase which may support the enhanced wind and stripped H-burning scenario of Ziolkowski (1977).

There are some additional unidentified *INTEGRAL* sources that have been suggested as HMXB candidates: the X-ray spectra of IGR J18325-0756, IGR J16283-4843 and IGR J18219-1347 show significant absorption; IGR J13186-6257 and XTE J1824-141 have periods of 20 days and 120 sec respectively. We decided not to include them here as the evidence for high mass companions remains too vague.

Table 1: Catalogue of High-Mass X-ray Binaries.

| Source name | RA | DEC | l | b | $P_s$ | $P_{orb}$ | S.T. | kpc | Type | $F_{17-60}$ |
|---|---|---|---|---|---|---|---|---|---|---|
| **Super-giant persistent systems** | | | | | | | | | | |
| 1A 0114+650 | 19.516 | 65.289 | 125.723 | 2.571 | 9475[1] | 11.59[2] | B0.5I[3] | 7.2[4] | cl,P | 9.40 |
| Vela X-1 | 135.531 | -40.555 | -96.933 | 3.930 | 283[5] | 8.964[6] | B0.5Ib | 1.7-2.1[7] | cl,P | 214.77 |
| 1E 1145.1-6141 | 176.870 | -61.956 | -64.509 | -0.027 | 297[8] | 14.365[9] | B2Iae[10] | 8.2[11] | cl,P | 18.99 |
| GX 301-2 | 186.651 | -62.772 | -59.891 | -0.031 | 675-700[12] | 41.492[13] | B1 Ia+[14] | 3-4[15] | cl,P | 181.21 |
| 4U 1538-522 | 235.600 | -52.385 | -32.581 | 2.177 | 528-530[16] | 3.728[17] | B0I[18] | 5.5[19] | cl,P | 16.38 |
| IGR J16318-4848 | 247.951 | -48.817 | 335.616 | -0.448 | | | sgB[e][20] | 0.9-6.2[21] | abs,cl | 24.63 |
| IGR J16320-4751 | 248.007 | -47.875 | 336.329 | 0.169 | 1309[22] | 8.986[23] | O8I[24] | 3.5[25] | abs,cl,P | 14.23 |
| IGR J16393-4641 | 249.772 | -46.704 | 336.001 | 0.075 | 912.0[26] | 4.24[27] | OB?[28] | >10?[29] | abs,cl,P | 5.9 |
| IGR J16493-4348 | 252.362 | -43.819 | -18.629 | 0.603 | 1093[30] | 6.782[31] | B0.5 Ib[32] | >6[33] | cl,P | 1.81 |
| OAO 1657-415 | 255.199 | -41.656 | -15.631 | 0.324 | 38.2[34] | 10.448[35] | Ofpe/WN9[36] | 4-8[37] | cl,P | 65.44 |

[1] Wang (2011) [2] Crampton et al. (1985) [3] Crampton et al. (1985) [4] Reig et al. (1996) [5] Quaintrell et al. (2003) [6] Quaintrell et al. (2003) [7] Nagase et al. (1986) [8] White et al. (1980) [9] Ray and Chakrabarty (2002) [10] Densham and Charles (1982) [11] Densham and Charles (1982) [12] Kaper et al. (2006) [13] Koh et al. (1997) [14] Kaper et al. (2006) [15] Kaper et al. (2006) [16] Clark (2000) [17] Clark et al. (1994) [18] Reynolds et al. (1992) [19] Reynolds et al. (1992) [20] Filliatre and Chaty (2004) [21] Filliatre and Chaty (2004) [22] Lutovinov et al. (2005c) [23] Manousakis and Walter (2012) [24] Rahoui et al. (2008b) [25] Rahoui et al. (2008b) [26] Bodaghee et al. (2006) [27] Pearlman et al. (2011) [28] (Bodaghee et al., 2012a) [29] (Bodaghee et al., 2012a) [31] Pearlman et al. (2013) [32] Nespoli et al. (2010b) [33] Nespoli et al. (2010b) [34] White and Pravdo (1979) [35] Mason et al. (2012) [36] Mason et al. (2012) [37] Mason et al. (2012)

Table 1: Catalogue of High-Mass X-ray Binaries.

| Source name | RA | DEC | l | b | $P_s$ | $P_{orb}$ | S.T. | kpc | Type | $F_{17-60}$ |
|---|---|---|---|---|---|---|---|---|---|---|
| 4U 1700-37 | 255.986 | -37.844 | 347.7544 | 2.173 | - | 3.412[1] | O6.5 Iaf+ [2] | 1.9[3] | cl,BH? | 209.4 |
| EXO 1722-363 | 261.297 | -36.283 | 351.497 | -0.354 | 413.89[4] | 9.742[5] | B0-B1Ia[6] | 6-10.5[7] | abs,cl,P | 7.72 |
| IGR J18027-2016 | 270.666 | -20.283 | 9.430 | 1.039 | 139.612[8] | 4.469[9] | B1 Ib[10] | 12.4[11] | abs,cl,P | 4.21 |
| XTE J1855-026 | 283.870 | -2.601 | 31.082 | -2.085 | 360.7[12] | 6.0724[13] | B0 Iaep[14] | | cl,P | 10.34 |
| 4U 1907+097 | 287.406 | 9.833 | 43.752 | 0.488 | 437-440[15] | 8.375[16] | O8.5Iab[17] | < 8[18] | cl,P | 13.36 |
| 4U 1909+07 | 287.701 | 7.595 | 41.896 | -0.810 | 605[19] | 4.4[20] | O7.5-9.5sg[21] | 7[22] | cl,P | 12.68 |
| IGR J19140+0951 | 288.526 | 9.885 | 44.298 | -0.461 | | 13.55[23] | B0.5Ia/d[24] | 2-5[25] | cl,P | 8.93 |
| **Super-giant fast X-ray transients** | | | | | | | | | | |
| IGR J08408-4503 | 130.199 | -45.058 | 264.040 | -1.950 | - | - | O8.5Ib(f)[26] | 2.7[27] | ? | 0.4 |
| IGR J11215-5952 | 170.445 | -59.863 | 291.893 | 1.073 | 186.78[28] | 165[29] | B0.5Ia[30] | 6.4-8.0[31] | e,cl,P | 0.15 |

[1] Corbet et al. (2010c) [2] Jones et al. (1973) [3] Ankay et al. (2001)
[4] Manousakis and Walter (2011) [5] Manousakis and Walter (2011) [6] Mason et al. (2009)
[7] Mason et al. (2009) [8] Mason et al. (2011) [9] Mason et al. (2011)
[10] Masetti et al. (2008a) [11] Torrejón et al. (2010) [12] Corbet et al. (1999b)
[13] Corbet and Mukai (2002) [14] Negueruela et al. (2008a) [15] in 't Zand et al. (1998b)
[16] in 't Zand et al. (1998b) [17] Nespoli et al. (2008a) [18] Nespoli et al. (2008a)
[19] Morel and Grosdidier (2005) [20] Wen et al. (2000) [21] Morel and Grosdidier (2005)
[22] Morel and Grosdidier (2005) [23] Corbet et al. (2004) [25] Hannikainen et al. (2007)
[26] Barba et al. (2006) [27] Leyder et al. (2007) [28] Romano et al. (2009c)
[29] Romano et al. (2009c) [30] Negueruela (2010) [31] Negueruela (2010)

Table 1: Catalogue of High-Mass X-ray Binaries.

| Source name | RA | DEC | l | b | $P_s$ | $P_{orb}$ | S.T. | kpc | Type | $F_{17-60}$ |
|---|---|---|---|---|---|---|---|---|---|---|
| IGR J16195-4945 | 244.884 | -49.742 | 333.557 | 0.339 | - | 16(?)[1] | O,B,A[2] | 5[3] | cl | 1.66 |
| IGR J16207-5129 | 245.193 | -51.502 | 332.459 | -1.050 | - | 9.726(?)[4] | B0I[5] | 6[6] | cl | 2.74 |
| IGR J16328-4726 | 248.158 | -47.395 | 336.749 | 0.422 | - | 10.07[7] | O8Iaf[8] | - | cl | 3[9] |
| IGR J16418-4532 | 250.462 | -45.540 | 339.189 | 0.489 | 1212[10] | 3.73886[11] | O8.5[12] | 13[13] | cl,abs,P | 3.40 |
| IGR J16465-4507 | 251.647 | -45.118 | 340.053 | 0.135 | 228[14] | 30.32[15] | B0.5-O9.5Ia[16] | 3.8-23.6[17] | cl, P | 4.64 |
| IGR J16479-4514 | 252.027 | -45.202 | 340.163 | -0.124 | - | 3.3193[18] | O8.5Ib[19] | 1.1-7.7[20] | SFXT | 3.62 |
| IGR J17354-3255 | 263.854 | -32.938 | 355.447 | -0.269 | - | 8.452[21] | - | - | cl | 1.01 |
| XTE J1739-302 | 264.798 | -30.344 | 358.068 | 0.445 | - | 51.47[22] | O8.5Iab(f)[23] | 2.3[24] | eSFXT | 0.945 |
| IGR J17544-2619 | 268.605 | -26.331 | 3.236 | -0.336 | - | 4.926[25] | O9Ib[26] | 2.1-4.2[27] | eSFXT, P | 0.68 |
| SAX J1818.6-1703 | 274.658 | -17.047 | 14.080 | -0.704 | - | 30.0[28] | B0.5Iab[29] | 2.0-2.2[30] | eSFXT | 1.31 |
| AX J1820.5-1434 | 275.125 | -14.573 | 16.473 | 0.068 | 152.26[31] | 54.0[32] | O9.5-B0Ve[33] | - | ?, P | 1.29 |

[1] Morris et al. (2009a), inferred from the fraction of time spent in outburst
[2] Tomsick et al. (2006a) [3] Tomsick et al. (2006a) [4] Jain et al. (2011), not confirmed
[5] Negueruela et al. (2007a) [6] Nespoli et al. (2008b) [7] Corbet et al. (2010a)
[8] Coleiro et al. (2013) [9] Bird et al. (2010a), in the 20-40 keV band [10] Sidoli et al. (2012)
[11] Levine et al. (2011) [12] Rahoui et al. (2008a) [13] Rahoui et al. (2008a) [14] Lutovinov et al. (2005a) [15] Clark et al. (2010) [16] Nespoli et al. (2008b); Rahoui et al. (2008a)
[17] Heras and Walter (2004) [18] Romano et al. (2009b) [19] Nespoli et al. (2008b)
[20] Nespoli et al. (2008b) [21] D'Ai' et al. (2010) [22] Drave et al. (2010)
[23] Negueruela et al. (2006a) [24] Negueruela et al. (2006a) [25] Clark et al. (2009)
[26] Pellizza et al. (2006) [27] Pellizza et al. (2006) [28] Bird et al. (2009)
[29] Torrejón et al. (2010) [30] Torrejón et al. (2010) [31] Kinugasa et al. (1998)
[32] Segreto et al. (2013) [33] Liu et al. (2006), counterpart could be two blended sources

Table 1: Catalogue of High-Mass X-ray Binaries.

| Source name | RA | DEC | l | b | $P_s$ | $P_{orb}$ | S.T. | kpc | Type | $F_{17-60}$ |
|---|---|---|---|---|---|---|---|---|---|---|
| AX J1841.0-0536 | 280.252 | -5.596 | 26.764 | -0.239 | - | 6.45[1] | B1Ib[2] | 1.7-5.2[3] | SFXT | 0.94 |
| AX J1845.0-0433 | 281.259 | -4.565 | 28.140 | -0.660 | - | 5.7[4] | O9Ia[5] | 6.4[6] | SFXT | 1.46 |
| IGR J18462-0223 | 281.553 | -2.375 | 30.223 | 0.079 | 997 | - | sg(?) | - | ?, P | 0.4[7] |
| IGR J18483-0311 | 282.071 | -3.171 | 29.750 | -0.745 | (21.05)[8] | 18.518[9] | B0.5Ia[10] | 3-4 | eSFXT, P | 4.11 |
| **Be systems** | | | | | | | | | | |
| 4U 0115+63 | 19.625 | 63.746 | 125.924 | 1.026 | 3.61[11] | 24.3[12] | B0.2Ve[13] | 7[14] | P | 125.50 |
| IGR J01363+6610 | 24.060 | 66.188 | 127.447 | 3.699 | | 160(?)[15] | B1IV-Ve[16] | 2.0 | | 15.44 |
| RX J0146.9+6121 | 26.744 | 61.351 | 129.553 | -0.785 | 1400[17] | | B1Ve[18] | 2.5[19] | P | 1.19 |
| IGR J01583+6713 | 29.577 | 67.223 | 129.352 | 5.188 | (469.2?)[20] | | B2IVe[21] | 4.0[22] | P(?) | 0.4[23] |
| V 0332+53 | 53.751 | 53.172 | 146.052 | -2.194 | 4.375[24] | 34.67[25] | O8-9Ve[26] | 7.0[27] | P | 393.29 |
| 4U 0352+309 | 58.849 | 31.036 | 163.084 | -17.144 | 835[28] | 250[29] | B0Ve[30] | 0.95[31] | P | 30.07 |
| RX J0440.9+4431 | 70.270 | 44.530 | 159.858 | -1.258 | 202.5[32] | 155[33] | Be[34] | 3.3[35] | P | 0.95 |
| A 0535+262 | 84.735 | 26.324 | -178.575 | -2.625 | 103[36] | 111.1[37] | B0IIIe[38] | 2[39] | P | 223.49 |
| IGR J06074+2205 | 91.861 | 22.097 | 188.385 | 0.814 | | | B0.5Ve[40] | 4.5[41] | | 0.3 |
| 2E 0655.8-0708 | 104.557 | -7.218 | -139.871 | -1.784 | 160.7[42] | 101.2[43] | O9.7Ve[44] | 3.9[45] | P | 3.26 |

[1] González-Galán (2015) [2] Nespoli et al. (2008b) [3] Nespoli et al. (2008b) [4] Goossens et al. (2013) [5] Coe et al. (1996) [6] Coleiro and Chaty (2013) [7] Bird et al. (2010a), in the 20-40 keV band [8] Sguera et al. (2007); Ducci et al. (2013a) [9] Romano et al. (2010b) [10] Rahoui et al. (2008a) [11] Rappaport et al. (1978) [12] Rappaport et al. (1978) [13] Negueruela and Okazaki (2001) [14] Negueruela and Okazaki (2001) [15] Corbet and Krimm (2010) [16] Reig et al. (2005b) [17] La Palombara and Mereghetti (2006) [18] Reig (2011) [19] Tapia et al. (1991) [20] Kaur et al. (2008) [21] Kaur et al. (2008) [22] Kaur et al. (2008) [23] Bird et al. (2010a), in the 20-40 keV band [24] Zhang et al. (2005) [25] Zhang et al. (2005) [26] Negueruela et al. (1999) [27] Negueruela et al. (1999) [28] White et al. (1976) [29] Delgado-Martí et al. (2001) [30] Telting et al. (1998) [31] Telting et al. (1998) [32] Reig (2011) [33] Tsygankov et al. (2012) [34] Motch et al. (1997) [35] Reig et al. (2005a) [36] Rosenberg et al. (1975) [37] Finger et al. (1996b) [38] Steele et al. (1998) [39] Steele et al. (1998) [40] Reig et al. (2010) [41] Reig et al. (2010) [42] Morgan et al. (2003) [43] Yan et al. (2012) [44] Pakull et al. (2003) [45] McBride et al. (2006)

Table 1: Catalogue of High-Mass X-ray Binaries.

| Source name | RA | DEC | l | b | $P_s$ | $P_{orb}$ | S.T. | kpc | Type | $F_{17-60}$ |
|---|---|---|---|---|---|---|---|---|---|---|
| GRO J1008-57 | 152.447 | -58.298 | -77.018 | -1.821 | 93.6[1] | 249.46[2] | B1-B2 Ve[3] | 5[4] | P | 13.62 |
| 3U 1022-55 | 159.401 | -56.801 | -74.647 | 1.496 | 860[5] | | B0 V-IIIe[6] | 5[7] | P | 8.66 |
| 1A 1118-615 | 170.238 | -61.917 | 292.499 | -0.892 | 405-407[8] | 24.0[9] | O9.5IV-Ve[10] | 3-7[11] | P | 4.75 |
| IGR J11305-6256 | 172.779 | -62.947 | 293.945 | -1.485 | - | - | B0IIIe[12] | 3[13] | | 2.58 |
| IGR J11435-6109 | 176.001 | -61.127 | 294.881 | 0.686 | 161.76[14] | 52.46[15] | B0Ve/B2Ve[16] | $\gtrsim$6-10[17] | P | 2.83 |
| 4U 1145-619 | 177.000 | -62.207 | -64.389 | -0.240 | 292[18] | 187.5[19] | B1Vne[20] | 0.5-3.1[21] | P | 2.30 |
| GX 304-1 | 195.322 | -61.602 | -55.897 | 1.247 | 272[22] | 132.5[23] | B2 Vne[24] | 2.4[25] | P | 1.38 |
| 2RXP J130159.6-635806 | 195.495 | -63.969 | -55.912 | -1.121 | $\sim$ 700[26] | | B0.5Ve[27] | 4-7[28] | P | 1.79 |
| 4U 1416-62 | 215.303 | -62.698 | -46.979 | -1.598 | 17.64[29] | 42.12[30] | B1Ve[31] | 1.4-11[32] | P | 0.78 |
| XTE J1543-568 | 236.011 | -56.748 | -35.036 | -1.450 | 27.12[33] | 75.56[34] | Be(?) | | P | 10.47 |
| AX J1700.2-4220 | 255.105 | -42.316 | 343.8034 | -0.030 | 54[35] | 44.0[36] | | | P | 1.61 |
| GRO J1750-27 | 267.300 | -26.647 | 2.368 | 0.508 | 4.45[37] | 29.806[38] | | | P | 1.23 |
| GS 1843+00 | 281.404 | 0.868 | 33.065 | 1.694 | 29.5[39] | | B0-B2 IV-Ve[40] | $\geq$ 10[41] | P | 3.23 |
| A 1845-024 | 282.048 | -2.426 | 30.395 | -0.404 | 94.8[42] | 241[43] | | | P | 9.66 |
| XTE J1858+034 | 284.673 | 3.437 | 36.820 | -0.066 | 221[44] | 380[45] | | | P | 52.68 |
| 4U 1901+03 | 285.917 | 3.207 | 37.187 | -1.248 | 2.763[46] | 22.58[47] | B0III?[48] | | P | 80.54 |

[1] Stollberg et al. (1993) [2] Kuehnel et al. (2012) [3] Coe et al. (2007) [4] Coe et al. (1994a) [5] Reig (2011) [6] Motch et al. (1997) [7] Motch et al. (1997) [8] Staubert et al. (2011) [9] Staubert et al. (2011) [10] Janot-Pacheco et al. (1981) [11] Janot-Pacheco et al. (1981) [12] Masetti et al. (2006a) [13] Masetti et al. (2006a) [14] Revnivtsev et al. (2005) [15] Corbet and Remillard (2005) [16] Negueruela et al. (2007b) [17] Negueruela et al. (2007b) [18] White et al. (1978) [19] Watson et al. (1981) [20] Jones et al. (1974) [21] Clark and Dolan (1999); Stevens et al. (1997) [22] Huckle et al. (1977) [23] Priedhorsky and Terrell (1983) [24] Parkes et al. (1980) [25] Parkes et al. (1980) [26] Chernyakova et al. (2005) [27] Coleiro et al. (2013) [28] Chernyakova et al. (2005) [29] Kelley et al. (1981) [30] Finger et al. (1996a) [31] Reig (2011) [32] Grindlay et al. (1984) [33] Marshall et al. (2000) [34] in't Zand et al. (2001) [35] Markwardt et al. (2010) [36] Corbet et al. (2010c) [37] Shaw et al. (2009) [38] Shaw et al. (2009) [39] Koyama et al. (1990a) [40] Israel et al. (2001) [41] Israel et al. (2001) [42] Makino and GINGA Team (1988a) [43] Zhang et al. (1996) [44] Takeshima et al. (1998) [45] Doroshenko et al. (2008) [46] Galloway et al. (2005) [47] Galloway et al. (2005) [48] Jones et al. (1974)

Table 1: Catalogue of High-Mass X-ray Binaries.

| Source name | RA | DEC | l | b | $P_s$ | $P_{orb}$ | S.T. | kpc | Type | $F_{17-60}$ |
|---|---|---|---|---|---|---|---|---|---|---|
| IGR J19294+1816 | 292.350 | 18.267 | 53.441 | 0.205 | 12.4[1] | 117.2[2] | Be(?) | | P | 11[3] |
| XTE J1946+274 | 296.410 | 27.366 | 63.206 | 1.399 | 15.8[4] | 169.2[5] | B0-1V-IVe[6] | 8-10[7] | P | 5.38 |
| KS 1947+300 | 297.397 | 30.211 | 66.099 | 2.092 | 18.7[8] | 40.415[9] | B0Ve[10] | 9.5[11] | P | 5.56 |
| EXO 2030+375 | 308.062 | 37.638 | 77.153 | -1.231 | 42[12] | 46.016[13] | B0 Ve[14] | 7.1[15] | P | 88.59 |
| SAX J2103.5+4545 | 315.901 | 45.753 | 87.134 | -0.681 | 358.6[16] | 12.68[17] | B0Ve[18] | 4.5-7[19] | P | 9.19 |
| IGR J22534+6243 | 343.365 | 62.723 | 109.875 | 2.881 | 46.67[20] | | Be[21] | | P | 0.60 |
| **Giant and main sequence systems** | | | | | | | | | | |
| IGR J00370+6122 | 9.286 | 61.386 | 121.242 | -1.468 | 359[22] | 15.663[23] | BN0.5II-III[24] | 3.3 | P | 0.47 |
| 1ES 1210-646 | 183.269 | -64.917 | -61.143 | -2.277 | | 6.7[25] | B2V[26] | 2.8[27] | | 0.83 |
| IGR J21343+4738 | 323.625 | 47.614 | 92.179 | -3.147 | | | B3 V[28] | | | 1.40 |
| 4U 2206+543 | 331.992 | 54.513 | 100.612 | -1.102 | 5559[29] | 9.568[30] | O9.5Vp[31] | 2.6[32] | P | 8.59 |
| **Roche-lobe overflow systems** | | | | | | | | | | |
| Cen X-3 | 170.306 | -60.628 | -67.905 | 0.352 | 4.82[33] | 2.087[34] | O6.5 II-III[35] | 5-8[36] | P | 47.58 |
| **Gamma-ray loud binaries** | | | | | | | | | | |
| LSI 61 +303 | 40.090 | 61.222 | 135.661 | 1.069 | | 26.496[37] | B0Ve[38] | 2.0[39] | | 1.22 |

[1] Rodriguez et al. (2009) [2] Corbet and Krimm (2009) [3] Turler et al. (2009), in the 20-40 keV band [4] Smith and Takeshima (1998) [5] Wilson et al. (2003) [6] Verrecchia et al. (2002a) [7] Verrecchia et al. (2002a) [8] Chakrabarty et al. (1995) [9] Galloway et al. (2004) [10] Negueruela et al. (2003) [11] Tsygankov and Lutovinov (2005b) [12] Parmar et al. (1985) [13] Stollberg et al. (1999) [14] Coe et al. (1988) [15] Wilson et al. (2002) [16] Hulleman et al. (1998) [17] Baykal et al. (2000) [18] Reig (2011) [19] Baykal et al. (2007) [20] Halpern (2012a) [21] Lutovinov et al. (2013a) [22] in't Zand et al. (2007) [23] in't Zand et al. (2007) [24] Reig et al. (2005b) [25] Corbet and Mukai (2008) [26] Masetti et al. (2009) [27] Masetti et al. (2009) [28] Masetti et al. (2009) [29] Reig et al. (2009) [30] Corbet and Peele (2001) [31] Blay et al. (2006) [32] Blay et al. (2006) [33] van der Meer et al. (2007) [34] van der Meer et al. (2007) [35] Day and Tennant (1991) [36] Day and Tennant (1991) [37] Casares et al. (2005a) [38] Casares et al. (2005a) [39] Frail and Hjellming (1991)

Table 1: Catalogue of High-Mass X-ray Binaries.

| Source name | RA | DEC | l | b | $P_s$ | $P_{orb}$ | S.T. | kpc | Type | $F_{17-60}$ |
|---|---|---|---|---|---|---|---|---|---|---|
| PSR B1259-63 | 195.699 | -63.836 | -55.816 | -0.992 | 0.047[1] | 1236.7[2] | B2e[3] | 2.3[4] | P | 0.82 |
| LS 5039 | 276.554 | -14.861 | 16.934 | -1.192 | | 3.906[5] | O6.5Vf[6] | 2.5[7] | | 0.64 |
| **Black-hole systems** | | | | | | | | | | |
| SS 433 | 287.957 | 4.979 | 39.698 | -2.235 | - | 13.1 | A3-7 | 5.5 | BH | 8.12 |
| Cyg X-1 | 299.588 | 35.202 | 71.342 | 3.068 | - | 5.6 | O9.7Iab | 1.86 | BH | 845.64 |
| Cyg X-3 | 308.108 | 40.959 | 79.851 | 0.709 | - | 0.2 | WNe+ | 11.3 | BH | 132.21 |
| **Unclear types** | | | | | | | | | | |
| IGR 08262-3736 | 126.557 | -37.620 | 256.438 | 0.285 | | | OBV[8] | | ? | 0.3[9] |
| IGR J10101-5654 | 152.529 | -56.914 | -77.765 | -0.676 | | | sgB[e][10] | | ? | 0.86 |
| AIGR J14331-6112 | 218.285 | -61.261 | 314.846 | -0.764 | - | - | BIII/BV[11] | - | ? | 0.64 |
| IGR J14488-5942 | 222.180 | -59.704 | 317.234 | -0.129 | - | 49.5(?)[12] | Oe/Be(?)[13] | - | ? | 0.4[14] |
| IGR J17200-3116 | 260.022 | -31.294 | -4.981 | 3.343 | 328.18[15] | | | | P | 1.67 |
| AX J1749.1-2733 | 267.275 | -27.550 | 1.583 | 0.062 | 132[16] | | B1-3[17] | 13-16[18] | ?, P | 1.19 |
| AX J1749.2-2725 | 267.292 | -27.421 | 1.701 | 0.116 | 217[19] | | B3[20] | 14[21] | ?, P | 1.00 |
| IGR J17586-2129 | 269.654 | -21.382 | 7.997 | 1.322 | - | | sgOB[22] | - | ? | 0.2 |
| IGR J18151-1052 | 273.790 | -10.880 | 19.111 | 2.964 | | | OB[23] | | CV(?)[24] | 0.45 |
| IGR J18179-1621 | 274.468 | -16.359 | 14.600 | -0.219 | 11.82[25] | | | | ?, P | $6-12$[26] |
| IGR J19173+0747 | 289.349 | 7.785 | 42.821 | -2.172 | | | early BV[27] | | ? | 0.56[28] |

[1] Johnston et al. (1992) [2] Johnston et al. (1994) [3] Johnston et al. (1994) [4] Negueruela et al. (2011) [5] Casares et al. (2005b) [6] Clark et al. (2001a) [7] Casares et al. (2005b) [8] Masetti et al. (2010c) [9] Bird et al. (2010a),in the 20-40 keV band [10] Coleiro et al. (2013) [11] Masetti et al. (2008b) [12] Corbet et al. (2010b) [13] Coleiro et al. (2013) [14] Bird et al. (2010a), in the 20-40 keV band [15] Nichelli et al. (2011) [16] Karasev et al. (2008) [17] Karasev et al. (2010a) [18] Karasev et al. (2010a) [19] Torii et al. (1998); Karasev et al. (2010a) [20] Karasev et al. (2010a) [21] Karasev et al. (2010a) [22] Coleiro et al. (2013) [23] Burenin et al. (2009) [24] Lutovinov et al. (2012b); Masetti et al. (2013) [25] Halpern (2012b) [26] Bozzo et al. (2012a), in the 20-50 keV band [27] Masetti et al. (2012b) [28] Pavan et al. (2011), in the 20-40 keV band

Table 1: Catalogue of High-Mass X-ray Binaries.

| Source name | RA | DEC | l | b | $P_s$ | $P_{orb}$ | S.T. | kpc | Type | $F_{17-60}$ |
|---|---|---|---|---|---|---|---|---|---|---|
| SWIFT J2000.6+3210 | 300.101 | 32.166 | 68.986 | 1.134 | 1056[1] | | BV/BIII[2] | 8[3] | ?, P | 2.15 |

[1] Morris et al. (2009b) [2] Masetti et al. (2008a) [3] Masetti et al. (2008a)

## 2.3 Corbet diagram

The Corbet diagram (Corbet, 1984), presenting HMXB as a function of spin and orbital periods, is a powerful tool to understand the nature and the evolution of the systems. Figure 5 displays the members of our catalogue for which both orbital and spin periods are available.

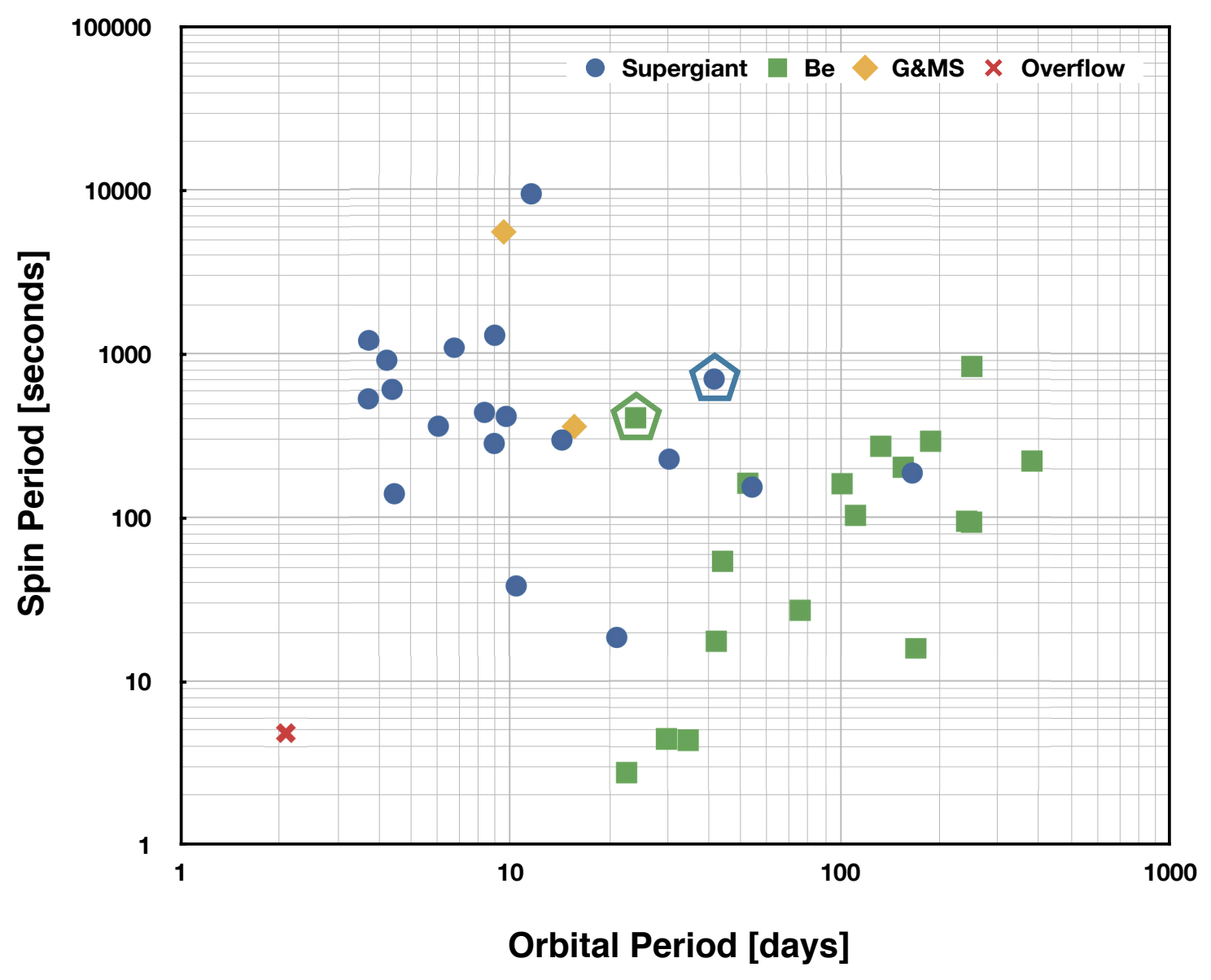

**Fig. 5** The Corbet diagram for the systems in our sample of HMXB with both orbital and spin periods available. Green squares are Be systems; blue circles are super-giant systems; yellow diamond are main sequence systems and the red cross is the Roche lobe overflow system Cen X-3. Pentagons identify systems discussed in the text.

The Be systems (green squares) are well aligned on the usual sequence (Corbet, 1984), excepting for the outlier 1A 1118-615 (green pentagon). Staubert et al. (2011) suggested that the long quiescence time between the outbursts of this system could cause the pulsar to spin down to a period characteristic of wind fed systems.

The super-giant systems (blue circles) have spin periods independent from their orbital periods, as expected for wind accretion. The supergiant with the longest orbital period, IGR J11215-5952 reaches the Be sequence. It features very regular outbursts and it has been suggested to be an evolutionary link with the Be systems (Liu et al., 2011). GX 301-2 (blue pentagon) remains persistently wind-fed by its hypergiant stellar companion, despite of its eccentric orbit. The few super-giant fast X-ray transients which cannot be explained as classical systems appear in two groups that will be further discussed in sect. 3.2: the short orbital period transient systems (for which no spin periods are available) and eccentric systems with orbital periods comparable to that of GX 301-2.

The few giant and main sequence systems (yellow diamonds), lacking emission lines, are in the wind-fed region of the diagram. The only galactic Roche-lobe overflow system Cen X-3 (red cross) has spun-up to very short period.

### 2.4 Expected X-ray luminosity of super-giant systems

The X-ray luminosity of an accreting neutron star (i.e. the mass accretion rate) is determined mainly by the density and velocity of the stellar wind near the compact object. Assuming a smooth stellar wind and a mass to luminosity conversion factor of 0.1 $mc^2$, the range of X-ray luminosities reachable by a system (with a specific companion and wind velocity) depends mostly on the orbital period and eccentricity (see, e.g., Castor et al., 1975; Lamers and Cassinelli, 1999; Vink et al., 2000) as schematized in Fig 6. The main secondary parameter driving the luminosity is the wind velocity. An increase of the terminal velocity by a factor of 3 pushes the red lines in Fig. 6 downwards by a factor of 50, and could explain part of the outlier luminosities.

Persistent systems (i.e. $L_X > 10^{35}$ erg/s) are expected at short orbital periods. Eccentric systems generate variations by factors up to 100 and can appear as transitory. Systems with short orbital periods and reaching low luminosities require a mechanism quenching accretion. Hydrodynamical effects of the neutron star on the stellar wind (Blondin et al., 1991; Manousakis and Walter, 2015a) can generate variability by a factor $> 100$. Intrinsic clumping of the stellar wind (Walter and Zurita Heras, 2007a) or magnetic gating mechanisms (Bozzo et al., 2008c) can have even larger effects.

## 3 Types of High-Mass X-ray Binaries

### 3.1 Persistent super-giant systems

*INTEGRAL* discovered 13 new persistent sgHMXB in addition to the ten classical wind-fed systems previously known in the Galaxy. Six of them, featuring absorbing column densities persistently $\gtrsim 10^{23}$ cm$^{-2}$, are known as "obscured systems". The classical systems also display strong absorption close to eclipse but are less absorbed ($N_H \sim 10^{22}$ cm$^{-2}$) at the inferior conjunction. Obscured and classical systems are very similar and the distinction between them is mostly due to the fact that the former were first identified at hard X-rays. One of the obscured system, IGR J16318-4848, is peculiar and deserves a special category. Note that several SFXTs (see section 3.2) turn out to be classical systems as well.

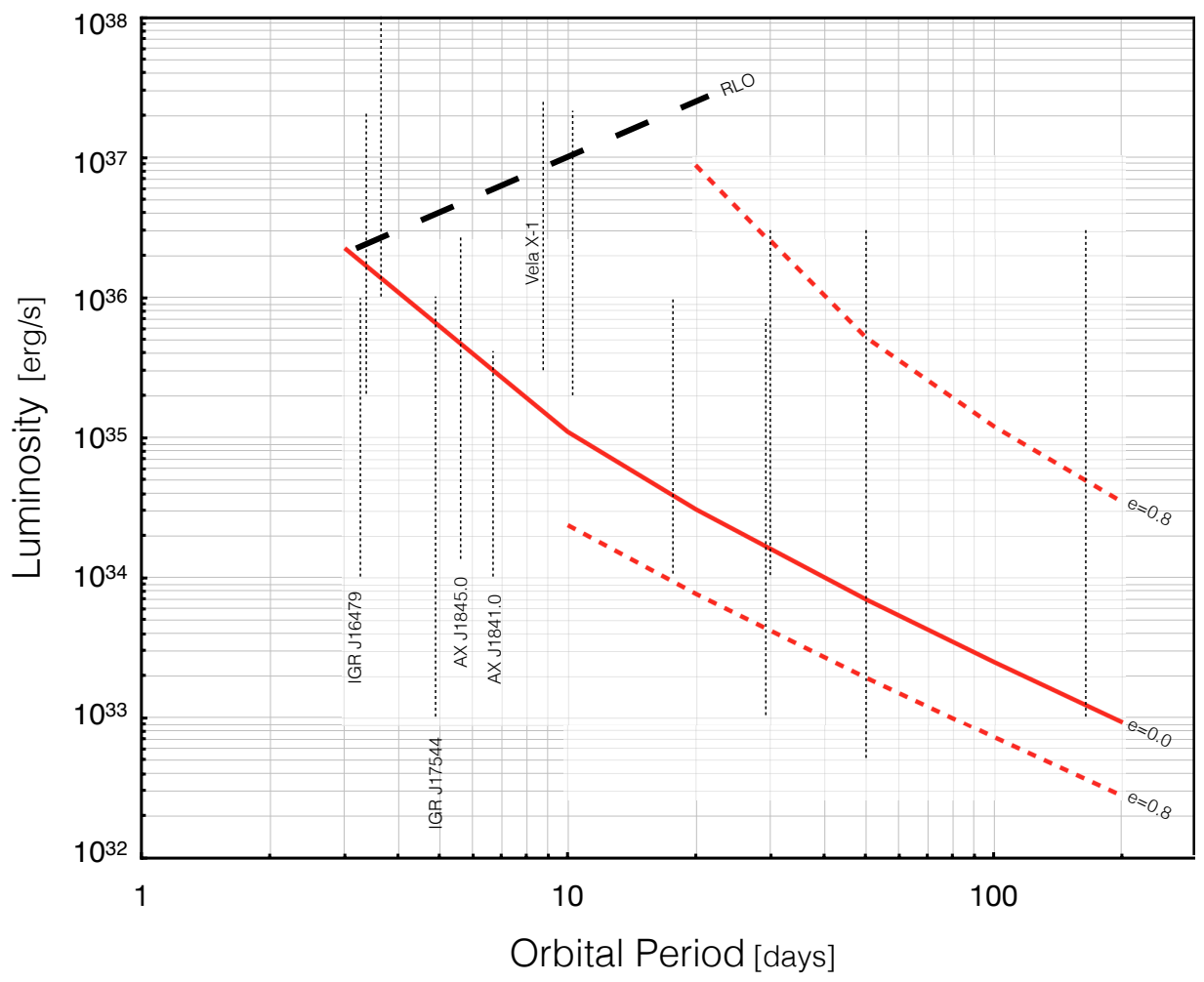

**Fig. 6** X-ray luminosity expected from a sgHMXB for a smooth stellar wind ($V_\infty = 10^3$ km/s, $\beta = 0.8$, $\dot{M} = 10^{-6}$ $M_\odot$/y, $M_\star = 20$ $M_\odot$) as a function of the orbital period for eccentricities of 0 (continuous line) and 0.8 (short dashed lines). The long dashed line indicates the Roche-lobe overflow limit. The range of observed variability (minimum and maximum connected by dotted lines) is indicated for a number of sources discussed in this review.

### *3.1.1 Classical Super-giant Systems*

Several of the classical sgHXMB are bright enough to allow long and meaningful lightcurves to be obtained at hard X-rays:

- Vela X-1 is the prototype of the classical sgHMXB. It has been observed continuously by *INTEGRAL* with several orbits at high temporal resolution. Its 17-80 keV luminosity (outside of eclipses) varies in the range $(0.6-25)\times10^{36}$ erg/s for an average of $1.4\times10^{36}$ erg/s. The brightest flares are short (down to 0.5h) and sequences of flares, separated by $\sim$ 2h have been observed. The pulsed fraction does not vary significantly during the flares, indicating that the mass inflow rate through the accretion column varies considerably. The flare rate is decreasing smoothly with luminosity (Fig. 7) suggesting that the variability is driven by a single mechanism. Low luminosities are observed during short (fraction of an hour) periods. Even if they have been named "off-states", accretion goes on but at a reduced rate (during the five off-states presented by Kreykenbohm et al.

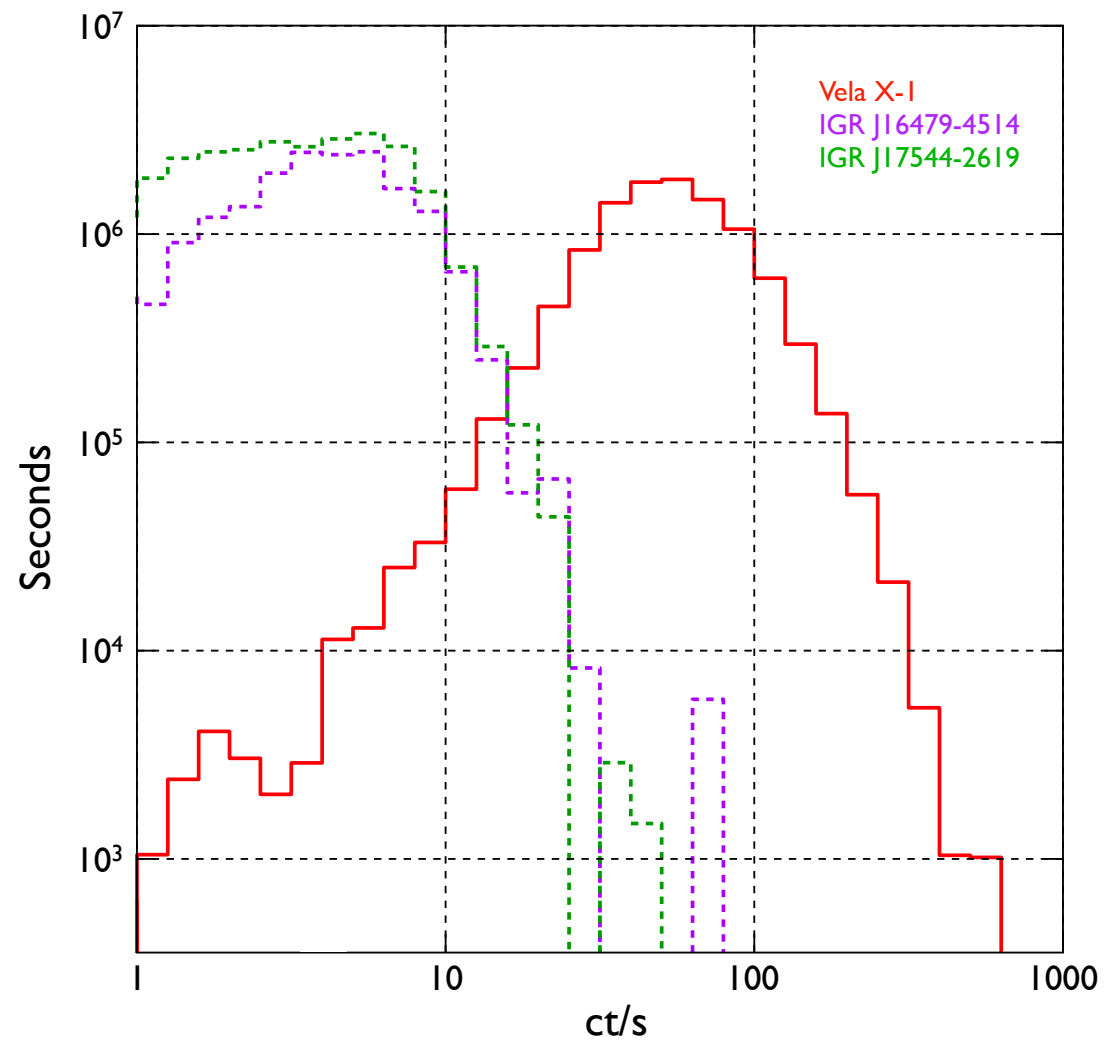

**Fig. 7** Histogram of the effective time during which a given count rate is observed for Vela X-1 (*INTEGRAL/ISGRI* 17-80 keV). Eclipses have been removed and the distribution has been corrected for the statistical noise. The total exposure time is 11.8 Msec; time bins span from 24 to 720 sec depending on the source strength; the average count rate statistical uncertainty is 4 ct/s. The distributions for two bright SFXTs are shown as well.

(2008a), the average *INTEGRAL/ISGRI* count rate was $10 \pm 0.7$ ct/s, i.e. $3 \times 10^{35}$ erg s$^{-1}$). Fig. 7 indicates that the luminosity distribution extends smoothly towards low values before slightly bending up, suggesting that a distinct variability mechanism is required. Suzaku observations confirmed that this bending is indeed related to the "off-states" (Doroshenko et al., 2011).

- 4U 1700-37 is characterised by very short flares (with duration down to 250 sec) reaching $\gtrsim 10^{37}$ erg/s. *XMM-Newton* observed it in quiet state at $2 \times 10^{35}$ erg/s (van der Meer et al., 2005). Its luminosity distribution follows an asymmetric log normal, peaking at $10^{36}$ erg/s (Fig. 8).
- The variability of OAO 1657-415 (Fig. 8) is shaped as an highly asymmetric log-normal distribution. Periods of enhanced activity are very long (10 to 120 days, i.e. 1-12 orbits) reaching $\gtrsim 10^{37}$ erg/s. Periods of low activity ($\lesssim 2 \times 10^{35}$ erg/s) are also relatively long (several days). The variability is dominated by stellar wind density/velocity variations that extend over the complete orbit ($\sim 2R_\star$) and varies on time scales of months or by low velocity clumps corotating with the neutron star. It is interesting to note that the companion is a peculiar O star, (possibly a Wolf-Rayet), that can generate highly structured winds.
- Fig. 8 shows clearly that an additional component is required in GX 301-2 to explain its high flux activity: the dense accretion stream forming close to periastron. Short flares (fraction of an hour) are superimposed. In about half of the orbits long secondary flares can be observed during the less active part of the orbit, indicating that the tidal stream generates a spiral structure.

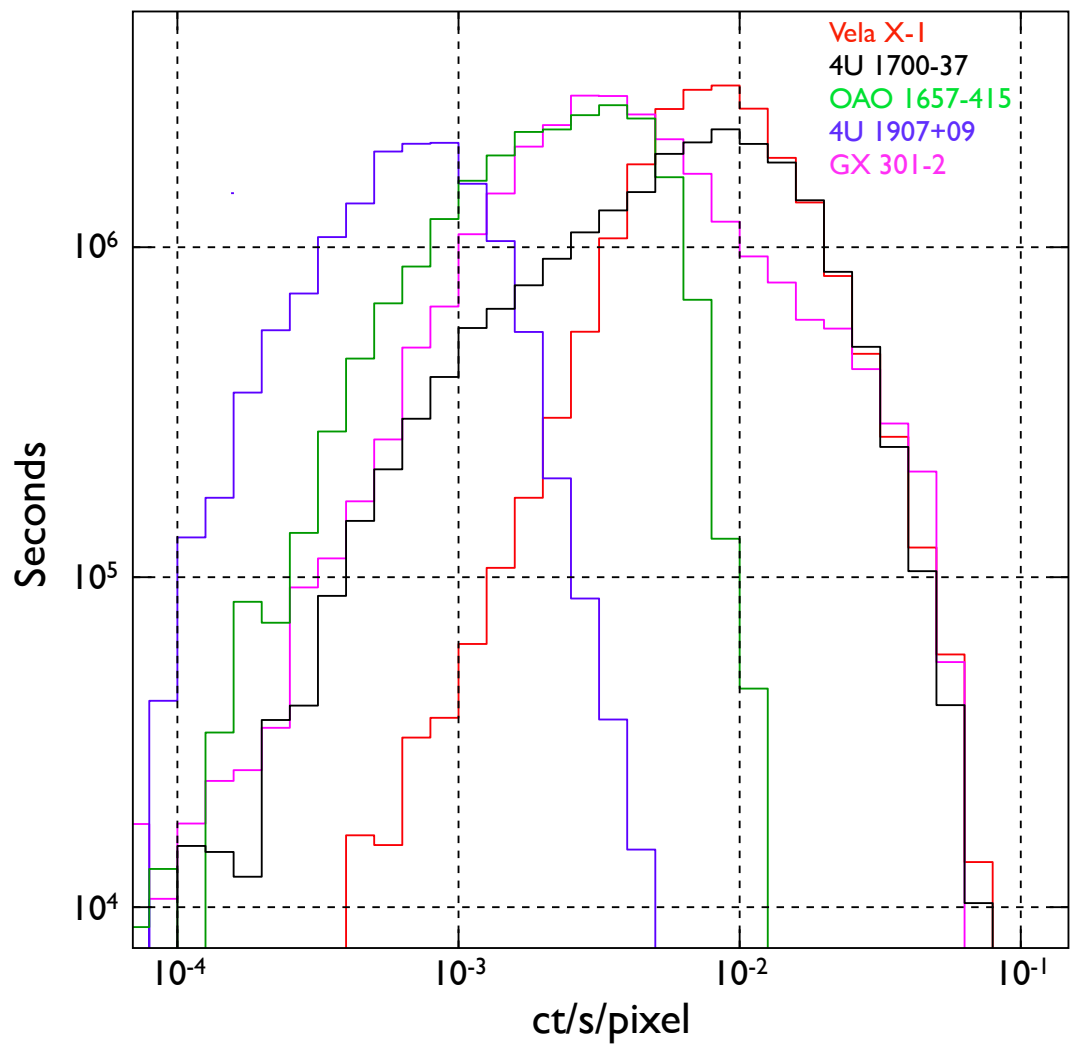

**Fig. 8** Histogram of the effective time during which a given count rate is observed for four classical HMXB (*Swift/BAT*). Eclipses have been removed. The right-and side of the distribution is much steeper for OAO 1657-415, which is dominated by very long activity periods rather than by narrow spikes as observed in the other sources.

Short "off-states" have been observed in Vela X-1, GX 301-2 (Göğüş et al., 2011) and 4U1907+09 (Doroshenko et al., 2012a). The "off-states" of 4U1907+09 are particularly frequent ($\gtrsim 20\%$ duty cycle) and are missing close to periastron. Fig. 8 indicates that classical sgHMXB, such as 4U 1700-37 and OAO 1657-415 feature an intense activity at low count rate, similar to that observed in 4U 1907+09, and that Vela X-1 is more rarely is such a state.

*Variability amplitude and off-states*

The variability of the accretion rate by a factor $10 - 100$ in wind-fed systems in circular orbits was successfully explained by hydrodynamical simulations (Blondin et al., 1990). When the system is close to Roche-lobe overflow, the tidal stream further increases the wind density in the direction of the compact object (Blondin et al., 1991), explaining enhanced variability in eccentric systems (such as GX 301-2). Photoionisation of the wind by the compact object also generates wind inhomogeneities in the form of additional streams (Blondin, 1994) and obscuration at late orbital phases. Large and rapid variations of the mass accretion rate have interesting consequences for the formation of the hard X-ray spectrum that can be probed on short time scales with *NuStar* (Fürst et al., 2014b). A number of explanations was put forward to explain variability factors as large as $\sim 10^3$ in classical sgHMXB and in particular in Vela X-1 (Fig. 8):

- *Wind clumping:*

  Line driven instability can in principle generate huge density variations in the stellar wind of massive stars but the details and the geometry are not yet understood. Besides multiple observational evidence (Bouret et al. 2005; Fullerton et al. 2006; Prinja & Howarth 1986; L'epine & Moffat 1999; Markova et al. 2005; Lupie & Nordsieck 1987; Davies et al. 2007; Cassinelli & Olson 1979; Oskinova et al. 2006), wind clumping is still poorly constrained. If huge density variations can in principle be accounted for by wind clumps (in't Zand, 2005a; Walter and Zurita Heras, 2007b), it is unclear if the density contrasts will propagate to the magnetosphere, how clumping interplays with the hydrodynamic effects in the wind induced ba the presence of the compact object and if a reasonable clump model can generate the observed luminosity distribution (Fig. 8).

- *Hydrodynamics:*

  Manousakis and Walter (2015a) have included the effect of photo-ionisation on the wind acceleration in the hydrodynamical model of Vela X-1. Even with a very simplified treatment, the model allowed to probe the dynamics of the region surrounding the neutron star and in particular the collision between the primary stellar wind, slowed down by photo-ionisation and flowing outwards and a gas stream flowing inwards from the tidal stream towards the neutron star. A shock front is generated and moves inwards and outwards regularly creating low density bubbles expanding to $\sim 10^{11}$ cm before crashing on the accretion radius. This "breathing" mechanism generates instantaneous accretion rates 10 times lower than predicted previously, global luminosity variations by a factor of $10^3$, and transient modulations with a characteristic time-scale of $\sim$ 6500 sec (for the geometry of Vela X-1). Interestingly such transient modulations have been detected in Vela X-1 (Kreykenbohm et al., 2008a). The model predicts a luminosity distribution that is slightly too narrow when compared to the observations. The identification of a mechanism that can explain both the observed variability and quasi-periods is, however, encouraging.
  Shakura et al. (2013) have shown that two regimes of subsonic accretion are possible at the boundary of the magnetosphere depending on whether or not the plasma is cooled by Compton processes (high vs low accretion rate). The different cooling times determine the fall-down velocity i.e. the accretion rate at the boundary of the magnetosphere. At low luminosity the X-ray photons are directed perpendicular to the neutron star surface, inverse Compton cooling is less efficient and a change of the pulse profile could be observed (Doroshenko et al., 2011). This mechanism increases the luminosity ratio produced by an externally driven mass accretion variability.
These two mechanisms will work together. The breathing mode that occurs high above the magnetosphere will be amplified by the change of geometry of the accretion column and of the cooling mechanism. The amplification might not be so effective, nor needed, if the seed density variations are strong enough.

- *Magnetic gating:*

  Doroshenko et al. (2011) have investigated the possibility for the variability of Vela X-1 to be generated by Kelvin-Helmoltz instability at the magnetospheric boundary, leading to "magnetic gating" of the accretion (opening and closing the gate) (Bozzo et al., 2008c). The required magnetic field of $(2-10) \times 10^{13}$ G can in principle be accommodated if the CRSF would be generated close to the top of the accretion column at high flux level. However, *NuStar* observations (Fürst et al., 2014c) recently revealed that the CRSF harmonic energy is correlated to the X-ray luminosity down to $10^{36}$ erg/s (this is not the case for the fundamental), which was interpreted with a surface magnetic field of $2 \times 10^{12}$ G. The spectrum of an off-state of Vela X-1, presented in the same paper, did not show any CRSF possibly pointing to a higher magnetic field, but the relatively low signal to noise obtained is not yet conclusive.

The X-ray variability of classical sgHMXB systems is complex but most of the behaviour seems to be reproducible by hydrodynamical effects (even if this has not been done effectively for all systems). It is not obvious that additional physical mechanisms such are clumping or magnetic gating are required to explain the observations. OAO 1657-415 features variability on very long ($>> P_{orb}$) time scales that can only be related to global wind structures but these variations have not been studied in detail so far.

### *3.1.2 Obscured super-giant systems*

The five super-giant HMXBs featuring persistently high obscuration ($\gtrsim 10^{23}$ cm$^{-2}$) harbour pulsars orbiting in 3.7-9.7 days around O8-B1 companions. In three of them (IGR J16393-4611, IGR J16418-4532, IGR J18027-2016), the orbital periods are very short (<4.4 days) and the pulsars orbit close to the surface of their companion stars. Two classical sgHXMB have similarly short periods:

– 4U1700-37: *EXOSAT* spectra obtained along the orbit have shown a phase dependent absorbing column density with a minimum $\sim 0.5 \times 10^{23}$ cm$^{-2}$ (Haberl et al., 1989). High Fe K$\alpha$ equivalent width and important scattered and soft X-ray excess emission (van der Meer et al., 2005) indicate that the absorbing column density was underestimated. A minimum value of $\sim 2 \times 10^{23}$ cm$^{-2}$ was reported, matching our definition for an obscured source.

– 4U 1909+07: A low absorption was reported but the spectrum is fairly complex showing an Iron K$\alpha$ line and a soft X-ray excess. The value of the absorbing column density in this object is not settled but the combined spectrum built from *RXTE, INTEGRAL* and *BAT* data is reasonably represented with $N_H \sim 1.3 \times 10^{23}$ cm$^{-2}$ and an Fe K$\alpha$ equivalent width of 100 eV (Fig. 9).

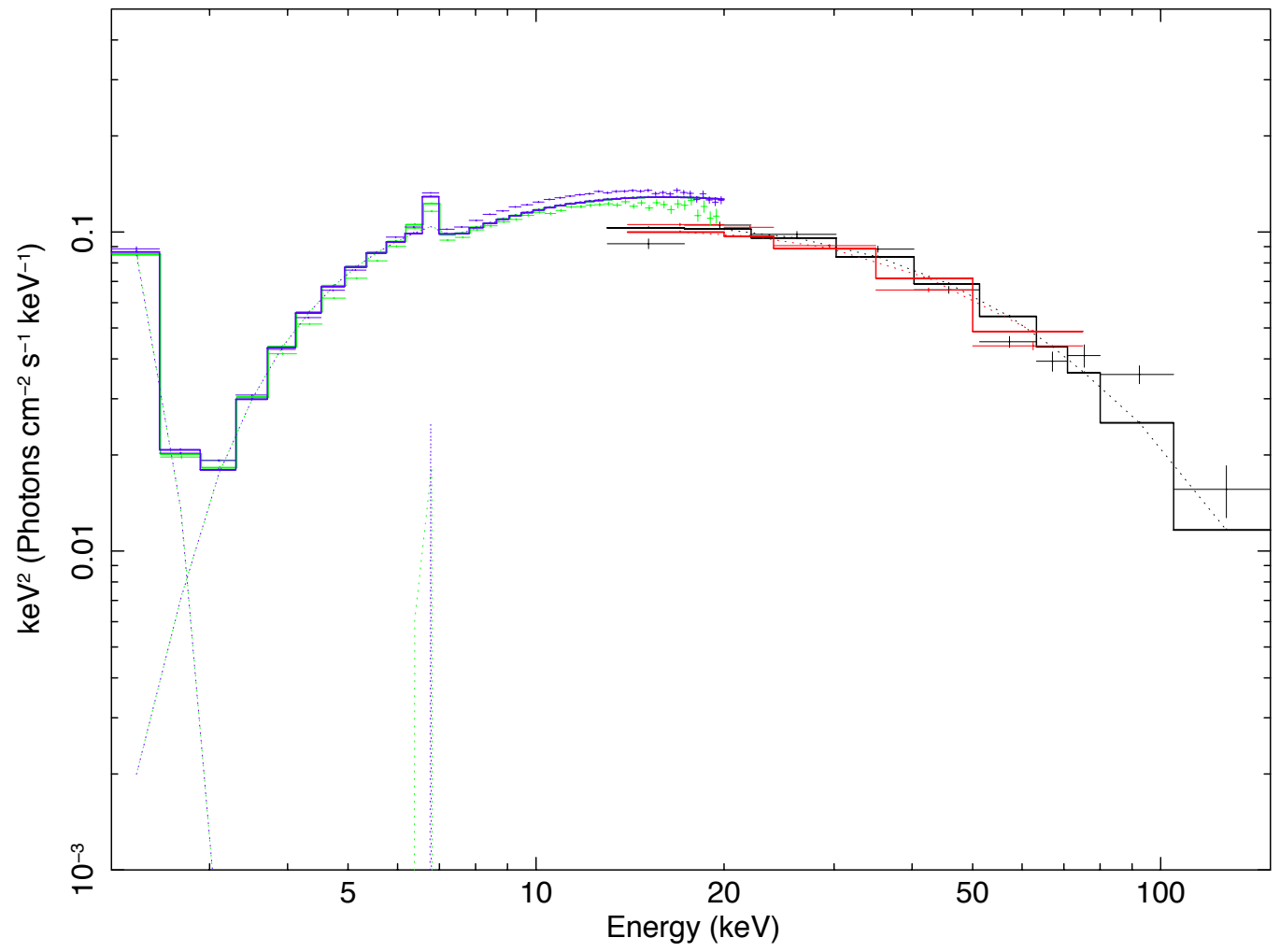

**Fig. 9** *RXTE PCA, INTEGRAL/ISGRI* and *Swift/BAT* spectra of 4U 1909+07 fitted with an absorbed cutoff powerlaw plus Fe K$\alpha$ and soft X-ray excess.

It is therefore plausible to assume that all persistent systems with $P_{orb} < 5$ days become obscured and are in transition towards Roche lobe overflow. As a matter of fact they all have L$\gtrsim 10^{36}$ erg/s with IGR J16418-4532 reaching up to $10^{38}$ erg/s (see also sect. 3.2). The two remaining obscured systems have longer orbital periods ($\approx$ 10 days) and other explanations have been found for their obscuration:

– EXO 1722-363: Comparison of observations with hydrodynamic simulations indicate that the large absorbing column density and its variability with orbital phase can be understood if the wind terminal velocity is low and if the neutron star is massive enough ($> 1.8$ M$_\odot$) to strongly perturb the stellar wind (Manousakis et al., 2012b).

– IGR J16320-4751: The absorbing column density is pretty constant at $\approx 10^{23}$ cm$^{-2}$ (the exact value is model dependent; Rodriguez et al., 2006) along the orbit and increased to $5 \times 10^{23}$ cm$^{-2}$ at one occasion (Zurita-Heras et al., 2009). The infrared reddening towards IGR J16320-4751 is exceptional and significantly larger than what can be expected from the 21 cm measurements (Chaty et al., 2008). This indicates the presence of large amount of dust in the vicinity of the source that can explain a fraction of the constant X-ray obscuration. IGR J16320-4751 might well be a classical system obscured by the environment and not by intrinsic processes.

Obscured sgHMXBs can therefore be understood as classical systems in transition to Roche lobe overflow or with relatively low velocity winds. As the neutron stars can cut-off wind acceleration via ionisation (Stevens and Kallman, 1990), the wind can be slower in binaries than in isolated stars. For instance GX 301-2, 4U 1907+09 and EXO 1722-363 feature high absorbing column densities and low wind terminal velocities of 500, 1000 and 600 km/s (Kaper et al., 2006; Kostka and Leahy, 2010; Manousakis et al., 2012b). Even the companion of Vela X-1 has a wind terminal velocity less than what would be expected from its high luminosity (Kudritzki and Puls, 2000). Once the companion is close to overflowing its Roche lobe, deep spiral-in is unavoidable (van den Heuvel and De Loore, 1973) and results in a Common Envelope phase (Taam et al., 1978).

Obscured systems account for $\sim 20\%$ of the persistent sgHMXBs detected at hard X-rays. This suggests that the systems remain, on average, for about 20000 years close to Roche lobe overflow.

#### *3.1.3 IGR J16318-4848*

IGR J16318-4848, the most obscured persistent sgHMXB, is almost Compton thick with an absorbing column density varying in the range $(1.2-2.2) \times 10^{24}$ cm$^{-2}$ (de Plaa et al., 2003; Ibarra et al., 2007). The X-ray absorption is much larger than that of the infrared counterpart (Revnivtsev et al., 2003). Walter et al. (2003) and Barragán et al. (2009) did not find any significant Fe K$\alpha$ Compton shoulder indicating that the absorbing column density averaged isotropically is several times lower than observed on the line of sight. IGR J16318-4848 has been detected continuously with *INTEGRAL/ISGRI* and *Swift/BAT* for more than 10 years. During this period, the hard X-ray luminosity, averaged over two months or over a year, has shown variability by a factor of only three, respectively two, around an average value of $10^{35}$ erg/s. This corresponds to the typical behaviour of a classical system with a close to circular orbit and excludes scenarios involving a high eccentricity or a Be system. Flares and low flux states reaching 100 mCrab, and $< 2$ mCrab, respectively, are observed on time scales of some days very regularly. No period is detected.

Walter et al. (2006a) suggested that the compact object is orbiting within the dense equatorial outflow of its B[e] super-giant companion. The thickness of the disk was evaluated as $\sim 0.7$ $R_*$ (Chaty and Rahoui, 2012) and densities $> 10^{13}$ cm$^{-3}$ are mentioned in such disks (Levesque et al., 2014), which would correspond to a Hydrogen column density through the disk of $\sim 10^{24}$ cm$^{-2}$. If this interpretation is correct the inclination angle of the system on the line of sight should be $\sim 15$ degrees to explain the absence of a Compton shoulder. Such a geometry does not generate any eclipse if the orbital period is $\gtrsim 40$ days. Thanks to the high density wind, the accretion rate on the compact object remains large enough even far away from the companion star. The variability is probably related to hydrodynamic instabilities that the compact object will not fail to be produce. The fate of IGR J16318-4848 is unclear. Chaty and Rahoui (2012) estimated the size of the infrared emitting disk to $\sim 70$ $R_*$. If the compact object orbits in the external regions of that disk, the system may end up in a BH/NS binary (Taam and Sandquist, 2000).

### 3.2 Super-giant fast X-ray transients

Super-giant fast X-ray transients (SFXTs) were identified as a new class of sources in 2005 (Sguera et al., 2005; Negueruela et al., 2006b) thanks to the long term monitoring program of the Galactic plane carried out with *INTEGRAL*. These hard X-ray transients produce short and bright flares with typical durations of a few hours and peak fluxes of few tens to hundred mCrab (in the energy band ∼20-100 keV). Given the short and sporadic nature of these events, the large field of view of the *IBIS/ISGRI* imager on-boad *INTEGRAL* proved to be particularly well suited to search for SFXT sources (Sguera et al., 2006a; Walter and Zurita Heras, 2007a). So far, about 15 objects have been identified among the SFXTs (Falanga et al., 2011). Outside the short bright events, these sources are hardly detectable with *INTEGRAL*. Their average persistent X-ray flux is a factor of $\sim 10^2$-$10^5$ lower than the one at the peak of the bright flares. This is much below the sensitivity level of any presently available large FoV X-ray instrument, and thus deep pointed observations with focusing high sensitivity X-ray telescopes are required to study their persistent emission (e.g., *XMM-Newton, Chandra, Suzaku, Swift/XRT*; Romano et al., 2009b; Sidoli et al., 2008; Romano et al., 2010c; Bozzo et al., 2010; Sidoli et al., 2010; Bodaghee et al., 2011a; Bozzo et al., 2012b; Sidoli et al., 2013a).

Since 2005, SFTXs have been monitored regularly in the X-rays and a relatively large effort was devoted to perform observations of these sources in different energy domains, spanning from the far IR to $\gamma$-rays (Walter, 2007) and up to the very high energies (∼GeV; Sguera et al., 2009, 2011). It was soon understood that all SFXT systems were hosting a compact object accreting from the wind of a massive companion, typically a super-giant O-B star (Tomsick et al., 2006a, 2008; Chaty et al., 2008; Masetti et al., 2008b; Negueruela et al., 2008b; Tomsick, 2009; Tomsick et al., 2009b; Chaty, 2010; Bodaghee et al., 2012a). SFXTs were thus classified as a subclass of wind-accreting super-giant X-ray binaries. Accurate spectroscopic classifications of super-giant stars in SFXTs made it possible to establish in a few cases the mass and radius of the star, together with its wind properties (i.e. mass loss rate and terminal velocity; see, e.g., Rahoui et al., 2008a, and references therein). The detection of X-ray pulsations in a few sources, with periods ranging from few up to thousand seconds, led to the conclusion that compact objects in SFXTs should be relatively young neutron stars, with magnetic field (at least) as high as $10^{11}$-$10^{12}$ G (Grebenev and Sunyaev, 2007a; Bozzo et al., 2008c; Sguera et al., 2010). In several SFXTs, long term observations carried out with *INTEGRAL* and *Swift* also permitted to measure their orbital periods. Reported values span from 3.3 up to 57 days, the only exception being the source IGR J11215-5952 with an orbital period of ∼168 days (see Table 1). The similarity between sgHMXBs and SFXTs in terms of constituent stars and orbital properties make it difficult to understand the peculiar behaviour displayed by the latter in the X-ray domain (Bozzo et al., 2013).

A large number of X-ray flares has been recorded so far from the SFXTs and thus the flaring state of these sources is known in fairly good details (see, e.g., Romano et al., 2013, for recent reviews). *INTEGRAL* and *Swift* observations permitted to carry out broad band spectral analysis of these events and it is now established that flares can occur at different luminosity levels, spanning from a few times $10^{35}$ to $10^{37}$ erg/s. The brightest flares (peaking at $>10^{36}$ erg/s) are sometimes called "outbursts" to distinguish them from the lower luminosity events. In four sources flares and outbursts showed evidence of clustering at preferred orbital phases. In the other sources they have been detected at any time during the neutron star revolution around the companion.

The spectral model generally used to fit X-ray spectra of the flares is an absorbed cutoff power-law (e.g. Romano et al., 2011a). The measured parameters are on average very similar to those observed in other classes of young accreting X-ray pulsars: (i) the absorption column density is higher than the Galactic value in the direction of the source due to locally distributed dense material from the stellar wind; (ii) the power-law photon index ranges from 0.5-2.0; (iii) the cutoff energy (if any) is between 10 and 30 keV (Sidoli et al., 2009b,c; Ducci et al., 2010). Some flares are accompanied by remarkable increases in the absorption column density, indicative of possible local enhancement in the accreting material around the compact object. Many flares, however, do not show such a feature and are accompanied by relatively modest variations (if any) in the spectral photon index. Thermal spectral components during SFXT flares are rare, at odds with other classes of highly accreting neutron stars. So far, the best examples are these of IGR J08408-4503 (Sidoli et al., 2009a) and AX J1845.0-0433 (Zurita Heras and Walter, 2009a), where prominent black-body spectrum were observed with temperature and emission radius comparable to those expected for a hot spot on the neutron star surface, similar to what is detected in other classes of accreting X-ray pulsars. Long term observations with wide field instruments also permitted to accurately investigate the duty cycle of SFXTs. The general finding is that these sources spend only a small fraction of their time ($\lesssim$5-10%) in the flaring states (i.e. at luminosities $\gtrsim 10^{35}$ erg/s; Paizis and Sidoli, 2014), and on-average display a much lower persistent luminosity, that ranges from $10^{32}$ (very low state) to $10^{34}$ erg/s (intermediate state).

In contrast with the flaring state, the intermediate and the very low luminosity states of the SFXTs are still poorly known. In these states, the low X-ray luminosity of the SFXTs implies that deep pointed observations lasting several tens of ks (typically about 10-30 ks with *XMM-Newton*) are required to measure accurately the spectral properties and investigate their time variability with sufficient accuracy. Such long integration times challenge our understanding of processes occurring on the most relevant time scales that are comparable to dynamical processes occurring close to the neutron stare magnetosphere and typically range from a few to hundred seconds. These observations are usually also limited in time to a maximum total exposure time of $\ll$100 ks per source, and they can only probe a relatively small fraction of the neutron star orbit around the companion. The picture that was achieved so far of the low emission states of SFXTs thus remains fragmented.

A XMM-Newton observation of IGR J16479-4514 revealed in 2008 that part of the X-ray variability of this source was due to an extended X-ray eclipse, lasting about 0.6 day (Bozzo et al., 2008d). X-ray eclipses were later discovered in IGR J16418-4532 (Drave et al., 2013) and possibly in IGR J16207-5129 and IGR J17354-3255 (Bodaghee et al., 2010; Ducci et al., 2013b). *XMM-Newton* and *Suzaku* observations of XTE J1739-302, IGR J17544-2619, IGR J16328-4726, and IGR J08408-4503 revealed the presence of pronounced X-ray variability also during "quiescence". This variability comprises small flares that occur on the same time scales as the brightest outbursts but reaches peak luminosities that are a factor of $10^2$-$10^3$ lower. Some of these flares are also accompanied by modest changes in the spectral slope and/or in the value of the local absorption. X-ray dips have been observed in two sources (Bozzo et al., 2012b; Drave et al., 2013). Due to the relatively low statistics of the corresponding data, their nature is still debated but they seem to have a different origin with respect to dips usually observed in low mass X-ray binaries. The latter are usually ascribed to the presence of geometrically thick material on the rim of the accretion disk surrounding the neutron star (Kuulkers et al., 1998). Soft spectral components, dominating the X-ray emission at energies $\lesssim$2 keV, have been detected in SFXTs much more often in quiescence than during flaring states (Zurita Heras and Chaty, 2009a; Bozzo et al., 2010; Sidoli, 2010). In the case of XTE J1739-302 and AX J1845.0-0433 these components have been mainly ascribed to the reprocessing of the X-ray emission from the neutron star by the surrounding wind material (Hickox et al., 2004), but in the case of IGR J08408-4503 it was argued that the soft X-ray emission could have been produced within the super-giant wind itself. The temperature and luminosity of this component resembled, indeed, that of close-by isolated super-giant stars (see, e.g., the case of $\zeta$-Puppis; Nazé et al., 2012, and references therein). If confirmed, this would suggest that accretion during the lowest luminosity periods displayed by some SFXTs might be strongly inhibited: X-ray observations of SFXTs in these states could then be used to directly probe the properties of their super-giant companions' wind.

Interesting spectral and timing behaviours have thus been revealed from "quiescent" SFXTs (i.e. outside the flaring states), but it is not clear yet if such phenomena occur in all sources or if they are peculiar of a few specific systems. As we argue later in this section, the latter seems so far the most reliable conclusion and thus SFXTs might need to be divided in a number of sub-classes.

Early models proposed to interpret the peculiar X-ray variability of the SFXTs ascribed the fast flaring behaviour to the presence of very pronounced eccentricities coupled with inhomogeneous super-giant winds (Negueruela et al., 2008b; Chaty, 2010). This hypothesis was severely challenged already in 2009 when short orbital periods were measured in a few SFXTs (e.g., IGR J16479-4514 and IGR J17544-2619). As these systems display similar properties as those with much longer orbital periods (e.g. XTE J1739-302; Drave et al., 2010), it is unlikely that the separation between the neutron star and the companion is playing a central role in triggering the SFXT variability. In analogy with classical HMXBs, different possibilities have been considered to explain the SFXT behaviour, including large inhomogeneities in the wind ("clumps"), magnetic/centrifugal gates due to the magnetic field and rotation of the neutron star, and hydrodynamical effects. At odds with the classical HMXBs, all these possibilities require extreme values of the involved parameters in order to match the SFXT dynamical range in the X-ray luminosity. The orbital characteristics of these sources mostly affect the way in which different effects combine to give rise to the pronounced variability (see Sect. 4). We summarized in Table 2 the most relevant properties of all known confirmed and candidate SFXT sources. By taking advantage of all information published in the past ~10 years on these sources, we organized the SFXTs in the four sub-groups listed below.

1. *Classic-like systems:*

   These are variable systems behaving very much like classical sgHMXB:
   - IGR J16418-4532 is the most distant SFXT and the only one persistently detected above $10^{36}$ erg/s and reaching the Eddington luminosity during flares. This suggests that the system is close to Roche lobe overflow (Sidoli et al., 2012). The absence of strong orbital modulation indicates that its transient nature is likely related either to some hydrodynamic properties of the accretion stream (Manousakis et al., 2012b) and/or that a temporary accretion disk might form around the neutron star (see also Ducci et al., 2010, and references therein). The range of luminosity is similar to that observed in Vela X-1.

   - The five sources IGR J17354-3255, IGR J16207-5129, IGR J16328-4726, AX J16195-4945, and IGR J16465-4507, with intermediate orbital periods, feature a low variability amplitude with flares reaching a flux a few tens of times the average source level. The hard X-ray luminosities are in the range $10^{35-36}$ erg/s. These characteristics are comparable to those of some classical sgHMHB.

**Table 2** Super-giant fast X-ray transient candidates: variability of the hard X-ray luminosity and orbital parameters. The smallest luminosity given is an average value, lower instantaneous values are observed. The maximum luminosity corresponds to the peak of a flare.

| Source name | $P_{\rm orb}$ days | e | $L_{10keV}$ $10^{35}$erg/s | Remark |
|---|---|---|---|---|
| **Classic-like systems** | | | | |
| IGR J16418-4532 | 3.7 | - | $10-1000$ | Short flares, obscured, variations similar to these of Vela X-1. |
| IGR J17354-3255 | 8.4 | - | $2-20$ | The variability amplitude is less than a factor 100, excepting a deep minimum at a specific orbital phase, that could be an eclipse. |
| IGR J16207-5129 | 10 | - | $0.6-5$ | Low variability amplitude. |
| IGR J16328-4726 | 10 | - | $0.8-8$ | Low variability amplitude, only a few flares detected. A distance of 5 kpc was assumed. |
| AX J16195-4945 | 16 | - | $0.5-3$ | Low variability amplitude, flares are short, quiescent state at L$\approx$ 0.1. |
| IGR J16465-4507 | 30 | - | $0.1-7$ | Low variability amplitude. |
| IGR J11215-5952 | 168 | large | $0.01-30$ | Flares are long (days) and occur at periastron. Similar to GX 301-2. |
| **Fast transients reaching anomalously low luminosities** | | | | |
| IGR J16479-4514 | 3.3 | - | $0.1-6$ | Flares are short (hours) and frequent ($\sim$week). |
| IGR J17544-2619 | 5 | - | $0.1-30$ | Flares are clustered in orbital phase, rare low states with L$<$ 0.01 have been reported. |
| AX J18410-0536 | 6.4 | - | $0.1-10$ | Flares are short (hours). |
| AX J18450-0433 | 5.7 | <0.4 | $0.1-15$ | Flares are short and frequent ($\sim$weeks). |
| **Eccentric transients** | | | | |
| IGR J18483-0311 | 18 | 0.4 | $0.6-10$ | Flares are short and clustered in orbital phase. The luminosity far from periastron is $\sim$0.1. |
| SAX J18186-1703 | 30 | 0.4 | $0.01-30$ | Flares are short and clustered in orbital phase. L<0.001 has been measured once at apastron, an outlier value. |
| XTE J1739-302 | 51 | <0.8 | $0.07-30$ | Flares are short and frequent, not clustered in orbital phase. Minimum luminosity $\sim$ 0.005. |
| **Unclear systems** | | | | |
| IGR J08408-4503 | | - | $0.03-50$ | Flares are short, very rare ($\sim$year) and structured. |
| IGR J18462-0223 | | - | $0.04-3$ | Flares are short and very rare. |
| AX J18205-1434 | 54 | - | $0.2-2$ | Low variability amplitude, unknown distance, possibly not a super-giant. |

– The source IGR J11215-5952 is the only one displaying a long period of activity at periastron, most likely related to its anomalously large orbit and eccentricity (for an HMXB). The duration of the pronounced activity at periastron is much shorter than that usually observed from Be-systems, thus suggesting that accretion is never mediated through a stable accretion disk. The variability properties of IGR J11215-5952 more likely resemble those of GX 301-2, a classical sgHMXBs displaying a remarkably peaked X-ray activity around periastron.

2. *Fast transients reaching anomalously low luminosities:*

The four sources IGR J16479-4514, IGR J17544-2619, AX J18410-0536, and AX J18450-0433 display short orbital periods and large variability with average and minimal luminosities ($\lesssim 10^{34}$ erg/s) and typical flare luminosities $\gtrsim 10$ times lower (see Fig. 7) than expected in classical systems with such orbits (Oskinova et al., 2012).

– IGR J16479-4514: Sidoli et al. (2013b) analysed a Suzaku observation covering an almost complete orbit of the system. Apart from the eclipse related variability, the luminosity remained at a level of $\sim 10^{34}$ erg/s with a variability less than a factor of 10. Flares at two specific orbital phases and an $N_H \approx 10^{23}$ cm$^{-2}$ suggest the presence of accretion streams comparable to these found in obscured systems. Flares up to a level of $10^{36}$ erg/s were detected by *Swift/XRT* (La Parola et al., 2010b) while the source remained below $10^{34}$ erg/s for 20% of the time (Romano et al., 2014a). Note that the distance to the source is uncertain and that the luminosities quoted above could be significantly larger. No spin period is available.

– IGR J17544-2619: Drave et al. (2014) and Romano et al. (2014a) observed that its X-ray luminosity varies mostly in the range $10^{33-35}$ erg/s with some flares reaching few $10^{36}$ erg/s. The source activity shows a clear peak at periastron, reminiscent of the building up of a tidal stream, and a minimum at apastron. The average source luminosity of $10^{34}$ erg/s is well below the expectation for such a short orbital period. The detection of a CRSF indicating a magnetic field of $10^{12}$ G (Bhalerao et al., 2015) speaks against magnetic gating.

– AX J1845.0-0433: Zurita Heras and Walter (2009a) caught an outburst of the source with XMM-Newton and concluded that it was likely related to the accretion of a clump with a mass of $\sim 10^{22}$ g. The flare spectrum was steep and included a soft component with an absorption corrected luminosity (0.2-10 keV) of $10^{36}$ erg/s (and $\approx 10^{35}$ erg/s at hard X-rays). The spectra observed by *INTEGRAL/ISGRI* and *Swift/BAT* averaged over the missions are 10 times fainter. The X-ray lightcurves indicates that the source can shut down in a few minutes, corresponding to the free fall time at the accretion radius. A minimum luminosity of $0.5 \times 10^{34}$ erg/s (0.7-10 keV) was reported by Yamauchi et al. (1995a). No spin period is available.

– AX J1841.0-0536: Bozzo et al. (2011b) studied an outburst of the source well-sampled by XMM-Newton. A luminosity of the source (1-10 keV) was $\approx 4 \times 10^{32}$ erg/s in quiescence and reached $\approx 4 \times 10^{35}$ at the flare peak. The flare characteristics, in particular the evolution of the absorption, could be well explained by the ingestion of a wind clump with a mass of $\sim 10^{22}$ g. *Suzaku* detected a similar flare and a quiescent luminosity of $10^{34}$ erg/s (1-10 keV; Kawabata Nobukawa et al., 2012). The source is found to be 28% of the time below a luminosity of $10^{34}$ erg/s by *Swift/XRT* (Romano et al., 2014a). An average luminosity (20-100 keV) of $\sim 10^{34}$ erg/s can be derived from *ISGRI* and *BAT* data. No spin period is available.

The behaviour of this sub-class of SFXT systems could be related to abnormal low mass loss rates, high wind velocities or gating mechanisms (Bozzo et al., 2013). The flares with a duration of a few hours, are probably frequent ($\sim$ $P_{orb}$) but not that often detected (in particular in AX J18410-0536, where the hard X-ray flares are at the limit of sensitivity, but for which a flare was detected by chance when observed by *XMM-Newton*).

3. *Eccentric transients:*

   The three sources IGR J18483-0311, SAX J18186-1703, and XTE J1739-302 display large X-ray variability and short flares clustered at a specific orbital phase. The maximum luminosities reach few $10^{36}$ erg/s. The minimum (and average) luminosities detected decrease with the orbital period. The range of luminosities observed are not far from those expected by orbital modulation, if the intrinsic variability observed in classical systems is taken into account (Fig. 6). Note that the minimum observed in SAX J18186-1703 was detected only once and could be an outlier (Bozzo et al., 2008a; Zurita Heras and Chaty, 2009b).

4. *Unclear systems:*

   The last three sources IGR J08408-4503, IGR J18462-0233, and AX J18205-1434 are difficult to categorize, mostly because of a lack of good observations. IGR J08408-4503 and IGR J18462-0233 have unknown orbital periods and only a few flares were observed. Their average luminosities are very low, which may indicate eccentric orbits. AX J18205-1434 could be an eccentric classical sgHMXB, however, the high-mass nature of its companion was not firmly established yet.

According to our classification above, it appears that the SFXT class comprises seven systems with variability properties relatively similar to classical sgHMXBs (one in Roche-lobe overflow) and seven more extreme fast transients. The main peculiar property of the latter group is not the luminosity of the flares, but rather their low persistent luminosity which is on average much lower than expected when compared to classical systems. As we discuss further in Sect. 4, the spread in the properties of winds from super-giant stars and their intrinsic inhomogeneity can be invoked to interpret reasonably well the X-ray variability of the “classic-like systems”.

For the “fast transients reaching anomalously low luminosities”, featuring short orbital periods, additional mechanisms are required to explain their behaviour in the X-ray domain. As the average luminosity of these systems is significantly lower than expected (i.e. when comparing with classical systems with similar orbital periods), the additional mechanisms need to account for a substantial reduction of the mass accretion rate along the orbit of the compact object. In Sect. 4 we show that efficient “gating” mechanisms to inhibit the accretion onto the compact objects can be realized by taking into account the neutron star rotation and magnetic field. The need for gating mechanisms in the “eccentric transients” is somehow less critical than in the previous subclass, as the eccentric and elongated orbits of these systems also contribute to enhance their dynamic range in the X-ray luminosity and decrease its average value over each orbital revolution.

### 3.3 Be systems, X-ray pulsars and properties of cyclotron absorption lines

Binary systems with Be stars as secondaries constitute a substantial part of all HMXBs. By definition, Be stars are non super-giant B-type stars that have shown emission lines in their spectra, originating from a circumstellar disk expelled by a rapidly rotating star (Porter and Rivinius, 2003). A majority of these systems are transient sources exhibiting two type of outbursts. Type I outbursts are caused by the enhanced mass accretion rate close to periastron, last for 0.2–0.3 $P_{orb}$ and peak to $\sim 10^{37}$ erg s$^{-1}$. The rare type II outbursts, reaching the Eddington luminosity can last for several orbital periods and are probably related to events of stellar activity that may finally lead to the disappearance of the circumstellar disc.

Observing transient X-ray pulsars in bright outburst is essential to understand the physical processes at play close to the neutron star surface and in particular the response of the ”neutron star – magnetosphere” system to the variability of the mass accretion rate on different time scales.

As Reig (2011) wrote a detailed review of the observational properties of Be systems and related models, we are concentrating here only on some recent results obtained at hard X-rays.

Thanks to the number of wide field of view X-ray telescopes operating during the last decade (*RXTE/ASM, Swift, INTEGRAL, MAXI*), practically all ma-jor Be outbursts in this period could be detected and 8 new galactic BeXBs were discovered increasing the sample of these sources to 60; before INTEGRAL's launch this number was 52 (Liu et al., 2006).

The Be nature was confidently established for six of the newly discovered sources. Five systems feature pulsations with periods ranging from 12 to $\sim$ 700s (IGR J01583+6713, IGR J11435-6109, IGR J13020-6359, IGR J19294+1816, IGR J22534+6243) and orbital periods have been determined for four systems (IGR J01363+6610, IGR J11305-6256, IGR J11435-6109, IGR J19294+1816). More details on these systems can be found in table 1.

*INTEGRAL* was able to promptly observe dozens of bright type I and type II outbursts. As a result, comprehensive studies of spectral and timing properties of these transients were performed in a wide energy band for different time scales and source luminosities. In particular new CRSFs were discovered in the spectra of several X-ray pulsars (e.g., RX J0440.9+4431, EXO 2030+375, see Table 3).

Broad CRSF features are detected in a subset of the accreting X-ray pulsars. The first CRSF was detected in the spectrum of the X-ray pulsar Her X-1 (Truemper et al., 1978), a low-mass X-ray binary. By now cyclotron absorption lines were detected in the spectra of more than two dozens accreting pulsars. In four of them higher harmonics (up to the fifth!) were detected as well. Typical spectra detected by *INTEGRAL* for X-ray pulsars are shown on Fig. 10 for V 0332+53 (Tsygankov et al., 2006), which includes a CRSF with two higher harmonics, and LMC X-4 (Tsygankov and Lutovinov, 2005a) which does not.

The list of X-ray pulsars with confirmed cyclotron absorption lines in their spectra is presented in Table 3. Many CRSF were discovered with data from *Ginga* and *RXTE. INTEGRAL* contributed to new detections and to detailed studies of known lines thanks to its large effective area and high sensitivity in the energy range where most of CRSFs are located ($10 - 70$ keV).

The emission spectra of X-ray pulsars are usually approximated by phenomenological powerlaw models modified by an exponential cutoff at energies above 15-30 keV (White et al., 1983). Physical spectral models (see, e.g., Nagel, 1981; Meszaros and Nagel, 1985; Becker and Wolff, 2005, 2007) were constructed only for specific configurations of the emitting regions and are not able to explain in a self-consistent manner the variety of the observations.

The interaction of the radiation with the accreted matter in strong magnetic and gravitational fields is a complex problem. A number of authors attempted to simulate the shape of the continuum and CRSFs as a function of the pulse phase, source luminosity, geometry of the emission regions, etc (see, e.g., Araya and Harding, 1999; Araya-Góchez and Harding, 2000; Schönherr et al., 2007; Harding and Lai, 2006; Nishimura, 2008, and references therein).

The comparison of the model predictions with the observations, however, still fails to provide strong constrains on the physical parameters of the accretion regions due both to the limitations of the present-day hard X-ray telescopes and the complexity of the models.

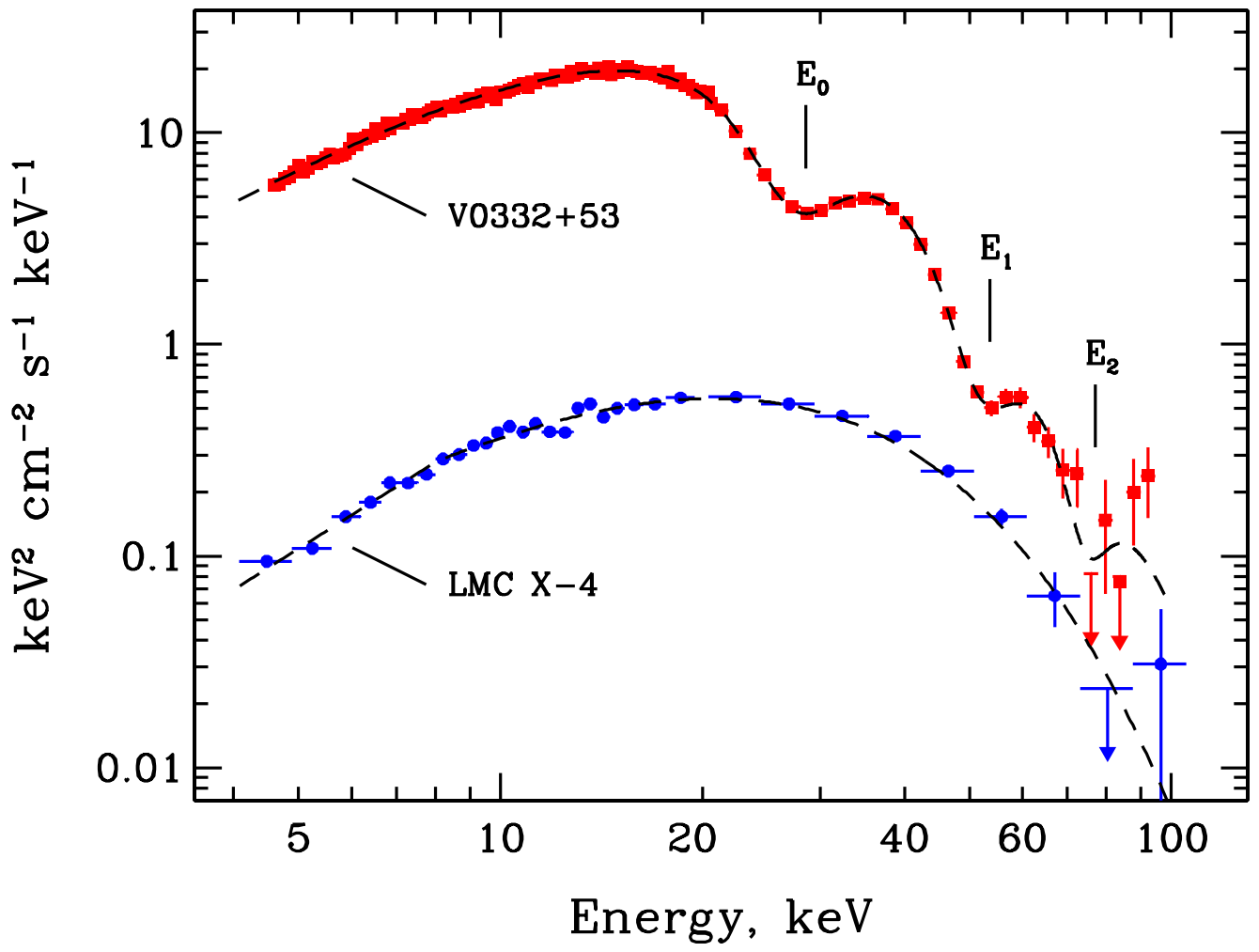

**Fig. 10** Energy spectra of two X-ray pulsars: V 0332+53 (red squares) with three harmonics of the cyclotron absorption line and LMC X-4 (blue circles) without such a feature (*INTEGRAL* data).

An important result of *INTEGRAL* is the discovery of an anti-correlation between the cyclotron energy and the X-ray luminosity in the transient X-ray pulsars V 0332+53 (Tsygankov et al., 2006; Mowlavi et al., 2006) and 4U 0115+63 (Nakajima et al. (2006); Tsygankov et al. (2007); but see Müller et al. (2013) and Boldin et al. (2013) for the influence of the continuum spectral shape). This result initiated a systematic study of the cyclotron energy properties as a function of the source luminosity.

This behaviour was interpreted with a change of the geometry of the accretion column, which is rising above the neutron star surface at luminosities higher than the critical one (Basko and Sunyaev, 1976; Mushtukov et al., 2015). Nishimura (2008, 2014) modelled the cyclotron line by the sum of the contributions emerging from individual line-forming regions along the accretion column with different magnetic field strength, temperature, and density. An increase of the mass accretion rate causes the emergence of additional line-forming regions with lower magnetic fields that lead to a decrease of the cyclotron energy.

Another recent model (Poutanen et al., 2013) suggests that a significant part of the accretion column radiation is intercepted and reflected by the neutron star surface because of relativistic beaming. Variations of the accretion column height lead to a shift of the illuminated part of the neutron star surface toward the equator where the magnetic field is weaker. This naturally causes the observed anti-correlation of the cyclotron line energy with luminosity. Moreover, this model is able to explain the amplitude of the cyclotron energy variability which is smaller than would otherwise be anticipated for the corresponding luminosity changes.

**Table 3** List of X-ray pulsars with known cyclotron lines (*: higher harmonics; ?: marginal detection).

| Source name | Cyclotron energy, keV |
|---|---|
| 4U 0115+63 | 11.5[1],20.1[2,*], 33.6[3,*], 49.5[4,*], 53[5,*] |
| V 0332+53 | 28[6], 53[7,*], 74[8,*] |
| 4U 0352+309 (X Per) | 29[9] |
| RX J0440.9+4431 | 32[10] |
| RX J0520.5-6932 | 31.5[11] |
| A 0535+262 | 50[12], 110[13,*] |
| MXB 0656-072 | 36[14] |
| Vela X-1 | 27[15],54[16,*] |
| GRO J1008-57 | 88[17,?], 75.5[18] |
| 1A 1118-61 | 55[19] |
| Cen X-3 | 28[20] |
| GX 301-2 | 37[21], 48[22] |
| GX 304-1 | 50.8[23] |
| 4U 1538-52 | 20[24], 47[25,*] |
| Swift J1626.6-5156 | 10[26] |
| 4U 1626-67 | 37[27] |
| Her X-1 | 42[28] |
| OAO 1657-415 | 36[29] |
| GRO J1744-28 | 4.7[30] |
| IGR J18179-1621 | 21[31] |
| GS 1843+00 | 20[32] |
| 4U 1907+09 | 19[33],40[34,*] |
| 4U 1909+07 | 44[35,?] |
| XTE J1946+274 | 36[36] |
| KS 1947+300 | 12.5[37] |
| EXO 2030+375 | 11[38,?], 36[39,?], 63[40,?] |
| Cep X-4 | 30[41] |

Further observations of X-ray pulsars during bright outbursts are needed to discriminate between the models.

For low-luminosity sources an opposite behaviour of the cyclotron energy with the luminosity has been observed (Staubert et al., 2007; Yamamoto et al., 2011; Klochkov et al., 2012). This has been explained by a squeeze of the emitting region towards the neutron star surface (where the magnetic field is higher) triggered by the ram pressure of the in-falling matter (Staubert et al., 2007).

The measurements of the cyclotron line energy as a function of luminosity are presented in Fig. 11 for the sources with known positive and negative correlations (shown by blue and red points, respectively). Finally, it is worth to note that for some transient pulsars no dependence of the cyclotron energy on luminosity has been detected for a wide range of luminosities (see, e.g., Caballero et al., 2013, for A 0535+26).

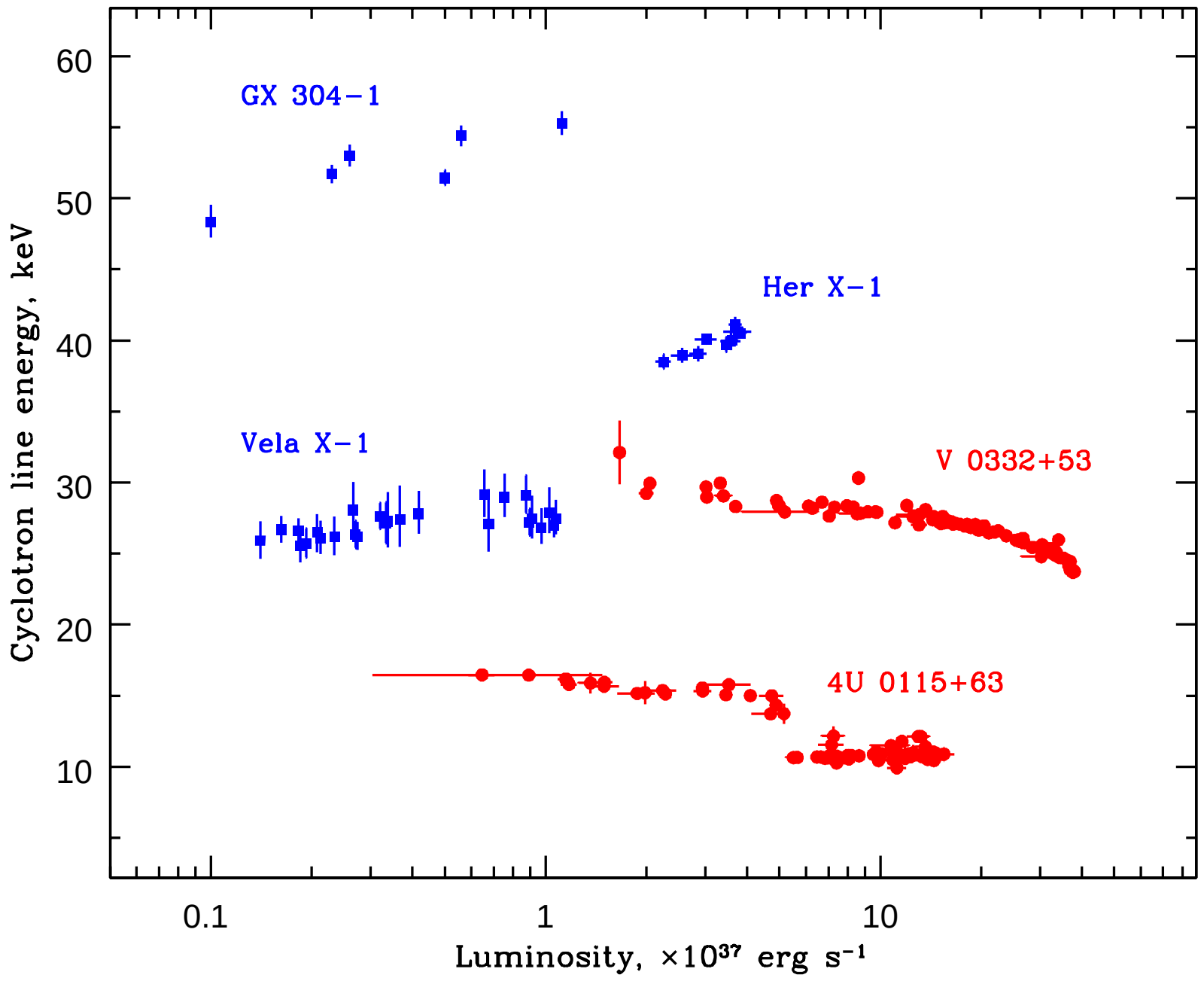

**Fig. 11** Cyclotron line energy dependence on the luminosity for the X-ray pulsars V 0332+53 (from Tsygankov et al., 2010), 4U 0115+63 (from Tsygankov et al., 2007), GX 304-1 (from Klochkov et al., 2012), Her X-1 (from Staubert et al., 2007) and Vela X-1 (the energy of the first harmonic divided by two is used; from Fürst et al., 2014b). Sources with positive and negative correlations of the cyclotron line energy with luminosity are shown by blue and red points, respectively.

Apart from the transient BeXBs, Reig and Roche (1999) pointed out the existence of persistent sources of the same Be/X type, but with low luminosities ($10^{34} - 10^{35}$ erg s$^{-1}$). Such objects are usually characterized by wide ($P_{orb} \gtrsim 200$ days) and low-eccentricity ($e \lesssim 0.2$) orbits, suggesting small natal kick (Pfahl et al., 2002), and by thermal excesses with a temperature of about $kT \simeq 1$ keV for a small emission region ($R < 0.5$ km). *INTEGRAL* detected very hard spectra in some of these systems. In particular, X Persei and RX J0440.9+4431 were detected significantly up to $\sim$ 160 keV (Doroshenko et al., 2012b; Lutovinov et al., 2012a) and $\sim$ 120 keV (Tsygankov et al., 2012), respectively. Broadband spectra of both sources are shown in Fig. 12 for illustration. Both cyclotron absorption lines and hard X-ray emission can be clearly seen.

## 4 Wind Accretion: a Chaotic Process

### 4.1 Slick winds

In the simplest approximation, the wind of a massive star is considered to be spherically symmetric and its properties are described by the so-called

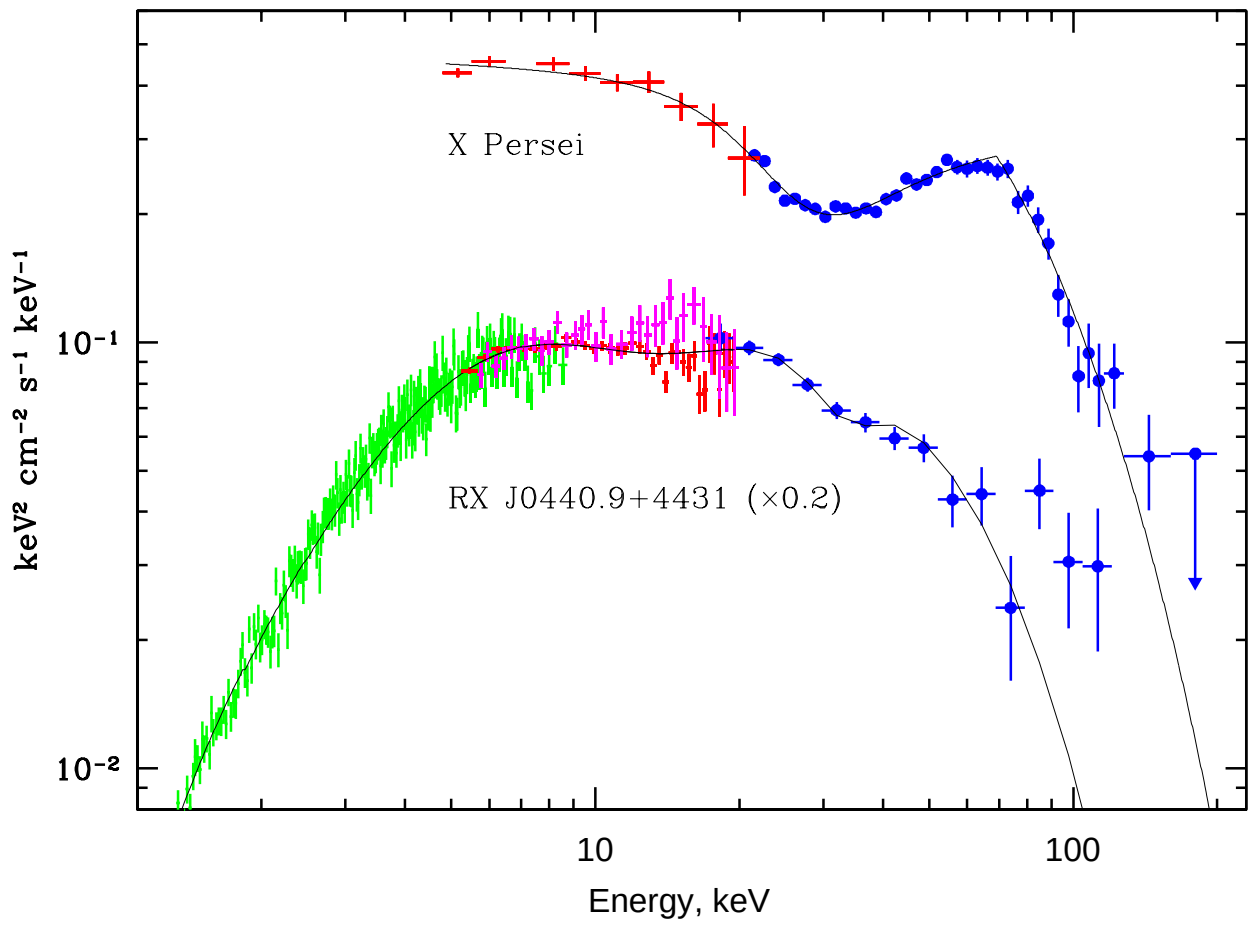

**Fig. 12** Broadband energy spectra of X Persei and RX J0440.9+4431 obtained with *JEM-X* (red crosses) and *IBIS* (blue circles) on board *INTEGRAL*, *RXTE/PCA* (magenta crosses) and *Swift/XRT* (green crosses). Broad cyclotron absorption lines $\simeq$ 30 keV and the hard X-ray emission above 100 keV are clearly visible in both sources. The spectrum of RX J0440.9+4431 is multiplied by a coefficient 0.2 for clarity.

"CAK" model (from the initials of its three inventors; Castor et al., 1975). This model predicts that symmetric and homogeneous winds stream outward from the stars as their atmospheres are not in hydrostatic equilibrium and gravity is overcome by gas and radiation pressure. The latter is generated by the high luminosity of the star, reaching $\sim 10^6$ $L_\odot$ in super-giants. The absorption of radiation in spectral lines provides the means to transfer energy and momentum to the out-flowing material and thus accelerates the wind up to velocities of $v_\infty \simeq$1000-3000 km s$^{-1}$, following a $\beta$-law (Lamers and Cassinelli, 1999):

$$v_w = v_\infty (1 - R_0/r)^\beta. \tag{1}$$

where $R_0$=$R_*[1-(v_0/v_\infty)^{1/\beta}]$, $R_*$ is the radius of the super-giant star, $v_0/v_\infty \simeq$ 0.01, and $v_\infty$ is the terminal wind velocity. Typical mass loss rates carried away by these fast winds are in the range $\dot{M}_w \simeq 10^{-7} - 10^{-5}$ $M_\odot$/yr.

According to the classical wind accretion scenario, the wind material flowing at supersonic velocities from the super-giant companion is shocked at a certain distance from the neutron star and then freely falls toward the surface of the compact object where it is accreted (see, e.g., Frank et al., 1992, and references therein). The distance of the (bow-) shock from the NS is usually termed the "accretion radius" and can be estimated as

$$R_{acc} = 2GM_{NS}/v_{rel}^2 \simeq 2GM_{NS}/v_w^2 = 3.7 \times 10^{10} v_8^{-2} \text{ cm}. \tag{2}$$

In the above equation we neglected the NS orbital velocity and approximated $v_w \sim v_\infty$ ($v_8$ corresponds $v_\infty$ in units of $10^8$ cm s$^{-1}$). A NS mass of 1.4 $M_\odot$ is considered throughout this section. $R_{acc}$ also defines the typical NS cross-section with respect to the wind material flowing around, and thus determines the effective fraction of the mass lost from the super-giant that the NS is able to capture at any time. If we assume as a first order approximation that the wind from the super-giant star is spherically symmetric, then we can express the mass loss rate at a distance $r$ from the star as $\dot{M}_w = 4\pi\rho_w(a)a^2v_w$ and the NS mass capture rate as $\dot{M}_{acc} = \pi\rho_w(a)R_{acc}^2v_w$, where $a$ is the orbital separation between the NS and its companion. It is thus clear that only a tiny fraction of the total mass loss rate from the super-giant star can be effectively accreted onto the NS:

$$\frac{\dot{M}_{acc}}{\dot{M}_w} \simeq \frac{1}{4}\left(\frac{R_{acc}}{a}\right)^2 \sim 2\times10^{-5}v_8^{-4}a_{10d}^{-2}. \tag{3}$$

We have assumed in the equation above a circular orbit, a binary orbital period of 10 days, and a total mass for the two stars of 30 $M_\odot$, i.e.

$$a = 4.2\times10^{12}P_{10d}^{2/3}M_{30}^{1/3} = 4.2\times10^{12}a_{10d}. \tag{4}$$

The accretion of all the captured material onto the neutron star gives rise to a total X-ray luminosity of

$$L_X = \frac{GM_{NS}\dot{M}_{acc}}{R_{NS}} \simeq 2\times10^{35}\dot{M}_{-6}a_{10d}^{-2}v_8^{-4}, \tag{5}$$

where $\dot{M}_{-6}$ is the super-giant mass loss rate in units of $10^{-6}$ $M_\odot$ yr$^{-1}$. This regime, in which all the mass captured by the NS is accreted onto its surface is usually called “direct accretion regime”.

Despite the initial success of the CAK model and the smooth wind accretion scenario described above (Vink et al., 2000), observational results proved in the past few years that these calculations are oversimplified as massive star winds are inherently inhomogeneous and the inhomogeneities play an important role in the accretion process.

### 4.2 Take the rough with the smooth

The most direct evidence for the presence of inhomogeneities in stellar winds is provided by the detection of peculiar features in the spectra of Wolf-Rayet and O-stars (Eversberg et al., 1998; Lépine and Moffat, 1999). Linear stability analyses already proved in the early 80s (Lucy and White, 1980) that line-driven winds are unstable for velocity perturbations. During the non-linear growth of the instability, high-speed material steepens into strong reverse shocks that compress most of the wind mass into finite dense ”clumps” and leave the surroundings filled-up with a lower density medium (Owocki et al., 1988). Initial 1D hydrodynamic simulations showed that clumps merge and grow in size while moving away from the stellar surface, leading to large variations in the local density (up to 4 orders of magnitude) and velocity (a factor of few). In these simulations, collisions between clumps were also shown to be able to produce a remarkable amount of X-rays (Feldmeier, 1995; Feldmeier et al., 1997; Cohen et al., 2010; Oskinova et al., 2011, 2012; Leutenegger et al., 2013). 2D hydrodynamical simulations later questioned the formation of large clumps, as in the multi-dimensional approach these structures are disrupted by the thin-shell and Rayleigh-Taylor instabilities (Dessart and Owocki, 2002, 2003, 2005). At present, a general agreement on the formation and characteristics of the clumps is still missing (Puls et al., 2008; Sundqvist et al., 2012; Šurlan et al., 2013).

The debate on the clump properties intensified in the early 2000s due to the suggestion that the enhanced density of these structures could be the main driver of the pronounced X-ray variability displayed by many high mass X-ray binaries. Starting from the initial investigations presented by Sako et al. (2003), several studies adopted this interpretation and used detailed X-ray timing and spectroscopic observations of classical super-giant HMXBs to infer the properties of clumps (i.e. mass, density, size, and velocity). The effect of clumps on the high energy emission from these sources is two-fold. Clumps simply passing in front of the X-ray source cause source dimming or even obscuration and display the signatures of photoelectric absorption. In addition to these phenomena, clumps that lead to increased accretion also give rise to large variations of the X-ray luminosity (qualitatively speaking, the instantaneous mass accretion rate onto the NS is proportional to the density of the surrounding wind material and thus the encounter with a clump can lead to an immediate increase of the X-ray luminosity by a factor of $\sim$10-100 for a few thousand seconds; see Eq. 5). Under these assumptions, the masses (radii) of clumps derived from the currently available X-ray data would be in the range $10^{18}$-$10^{20}$ g ($10^{10}$-$10^{11}$ cm), in agreement with what is expected from from simulations and observations of isolated super-giant stars (Kreykenbohm et al., 2008b). Fürst et al. (2010a) analysed in details a long INTEGRAL data-set of Vela X-1 and showed that the X-ray count-rate recorded from this source typically follows a log-normal distribution. They demonstrated through a MonteCarlo approach that such differential distribution would be expected in case wind accretion onto a neutron star occurs from a highly-structured clumpy medium. A similar result was found for a number of other classical sgHMXBs by exploiting the usage of cumulative luminosity distributions (Paizis and Sidoli, 2014). These studies thus seemed to provide a strong support in favor of clumps being the key ingredient triggering the X-ray variability displayed by classical sgHMXBs.

This conclusion is challenged by new hydrodynamic simulations of accreting neutron stars in sgHXMBs (Manousakis and Walter, 2015a), in which the required level of X-ray variability in Vela X-1 is reproduced by assuming only smooth winds and including the development of hydrodynamic instabilities and the effects of photo-ionisation to modulate the mass accretion rate onto the compact object. The collision between the primary stellar wind, slowed by photo-ionisation and flowing outwards and a gas stream flowing inwards from the tidal stream generates a shock front that moves inwards and outwards regularly creating transient low density bubbles. This "breathing" mechanism generates instantaneous accretion rates 10 times lower than predicted previously, log normal luminosity distributions with variations by a factor of $10^3$, and transient modulations. The identification of a mechanism that can explain both the observed variability amplitude and distribution and quasi-periods is encouraging. Log normal distributions are the signature of a self organised criticality. In our case the criticality condition is the angular momentum of the shock front discussed above which could alternatively lead or trail the neutron star.

When an accretion stream can develop in a classical system, the hydrodynamical effects of the neutron star are strong enough to explain the observed variability. An important question that is currently under investigation is whether intrinsically clumped winds would survive and have significant additional effect when compared to these of the neutron star. The observability through absorption of the presence of strong tidal streams matching the results of simulations based on CAK winds (Manousakis et al., 2012b) indicate that line driven instability plays a minor role in forming the global structure of the wind close to the surface of the star.

In 2005, the discovery of the super-giant fast X-ray transients (SFXT) opened new questions regarding physical processes at work during wind accretion onto NSs. As reported in Sect. 3.2, the SFXTs are far from being a homogeneous class of sources and thus we shall discuss them separately.

For the SFXTs that we classified as relatively similar to "classical systems", the observed variabilities are not larger than these observed in Vela X-1 (or GX 301-2, in the case of IGR J11215-5952 that has a very eccentric orbit). The X-ray dynamic range in these sources could tentatively be associated to hydrodynamically generated small scale inhomogeneities. The additional variability observed in "eccentric transients" can be accommodated for by the variation of the wind density along the orbit.

For the four SFXTs that we classified as "fast transients" the above explanations are not viable and other mechanisms have to be invoked. In these four systems, the compact objects orbit close to their companions and should generate tidal streams but feature anomalously low luminosities ($< 10^{34}$ erg/s) in quiescence. Despite the uncertainties still affecting our knowledge of the mass loss rates from OB super-giants (Puls et al., 2008; Vink et al., 2000), Eq. 5 shows that sgHMXBs with periods of 4-5 days should have typically an average luminosity of $\gtrsim 10^{36}$ erg s$^{-1}$. The flares therefore roughly reach the luminosities expected on average for smooth winds but the minimal luminosities are much too low (Romano et al., 2014b), suggesting a mechanism quenching accretion most of the time rather than generating inhomogeneities. The wind clump scenario (Walter and Zurita Heras, 2007b) can perhaps explain density ratios up to $10^{3-5}$ (Runacres and Owocki, 2005) between the clump and inter-clump medium. Such density contrasts are, however, predicted relatively far from the surface ($\sim 10\mathrm{R}_*$) of the companion and low and large densities are expected i.e. flares and low states. This is not matching the observations.

### 4.3 Magnetic gating

Grebenev and Sunyaev (2007a) and Bozzo et al. (2008d) proposed that such inhibition of accretion can occur due to centrifugal and/or magnetic gates related to the pulsar magnetic field and rotation. It is known since the early 70s that direct accretion onto a magnetized neuron star can occur only if the rotation of the compact object is slow enough to allow its magnetospheric boundary $R_m$ to reside within the so-called corotation radius:

$$R_{co} = 3.7 \times 10^9 P_{s2}^{2/3} \; cm \tag{6}$$

(here $P_{s2}$ is the NS spin period in units of 100 s). $R_{co}$ represents the distance from the NS at which a particle attached to its corotating magnetic field lines would reach a velocity comparable with the local keplerian velocity; the condition $R_m < R_{co}$ thus ensures that the accreting flow is not pushed outward (rather than accreted) by the rapidly rotating compact object. In case of wind accretion, the NS magnetospheric boundary $R_m$ can be roughly estimated by equating the magnetic to the free-fall pressure of the accreting material:

$$R_m = 3.3 \times 10^9 \dot{M}_{-6}^{-1/6} a_{10d}^{1/3} v_8^{-1/6} \mu_{30}^{1/3}. \tag{7}$$

Here, $\mu = B_{NS} R_{NS}^3$ is the neutron star magnetic moment and $\mu_{30} = \mu/(10^{30})$ G cm$^3$, for typical parameters (i.e. $R_{NS} = 10^6$ cm and $B_{NS} = 10^{12}$ G). By using Eq. (6) and (7), we thus conclude that direct accretion cannot occur in case of: (i) strongly magnetized ($\mu_{30} \gg 1$) and/or rapidly rotating ($P_{s2} \ll 1$) NSs, (ii) very slow wind velocities ($v_8 \ll 1$) and/or low mass accretion rates ($\dot{M}_{-6} \ll 1$). When $R_m \gtrsim R_{co}$, the centrifugal gate closes and the so-called "propeller" regime sets-in (Illarionov and Sunyaev, 1975), inhibiting a large fraction of the accretion. A precise estimate of the expected drop in the mass accretion rate is difficult to be provided, due to the occurrence of numerous physical processes and instabilities that cannot be taken into account in a simplified theoretical calculation. More sophisticated multi-dimensional simulations of the propeller regime have been carried out in the past years, supporting the above findings. However, these simulations could not include yet all relevant 3D magneto-hydrodynamic instabilities that dramatically affect plasma entry into the NS magnetosphere and thus the mass accretion rate (see, e.g. Toropin et al., 1999; Romanova et al., 2003, and references therein).

Accretion can also be inhibited by invoking a magnetic, rather than a centrifugal barrier. The magnetic barrier sets-in when $R_m \gtrsim R_{acc}$. In this condition, the inflowing material from the super-giant star cannot be gravitationally focused toward the compact object and it gets deflected away (rather than accreted) by the NS magnetosphere. For typical parameters, the expected drop in the mass accretion rate compared to the direct accretion regime can be as large as a factor of $\gtrsim 100$. By using the Eq. 2 and 7, the condition $R_m \gtrsim R_{acc}$ can be written as:

$$\dot{M}_{-6} \lesssim 4.5 \times 10^{-7} \mu_{30}^2 a_{10d}^2 v_8^{11}. \quad (8)$$

It can thus be easily deduced that the magnetic gating requires strong NS magnetic fields (B$\gg$$10^{12}$ G) to be applicable in the SFXT case.

The magnetic and centrifugal gates can also operate simultaneously when both the conditions $R_m \gtrsim R_{co}$ and $R_m \gtrsim R_{acc}$ are satisfied. As $R_{acc} \sim 10^{10}$ cm for typical parameters, the latter case is realized only when the corotation radius is also of the same order, i.e. in case of NS endowed with long spin periods ($\gtrsim$ 1000 s, see Eq. 6). If both magnetic and centrifugal gates are at work together, the lowest X-ray luminosity regime can be achieved with a total drop in the mass accretion rate by a factor of $10^4$-$10^5$. Gating models thus suggest that the peculiar X-ray variability of the SFXTs could be related to different values of the magnetic field and spin period of the NS hosted in these systems compared to classical sgHMXBs. In particular, the longer spin periods and more intense magnetic fields of the SFXTs could permit to achieve easily a dynamic range in the X-ray luminosity of $10^4 - 10^5$, by assuming only the presence of moderately dense clumps in the wind of the super-giant star.

Even though large magnetic fields are not always required for the gating models to be applicable to the SFXTs, the recent discovery of a possible cyclotron line at ~17 keV in the X-ray spectrum of one of the most highly variable SFXTs raised questions on the possibility of having very strongly magnetized NSs in these sources (Bhalerao et al., 2015). Such spectral feature would, indeed, indicate a NS magnetic field as low as $\sim 10^{12}$ G.

### 4.4 Cooling Switch

A different mechanism to halt the mass accretion flow in sgHMXBs and SFXTs was proposed by Shakura et al. (2012). These authors developed in details the previously proposed idea of the so-called "subsonic accretion regime" (Davies and Pringle, 1981; Ikhsanov, 2007). According to Elsner and Lamb (1977), the wind material halted at $R_{acc}$ is able to fall freely and accrete at the rate indicated by Eq. (3) only if it can be rapidly cooled below a critical temperature. The latter is determined by the operating condition of the RayleighTaylor instability (RTI), the main mechanism allowing material to penetrate the NS

magnetosphere and to get accreted onto the surface of the compact object. The wind material at the accretion radius is cooled by Compton scattering with lower energy photons produced close to the neutron star as a consequence of the on-going accretion. Shakura et al. (2012) demonstrated that systems endowed with an X-ray luminosity $\lesssim 4\times10^{36}$ erg s$^{-1}$ cannot cool rapidly enough the material at $R_{acc}$, and thus a hot envelope is formed around the NS in which the radial velocity of the inflowing material is significantly lower than the free-fall value. In these conditions, material can be cooled down sufficiently for the RTI to operate only close to the inner magnetospheric boundary $R_m$, and detailed calculations show that the reduced mass accretion rate corresponds to rougly 30% of the value given in Eq. (3).

In sources with even lower X-ray luminosities ($\ll 10^{36}$ erg s$^{-1}$), Compton cooling is not efficient enough to cool material located even in the closest proximity of the NS magnetospheric boundary and the system enters a radiatively (bremsstrahlung) cooling regime. In this case, only $\lesssim$10% of the mass flow rate given by the Eq. (3) is allowed to penetrate the NS magnetosphere and be accreted onto the surface of the compact object. On one hand, Shakura et al. (2013) suggested that a switch from the Compton to the radiatively cooling dominated settling regime could be invoked to explain the off-states displayed by several sgHMXBs (see also Sect. 3.1.1). Such switch would be caused by the change from the fan to the pencil-beam emission typically observed in young accreting X-ray pulsars at luminosities $\lesssim 10^{36}$ erg s$^{-1}$. Indeed, due to geometrical constraints, the pencil-beamed emission cannot illuminate sufficiently the inner boundary of the NS magnetosphere with the X-rays emitted from the compact objects, thus largely inhibiting the RTI and leading to the onset of the radiatively dominated settling accretion regime. On the other hand, Shakura et al. (2014) also suggested that a similar mechanism could be responsible for the peculiar X-ray variability displayed by the SFXTs. As these sources are typically characterized by an average X-ray luminosity $\lesssim 10^{34}$ rg s$^{-1}$, the authors proposed that SFXTs are in the radiatively dominated regime for most of the time. According to this interpretation, the bright SFXT flares/outbursts would correspond to peculiar episodes of enhanced accretion during which the hot envelope around the NS magnetiospheric boundary collapses and is accreted at once onto the NS. In their model, the collapse is induced by sporadic reconnections between rare magnetized clumps (transporting both the radial and tangential components of the super-giant star magnetic field) and the NS magnetic field lines close to $R_m$.

Although Shakura et al. (2014) showed that the accretion of the entire mass contained in the hot envelope would produce the required amount of X-rays to explain the emission recorded during SFXT flares/outbursts, the model still fails to explain why SFXTs should be characterized *a priori* by a lower averaged mass accretion rate than all other sgHMXBs. If no gating mechanism is at work to maintain an accretion rate low enough to sustain the formation of a hot envelope around the NS, the only remaining alternative to explain the low average luminosity of the SFXTs would be that their super-giant stars have

systematically faster and/or less dense winds compared to other OB super-giants in classical sgHMXBs. This hypothesis seems, however, unlikely given the fact that the spectroscopic classifications of OB super-giants in SFXTs and classical sgHMXBs show no systematic differences (Bozzo et al., 2013).

## 5 Populations of HMXBs

The properties of individual X-ray binaries in nearby galaxies have been studied for more than a decade, in particular after the launch of the *Chandra* X-ray observatory (see, e.g., Trudolyubov et al., 2001; Pence et al., 2001; Kong, 2003; Swartz et al., 2003). This requires to establish the nature of all X-ray sources, which remains difficult as the spatial resolution of *Chandra* and of the *Hubble Space Telescope* are not sufficient to unambiguously identify the counterparts and the nature of most sources. Therefore indirect methods, such as the construction of X-ray luminosity functions (LF), are needed to study of properties of populations of sources located in different regions of a galaxy (see, e.g., Gilfanov, 2004; Mineo et al., 2012a).

Observing our Galaxy and, for some aspects, the Large and Small Magellanic Clouds is therefore necessary to study the global properties of X-ray binaries. As it is impossible to track the evolution of individual sources, it is necessary to investigate the full population of X-ray binaries to understand their evolution, including its dependence on the companion mass or on the binary parameters. Catalogue of sources, collected with many different instruments (e.g., Liu et al., 2006), are also not well suited for statistical and physical studies of populations because of their non-uniformity.

A systematic survey of the Galaxy with *INTEGRAL* at hard X-ray energies ($> 17$ keV) with a moderate angular resolution ($\sim 12'$) allowed for the first time to overcome these difficulties and to obtain a virtually unbiased list of X-ray binaries in the Milky Way with an unprecedented sensitivity of $\simeq 3 \times 10^{-12}$ erg/s cm$^2$. An image of the inner ($|l| \lesssim 80°$) Galactic plane obtained by *INTEGRAL* is shown in Fig.1.

### 5.1 Distribution of HMXBs and its correlation with the spiral structure

High-mass X-ray binaries are a young galactic population and cannot migrate far from their birthplace, tracing regions of enhanced stellar formation. A spatial correlation between HMXBs and spiral arms was clearly established by Grimm et al. (2002), using data from *RXTE/ASM.*

As *INTEGRAL* observed the complete galactic plane and discovered many new high-mass X-ray binaries, several studies of their distribution were published

(Lutovinov et al., 2005a, 2007; Dean et al., 2005; Bodaghee et al., 2007, 2012c; Coleiro and Chaty, 2013).

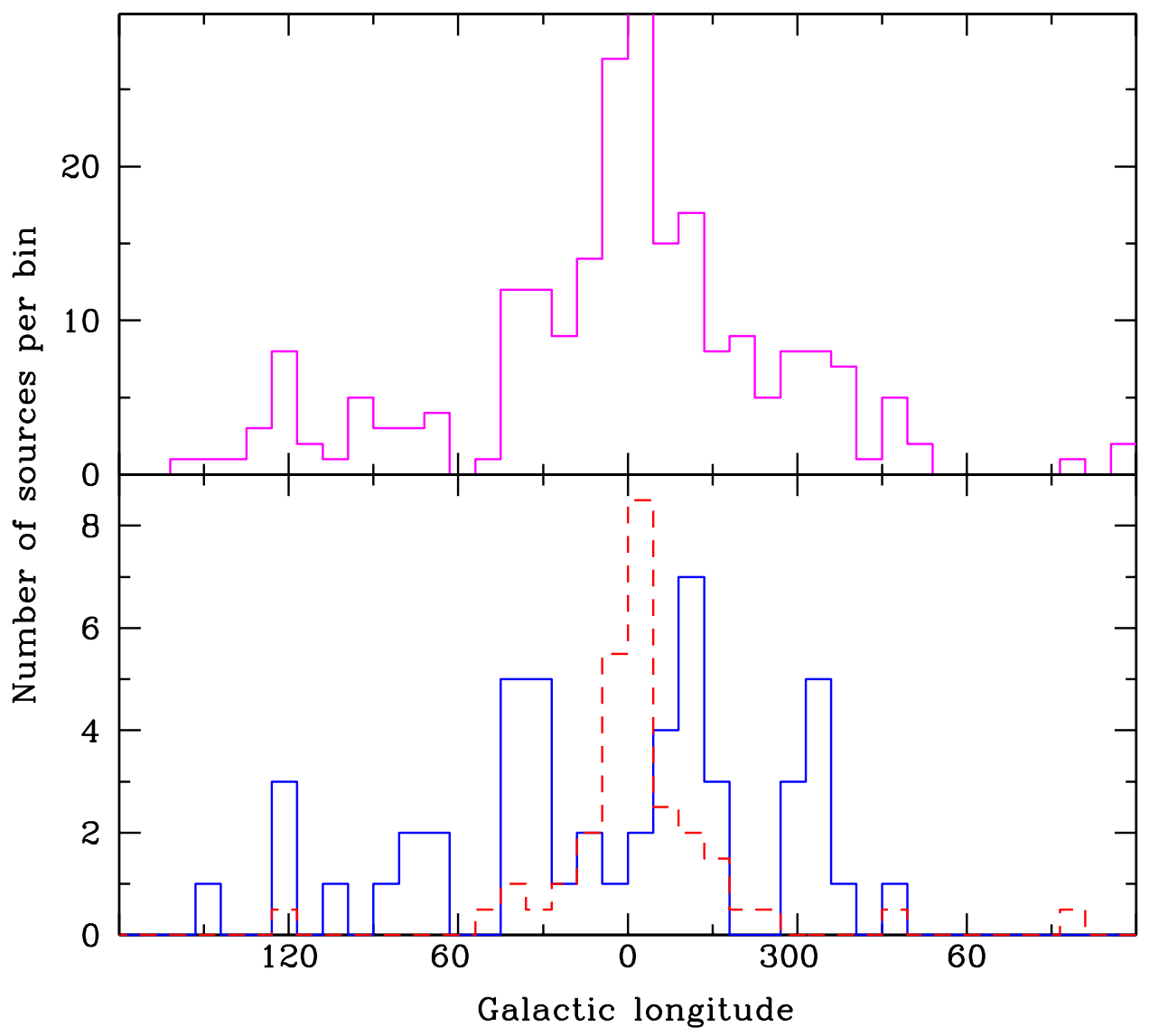


**Fig. 13** Distributions along the Galactic plane of all X-ray sources detected by *INTEGRAL* at low galactic latitude ($|b| < 5°$, *top*) and of high-mass (dark blue solid histogram) and low-mass (red dashed histogram; divided by two) X-ray binary systems (*bottom*).

The distributions of HMXBs and LMXBs along the Galactic plane are shown in Fig. 13. The overwhelming majority of the low-mass X-ray binaries is located in the Galactic bulge, while high-mass X-ray binaries are concentrated in the spiral arms. The HMXB distribution differs from a uniform or LMXB one with a probability $> 99.9\%$ (Lutovinov et al., 2005a, 2007).

A detailed comparison of the HMXBs distribution with the spiral structure shows that the correlation is not exact. In particular, the maxima of the HMXB angular distribution do not coincide with the tangents to the spiral arms. Although the distances of the majority of the systems and therefore their exact positions with respect to the spiral arms are uncertain, it has been argued that the displacement of the HMXB distribution when compared to the spiral arms is real and corresponds to a delay of several of $\sim 10^7$ years expected between the star formation and their appearance as bright X-ray sources.

Several observations support this interpretation. Galactic molecular clouds with very intensive star formation feature many young hot stars, but no high-mass X-ray binary systems (see, e.g. Feigelson et al., 2003; Nakajima et al., 2003). A small displacement between the massive X-ray binary systems and the position of the spiral arms was also detected in M83 (see Fig.17 from Soria

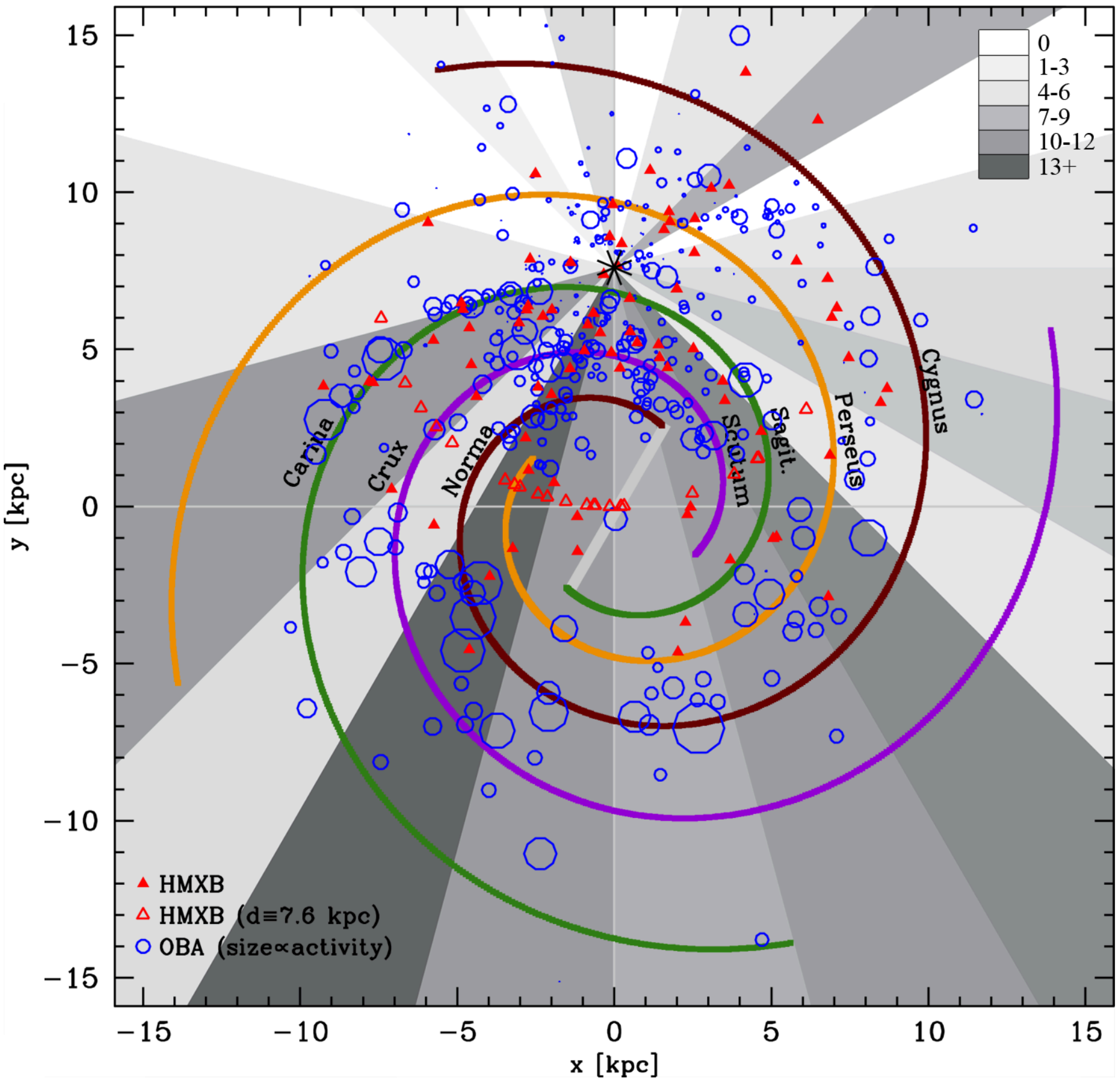


**Fig. 14** Galactic distribution of HMXBs (filled triangles – with known distances, open triangles – with unknown distances, placed at 7.6 kpc) and the locations of OB associations (circles, with a size proportional to the amount of activity in the association). As in Fig.13 the shaded sectors represent the distribution of HMXBs along the Galactic plane (Bodaghee et al., 2012c).

and Wu, 2003). Moreover, Shtykovskiy and Gilfanov (2005a) have shown, that the population of HMXBs does not correlate with the current regions of stellar formation in the LMC and found that they could be connected assuming an interval which can be estimated as $\simeq (1-2) \times 10^7$ years.

The spiral waves of the Galaxy (see, e.g. Lin et al., 1969) rotate with angular velocities varying between $\Omega \sim 20-60$ rad/Gyr, in the outer and inner galactic regions, respectively (Bissantz et al., 2003). The inner part of the spiral galactic structure is probably corotating with the stars up to a distance of $\sim 3.4$ kpc, corresponding approximately to the inner extremity of the Norma arm.

During the lifetime of massive stars and stars of average masses whose evolution can lead to the formation of HMXBs (see, e.g. Tutukov and Yungelson, 1973, 1993; Massevitch et al., 1976), the position of the spiral arms will change

considerably with respect to the stars and their tangent directions appear displaced with respect to the maxima of the HMXB population. The inner part of the Norma arm was at the position of the observed HMXB peak density approximately ∼15-20 million years ago which is in agreement with the model of Shtykovskiy and Gilfanov (2007).

A significant 2 dimensional clustering between HMXBs and OB associations was also found in the Milky Way (see Fig.14 and Bodaghee et al., 2012c). The two populations were found not perfectly aligned, confirming the above (1-D) analysis. An average offset of $0.4 \pm 0.2$ kpc was derived between a given HMXB and its nearest OB association, a distance consistent with natal kicks of $\sim 100{\pm}50$ km/s (Bodaghee et al., 2012c). The observed distribution of HMXBs in the Milky Way contains therefore information on the evolutionary history of massive binaries. Similar results were obtained by Coleiro and Chaty (2013), who found the correlation between HMXB distribution and the distribution of star forming complexes. Note that this was done using of a new approach for estimating of the distance and absorption for HMXBs, by spectral energy distribution fitting.

## 5.2 Luminosity function and surface density of HMXBs

The X-ray luminosity function is an important tool for the study of the formation and evolution of binary systems and of their dependence on the type of galaxy. The differential luminosity function of HXMBs in galaxies of different types is proportional to their star formation rate (SFR) (see, e.g. Grimm et al., 2002, 2003) and has an universal power law shape: $\frac{dN}{dL} \propto SFR \times L^{-\alpha}$, with an index of $\alpha \simeq (1.6 \pm 0.1)$ in a wide luminosity range $10^{35} - 10^{40}$ erg s$^{-1}$, that can be explained by the fundamental mass–luminosity and mass–radius relations for high-mass stars (Postnov, 2003). There are also some indications for a flattening of the HMXBs luminosity function at low luminosities both for sources in our Galaxy (Voss and Ajello, 2010) and for objects in the Small Magellanic Cloud (Shtykovskiy and Gilfanov, 2005b). The luminosity function at low luminosities is very important for the predictions of the number of sources that can be expected in future, more sensitive, surveys (Pavlinsky et al., 2009) and for estimating the contribution of HMXBs to the total X-ray luminosity of outer galaxies.

Luminosity functions can be straightforwardly constructed for outer galaxies as the distance to all sources is known and as focusing X-ray telescopes provide a rather uniform sensitivity. In the case of our Galaxy it is necessary to correct for the unequal sensitivity of the survey along the galactic plane. The simplest way to make such a correction is to assume a density distribution of HMXBs over the Galaxy. The latter can be done in different ways – in particular, Grimm et al. (2002) parametrised it as a disk with certain parameters, Voss

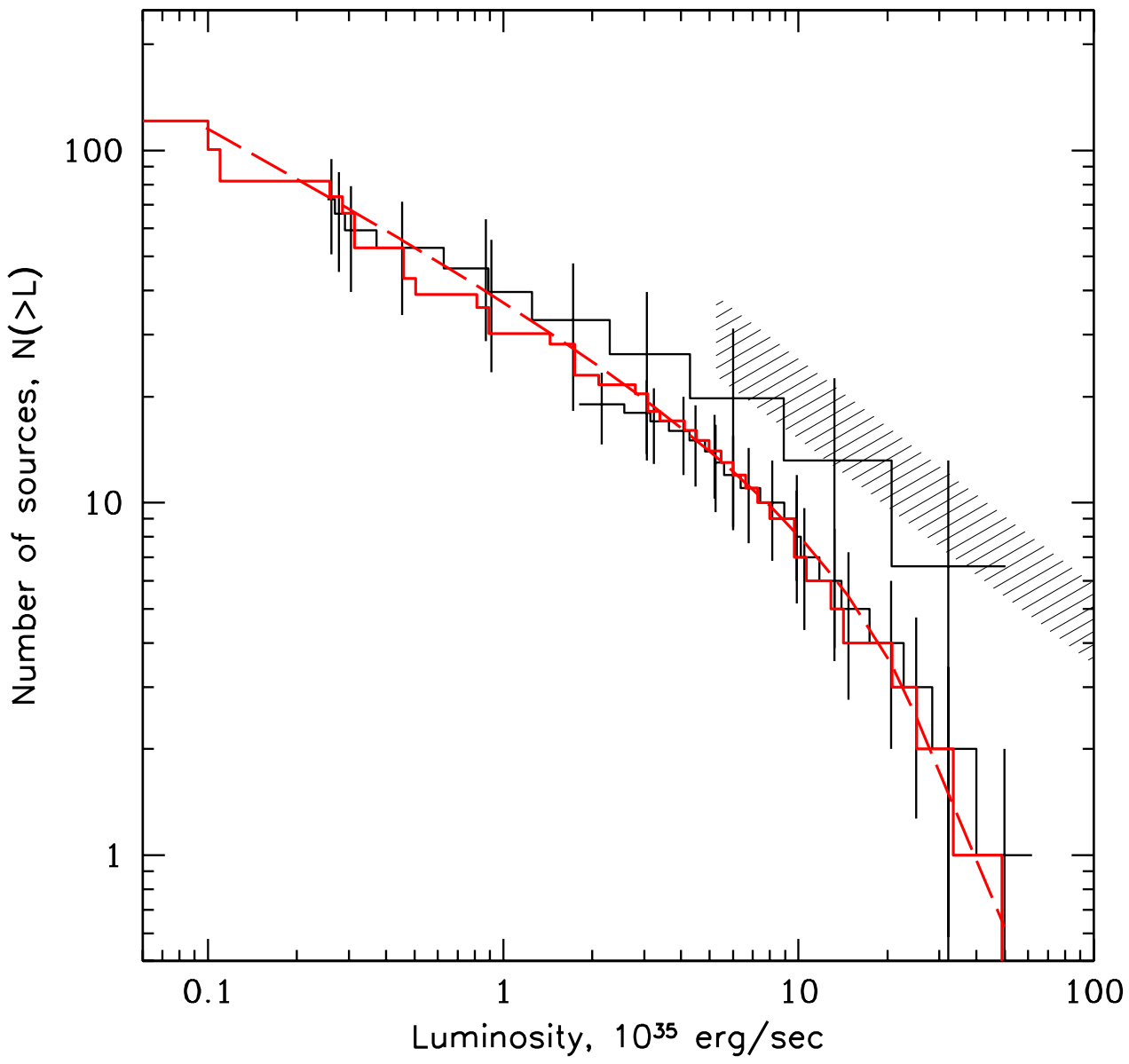


**Fig. 15** Luminosity functions of HMXBs accreting from the stellar wind (red histogram). Red dashed line represents the best fit model of the luminosity function with parameters from Table 4. Two black solid histograms represent luminosity functions within volume limited samples (see Lutovinov et al., 2013b, for details). Hatched area shows the number-luminosity function of all classes of HMXBs in our Galaxy from Grimm et al. (2002).

and Ajello (2010) expected that HMXBs are distributed like the stellar mass in the Galaxy.

*INTEGRAL* observations allowed us to measure the HMXBs density distribution and to calculate their luminosity function using fewer assumptions (Lutovinov et al., 2013b). It was first shown that the most numerous population of persistent HMXBs in our Galaxy are the wind-fed systems as other types of HMXBs indeed have only a few representatives. Then an axisymmetric distribution of HMXBs was assumed, i.e. that the Galaxy could be divided into several annuli of constant HMXBs surface density and luminosity function. A model of the latter in the form of a broken power law (with slopes $\alpha_1$ and $\alpha_2$ below and above the break at the luminosity $L_*$) was then adjusted to the data. The best fit luminosity function is presented in Fig.15 and the parameters are listed in Table 4.

It is clearly seen that the luminosity function of HMXBs in a wide range of luminosities ($10^{34} - 10^{37}$ erg s$^{-1}$) cannot be described by a simple power law. It features a break or a curvature at luminosities around $(0.4 - 2) \times 10^{36}$ erg s$^{-1}$and a flattening at low luminosity, confirming previous results (Shtykovskiy and Gilfanov, 2005b; Voss and Ajello, 2010).

The normalizations of the luminosity function can be used to calculate the surface density of HMXBs in each annulus. The results are presented in Table 4

**Table 4** Best fit parameters of the luminosity function of HMXBs and their spatial density distribution

| Parameter | Value and $1\sigma$ error |
|---|---|
| $\alpha_1$ | $1.40 \pm 0.13$(stat.)$\pm 0.06$(syst.) |
| $\alpha_2$ | $> 2.2$ |
| $L_*$, $10^{36}$ erg s$^{-1}$ | $2.5^{+2.7}_{-1.3}$(stat.)$\pm 1.0$(syst.) |
| $R_g$, kpc | $N(L > 10^{35}$ erg s$^{-1})$ kpc$^{-2}$ |
| 0-2 | $0.0 \pm 0.05$(syst.) |
| 2-5 | $0.11^{+0.05}_{-0.04}$(stat.)$\pm 0.02$(syst.) |
| 5-8 | $0.13^{+0.04}_{-0.03}$(stat.)$\pm 0.01$(syst.) |
| 8-11 | $(3.8^{+2.1}_{-1.2}) \times 10^{-2}$(stat.)$\pm 6.5 \times 10^{-3}$(syst.) |
| 11-14 | $(6.2^{+7.2}_{-4.3}) \times 10^{-3}$(stat.)$\pm 4.8 \times 10^{-3}$(syst.) |

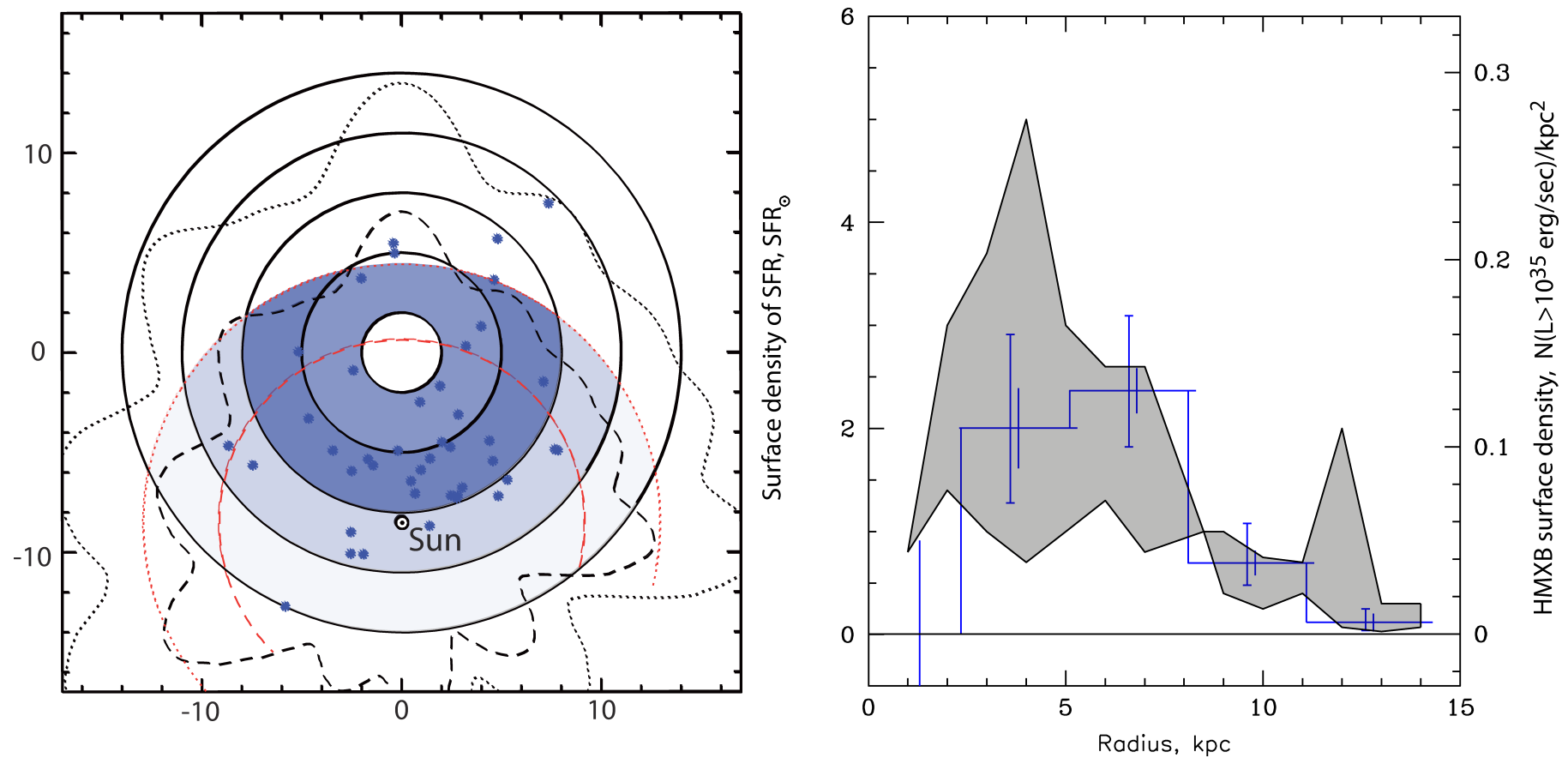


**Fig. 16** (*left*) Surface density of HMXBs in the Galaxy (a darker color of the annulus corresponds to a higher surface density of HMXBs, see Table 4). Blue points indicate positions of persistent HMXBs. Different lines correspond to different sensitivity levels of the *INTEGRAL* survey (Lutovinov et al., 2013b). (*right*) Dependence of the HMXBs surface density (histogram, right axis) and star formation rate surface density (upper and lower bounds, solid curve, left axis) on the galactocentric distance.

and Fig.16. The distribution of the surface density of HMXBs in the Galaxy has a maximum at galactocentric distances of $2 - 8$ kpc, as is also observed for the galactic SFR.

A comparison of the surface density of HMXBs with that of the star formation rate in the Galaxy (Guesten and Mezger, 1982; Lyne et al., 1985; Chiappini et al., 2001) shows a very good correlation that can be expressed as

$$N(HMXB, L_{\rm x} > 10^{35} {\rm erg\ s^{-1}})/{\rm kpc}^2 \approx 5.5 \times 10^{-2}\ SFR/SFR_{\odot}$$

where $SFR_{\odot}$ is the star formation rate near the Sun.

Finally the observations from *INTEGRAL* allow us to calculate the scale of the HMXBs vertical distribution as $\simeq 85 - 90$ pc which is significantly larger (by about $\sim$ 50 pc) than that of massive stars. This indicates that HMXBs should have travelled some distance from their birth places, similar to what was discussed above for the spatial correlation between HMXBs and OBAs. Assuming that HMXBs receive a systematic kick 50-90 km $s^{-1}$ during supernova explosions, the kinematic age of the population of HMXBs with wind-fed neutron stars after the supernova explosion can be estimated as $\tau \simeq 0.5 - 1$ Myr.

## 6 Summary

Our knowledge of High-Mass X-ray Binaries, and in particular of super-giant ones, has improved significantly since the launch of the wide field of view hard X-ray imagers on board *INTEGRAL* and *Swift* in 2002 and 2004, respectively. The discoveries of 23 new super-giant systems, increasing their population in the Galaxy by a factor 2.6, and of new X-ray variability patterns came as a surprise, challenging our understanding of stellar wind accretion around neutron stars.

In this review we have tried to make some sense of the observed phenomenology, keeping in mind that wind accretion is a stochastic process (Sect. 2). The super-giant HMXB population was classified as follows:

*Classical super-giant systems* feature a low orbital eccentricity and variability by a factor of $\sim 10^3$ on time scales much longer than the free fall time at the accretion or Alfvén radia. It is likely that most of that variability can be explained by hydrodynamic effects driven by the gravitational field of the neutron star. This variability can be enhanced by magnetic gating or a cooling switch on short time scales but it is not yet clear if such mechanisms are operative or needed. Several SFXTs belong to this category.

*Obscured super-giant systems* are similar to classical system, but characterized by persistently high X-ray absorption ($\sim 10^{23}$ cm$^{-2}$). Most of them are luminous systems with orbital periods less than 5 days, in transition to Roche lobe overflow. Strong absorption can also be related to particularly slow stellar winds or by the presence of large amount of interstellar material on the line of sight. The extreme obscuration observed in IGR J16318-4848 has a different nature and probably originates in the equatorial outflow of its B[e] companion.

*Fast transients reaching anomalously low luminosities* (IGR J16479-4514, AX J18410-0536, AX J18450-0433 and IGR J17544-2619) have very short orbital periods (<5d) and display average and minimal luminosities of $\sim 10^{34}$ erg/s and typical flare luminosities ten times lower than expected in classical systems with similar orbits. Several mechanisms to quench the accretion have been discussed (low mass loss rates, high wind velocities, magnetic gating, cooling switch) but no univocal process has been identified. Note that no spin periods are available for any of these sources.

*Eccentric transients* (IGR J18483-0311, SAX J18186-1703, and XTE J1739-302) are SFXTs with orbits sufficiently eccentric to explain the range of observed X-ray fluxes. The short flares require specific hydrodynamic processes (or structured winds), possibly similar to these observed in classical systems.

So far, several attempts have been made to study either the combined effects of wind clumps, neutron star magnetic field/spin rotation or the effect of eccentricity on the accretion from a smooth wind. A more complete theoretical study including all these effects is still missing. Our currently poor knowledge of the orbital parameters of many SFXTs and the lack of spin periods and magnetic field measurements still make the comparison between the outcome of such extended study with the constraints obtained through the currently available data (Sect. 4) challenging.

Hard X-ray observations of *INTEGRAL* in combination with other observatories were also unique to probe the variations of the CRSFs and of the geometry of the accretion column as a function of the accretion rate. The impact of observing Be systems flares with sensitive hard X-ray instruments is very important and has led to several geometrical interpretations, new ideas and theories (Sect. 3) that should be tested in the future.

The clustering of HXMBs near star formation regions in the Galaxy, that could be determined for the first time thanks to deep observations of the Galactic plane, has allowed us to constrain their formation rate and, in addition, the average natal kicks of neutron stars (Sect. 5). Furthermore the fraction of HXMBs of different classes has allowed us to constrain some of the time scales and processes driving their evolution.

The low flux population of HXMBs remains undetected. The Spectrum-RG survey (Pavlinsky et al., 2009) should soon unveil it and help constraining further the evolution of these systems and populations. Thousands of normal galaxies will also be detected by Spectrum-RG opening a new window on their recent star formation and compact object population.

## A Notes on individual sources

*IGR J00370+6122* was discovered in 2003 during the deep observations of the Cassiopeia region with the *INTEGRAL* observatory (den Hartog et al., 2004). Studying the nature of the source Reig et al. (2005b) found that the optical counterpart is neither a Be star nor a supergiant star (the most adequate classification was to be B0.5 II-III at the distance of $\sim 3$ kpc) and so IGR J00370+6122 appears difficult to fit within the classical classification scheme of HMXBs (see, however, González-Galán et al. 2014 for the recent classification of the source as a BN0.7Ib and discussion its possible supergiant nature). Later den Hartog et al. (2006) and in't Zand et al. (2007) showed that the source is a recurrent transient X-ray pulsar (with a spin period of $P_{spin} \simeq 359$ s) in an eccentric orbit (with the orbital period $P_{orb} = 15.667$ days), demonstrating s flaring behavior within a dynamic range about 10-20.

*1A 0114+650* has been shown to be a rather unusual source. It was discovered by the *SAS-3* observatory during the galactic plane survey and was identified with a bright early type optical star (Dower et al., 1977), exhibiting properties consistent with both Be and supergiant X-ray binaries. The source nature was debated several years until Crampton et al. (1985) classified the optical star as B0.5 with the luminosity class I or II, i.e. as a supergiant at the distance 7.2 kpc (Reig et al., 1996). Using the optical data Crampton et al. (1985) determined also an orbital period of the system $P_{orb} \simeq 11.6$ days, which was confirmed later in X-rays by Corbet et al. (1999a). 1A 0114+650 is the X-ray pulsar with one of the longest known pulse periods $P_{spin} \simeq 2.65$ h (Farrell et al., 2008), which evolved on the time scale of several years (Wang, 2011). In addition to pulse and orbital variabilities in the system there is a superorbital periodicity with the period of 30.7 days (Farrell et al., 2006).

*IGR J01363+6610* was discovered with the *INTEGRAL* observatory during galactic plane scans (Grebenev et al., 2004b). The follow-up observations with the *XMM-Newton* observatory revealed a faint variable X-ray source inside the *INTEGRAL* error box. This source has a hard powerlaw spectrum with a photon index of $1.4 \pm 0.3$ and, based on the optical data, was associated with the Be-star as an optical companion in the binary system (Tomsick et al., 2011). The distance estimate $\sim 2$ kpc indicates a very low quiescent X-ray luminosity of the source $\simeq 10^{32}$ erg s$^{-1}$. A possible $\simeq 160$ days orbital period was found in the *Swift/BAT* data (Corbet and Krimm, 2010).

*RX J0146.9+6121* belongs to the class of low-luminosity persistent systems with Be-companions (see Reig, 2011, for details). Similarly other such systems its spectrum is characterized by a presence of the soft thermal component with the temperature of $kT \sim 1$ keV and the power-low tail at higher energies (La Palombara and Mereghetti, 2006). The source is the X-ray pulsar with a quite long pulse period $P_{spin} \simeq 25$ min, which was discovered by the *EXOSAT* observatory and erroneously related to the nearby source 4U 0142+614 (White et al., 1987; Reig, 2011). The system is located in the open cluster NGC 663 at a distance of 2.5 kpc (Tapia et al., 1991).

*4U 0115+63* was discovered by the *UHURU* observatory in the early 1970s by Giacconi et al. (1972); Forman et al. (1978). During the *SAS-3* observations in 1978, Cominsky et al. (1978) found a pulsation period of 3.61 sec. Rappaport et al. (1978) determined the binary's main parameters: orbital period $\sim 24.3$ days, orbital eccentricity 0.34, and projected semimajor axis of the relativistic object $a_x \sin i \sim 140$ light seconds (see also Tamura et al., 1992; Lutovinov et al., 1994, for an improvement of the parameters). The optical observations of the star V635 Cas (Hutchings and Crampton, 1981; Kholopov et al., 1981), which is the normal stellar companion of the X-ray source 4U 0115+63, performed by Negueruela and Okazaki (2001) allowed the star's spectral type to be improved, B0.2Ve, and the binary's distance to be estimated, 7–8 kpc. The X-ray pulsar 4U 0115+63 is unique in its spectral characteristics. At present, it is the only object in which spectrum the cyclotron line was detected with its forth higher harmonic. Properties of this cyclotron feature were studied in detail using data of many observatories (Wheaton et al., 1979; White et al., 1983; Mihara et al., 1998; Heindl et al., 1999; Santangelo et al., 1999; Lutovinov et al., 2000). Particularly Mihara et al. (1998, 2004) found that the position of the fundamental cyclotron line in the energy spectrum depends on the pulsar luminosity. Later this effect was confirmed using the *RXTE* and *INTEGRAL* data (Nakajima et al., 2006; Tsygankov et al., 2007; Klochkov et al., 2011; Boldin et al., 2013). Such a behaviour can be explained either by the modification of the emitting regions in the vicinity of the neutron star or by artificial effects due to poor knowledge of the spectral continuum (Ferrigno et al., 2009; Müller et al., 2013; Boldin et al., 2013).

*IGR J01583+6713* is a high mass X-ray binary with the Be companion star (type B2IVe) located at a distance about 4 kpc (Kaur et al., 2008). The sources was discovered by the *INTEGRAL* observatory during observations of the Cas A region in 2005 (Steiner et al., 2005). The follow-up observations with the *XRT* telescope of the *Swift* observatory revealed a strong absorption in the source spectrum $N_H \simeq 10^{23}$ cm$^{-2}$ (Kennea et al., 2005) and a presence of possible pulsations in its light curve with the period $P_{spin} \simeq 469.2$ s (Kaur et al., 2008). Note, that the latter result was not confirmed by Tomsick et al. (2011) based on the *XMM-Newton* and *Chandra* data.

*V 0332+53* was firstly detected by the Vela 5B observatory in 1973 (Terrell et al., 1982) during an outburst when its intensity reached $\sim 1.4$ Crab in the $3-12$ keV energy band. During subsequent outbursts in 1983–1984 and 1989, observed with the *EXOSAT* and *Ginga* observatories, respectively, the spin ($\sim 4.4$ s) and orbital ($\sim 34.25$ d) periods were determined by Stella et al. (1985). The cyclotron resonance scattering feature with an energy of $E_{cyc} = 28.5 \pm 0.5$ keV was detected in its spectrum Makishima et al. (1990). Based of the results of long-term monitoring campaign of Be/X-ray binaries with the Southampton/Valencia/SAAO, Negueruela et al. (1999) determined the spectral class of BQ Cam – the normal companion of the X-ray pulsar V 0332+53 – as O8-9Ve and estimated the distance to the system at $\sim 7$ kpc. The last leads to the maximum source luminosity of $\sim 4 \times 10^{38}$ erg s$^{-1}$ observed during outbursts, that make it one of the brightest X-ray sources in the Galaxy. The next powerful outburst of V 0332+53 began at the end of 2004 (Swank et al., 2004). An analysis of the follow-up observations performed with the *RXTE* and *INTEGRAL* observatories showed that beside the absorption feature at an energy of $\sim 26$ keV, there are two additional similar features in the source spectrum at energies of $\sim 49$ and $\sim 75$ keV, which were interpreted as the second and third harmonics of the main cyclotron frequency (Kreykenbohm et al., 2005; Pottschmidt et al., 2005). A good coverage of the whole outburst (including rising and declining parts) with the *RXTE* observations allowed to Tsygankov et al. (2006, 2010) to make a detailed spectral analysis and to show that the cyclotron line energy is not a constant but negatively correlated with the source luminosity and to obtain constraints on the magnetic field in the source as $B_{\rm NS} \simeq 3.5 \times 10^{12}$ G. Moreover, the line energy as well as its width and depth are also strongly variable on the pulse period scale (Lutovinov et al., 2015).

*4U 0352+309/X Persei* is a classical persistent Be/X-ray binary system, consisting of an X-ray pulsar and a Be-star companion optically identified with the star HD 24534 (spectral type B0Ve). It was discovered during a high X-ray intensity state in 1972, and pulsations with the period of $P_{spin} \simeq 835$ s were detected by the *Copernicus* observatory (White et al., 1976). A distance to the source is estimated by different authors in the range of 700 to 1300 pc, but more often the value of $950 \pm 200$ pc is used (Telting et al., 1998). Adopting this distance the source peak luminosity $L_X \simeq 2 \times 10^{35}$ erg $s^{-1}$ was registered in 1975, 2003 and 2010 (Lutovinov et al., 2012a). Delgado-Martí et al. (2001) succeeded in determining orbital parameters for X Persei and showed that it is in a moderately eccentric orbit ($e \simeq 0.11$) with a very long $P_{orb} \simeq 250$ days orbital period. Deep observations performed with the *INTEGRAL* observatory allowed to detect the hard X-ray emission from X Persei up to $\simeq 160$ keV (Doroshenko et al., 2012b; Lutovinov et al., 2012a), that is non-typical for X-ray pulsars. In the X Persei spectrum there is also a strong absorption feature near the energy of $\simeq 30$ keV. It was discovered by Coburn et al. (2001) based on the *RXTE* data and interpreted as a cyclotron resonance scattering feature, that allowed to estimate a magnetic field on the neutron star surface $B_{NS} \simeq (2.4 - 2.9) \times 10^{12}$G, (Lutovinov et al., 2012a). This line was found to be significantly broader than is typically observed in X-ray pulsars (Coburn et al., 2002), that allowed other authors interpreted it as an artificial deficit of photons in the region where the different spectral components overlap (di Salvo et al., 1998; Doroshenko et al., 2012b). Due to a source proximity it is extremely bright in the optical and infrared wavebands ($m_{V,B} \simeq 6$), that is allowing to investigate and modeling the physical properties and behavior of Be-systems at different time scales (see, e.g., Roche et al., 1993; Clark et al., 2001b; Okazaki and Negueruela, 2001).

*RX J0440.9+4431* was found during the *ROSAT* Galactic plane survey with the optical companion BSD 24-491/LS V +44 17 classified as a Be star (Motch et al., 1997). Distance to the system was estimated as $3.3 \pm 0.5$ kpc (Reig et al., 2005a). For the first time RX J0440.9+4431 was detected in the hard X-ray energy band by the *INTEGRAL* observatory during the Type I outburst in September 2010 (Krivonos et al., 2010; Tsygankov et al., 2012). Before this the source belonged to the small population of persistent low-luminosity binaries with Be companions and a slowly rotating neutron star (the pulse period is $\sim 202.5$) . Based of a set of equally spaced in time Type I outbursts (each of them have the $3 - 100$ keV luminosity about $few \times 10^{36}$ erg s$^{-1}$) in 2010–2011 (Morii et al., 2010; Krivonos et al., 2010) it became possible to estimate the orbital period of RX J0440.9+4431 as $\sim 155$ days (Tsygankov et al., 2011). The spectral analysis of the *INTEGRAL* data revealed a $\sim 32$ keV cyclotron resonant scattering feature in the source spectrum, that corresponds to the magnetic field strength of the neutron star surface $B \simeq 3.2 \times 10^{12}$ G (Tsygankov et al., 2012). Moreover, the source spectrum is rather hard and its emission is clearly detected above 100 keV (Krivonos et al., 2015).

*A 0535+262* is a typical Be/XRP transient discovered with *Ariel V* during a giant (Type II) outburst (Rosenberg et al., 1975). Besides giant outbursts not related to the specific orbital phase, the source demonstrates also normal (Type I) outbursts linked to the periastron passages of the neutron star (see, e.g. Giovannelli and Graziati, 1992). The binary system consists of a B0IIIe star HDE 245770 at the distance of $\sim 2$ kpc (Steele et al., 1998) and a neutron star rotating with a spin period $\sim 103$ s. The orbit is highly eccentric ($e \sim 0.47$) with a period of $\sim 111.1$ days (Finger et al., 1996b). The energy spectrum of A 0535+262 is modified by two absorption features at $\sim 45$ keV and $\sim 100$ keV, which are interpreted as a cyclotron absorption line and its first harmonic. The magnetic field strength on the neutron star surface can be estimated as $B \sim 4 \times 10^{12}$ G (Kendziorra et al., 1994; Grove et al., 1995). A comprehensive analysis of the *INTEGRAL*, *RXTE*, and *Suzaku* spectral data does not reveal variations of the cyclotron energy during outbursts (Caballero et al., 2013).

*IGR J06074+2205* was discovered in 2003 with the *JEM-X* telescope on board the *INTEGRAL* observatory (Chenevez et al., 2004). A number of studies were dedicated to the search of the optical counterpart of this source (Halpern and Tyagi, 2005; Tomsick et al., 2006b; Masetti et al., 2006a; Reig and Zezas, 2009; Reig et al., 2010). Finally Reig et al. (2010) identified it with a relatively bright ($V = 12.3$) B0.5Ve star located at a distance of $\sim 4.5$ kpc.

*2E 0655.8-0708* (better known as MXB 0656-072) is a transient X-ray source in the Galactic plane discovered in 1975 with the *SAS-3* observatory (Clark et al., 1975). Pulsations of the X-ray flux with a period $\sim$ 160.7 s were found with *RXTE/PCA* (Morgan et al., 2003). An optical companion was identified with a O9.7Ve star (Pakull et al., 2003) at the distance of $3.9 \pm 0.1$ kpc (McBride et al., 2006). An orbital period was estimated from *SWIFT/BAT* and *RXTE/ASM* data as $\sim$ 101.2 days (Yan et al., 2012). During the strong Type II outburst in 2003 the detailed spectral analysis of MXB 0656-072 was performed using *RXTE* data (Heindl et al., 2003). To describe the source spectrum the standard model of power law with the high energy cutoff was modified by an iron and cyclotron absorption lines. Best fit parameters were: photon index $1.09 \pm 0.01$, cutoff energy $E_{cut} = 16.8 \pm 0.1$ keV, exponential folding energy $E_{fold} = 11.5 \pm 0.3$ keV, cyclotron line energy $E_{cycl} = 36 \pm 1$ keV. A slightly different parametrization of the spectral model gives the cyclotron line energy $32.8^{+0.5}_{-0.4}$ keV which is stable through the outburst and over the pulsar spin phase (McBride et al., 2006).

*IGR J08408-4503* was discovered in the Vela region on 2006 May 15 with INTEGRAL during a short flare lasting less than 1000 s (Götz et al., 2007). Its optical counterpart was later identified as the supergiant star HD 74194 located at 3 kpc, thus confirming that this source belongs to the SFXT class (Götz et al., 2007; Masetti et al., 2006a). IGR J08408-4503 was observed in outburst several times with *INTEGRAL* and *Swift* (Götz et al., 2007; Leyder et al., 2007; Sidoli et al., 2009c; Barthelmy et al., 2009). Observations of the source in the lower X-ray activity state were performed with *Swift*, *XMM-Newton*, and *Suzaku* (Kennea and Campana, 2006; Bozzo et al., 2010; Sidoli et al., 2010). These revealed the presence of a peculiar soft (<2 keV) spectral component possibly associated with the X-ray emission from the supergiant wind itself. IGR J08408-4503 is the best suited SFXT to study soft spectral components as it is on average the less absorbed one.

*Vela X-1* is the archetype of the persistent classical sgHMXBs. The pulsar (1.86 $M_\odot$; spin 283 sec) orbits a B0.5Ib supergiant in 8.964 days on an almost circular ($e \approx 0.09$; R=1.76 $R_*$) trajectory (Quaintrell et al., 2003), see however Koenigsberger et al. (2012). The strong and continuous X-ray variability observed (Kreykenbohm et al., 2008a) has been explained by wind clumping (Fürst et al., 2010b), self-criticality of the wind-fed accretion flow (Manousakis et al., 2012a), magnetic gating (Doroshenko et al., 2011) or transition of cooling mechanism (Shakura et al., 2013). Vela X-1 is one of the few systems where the influence of photoionisation can be studied (Watanabe et al., 2006; Krtička et al., 2012; Manousakis and Walter, 2015b). The pulse profile changes with energy and variable cyclotron absorption features are detected (Doroshenko et al., 2011; Fürst et al., 2014b). Small variations of the spin period have been observed on various time scales (Bildsten et al., 1997). Vela X-1 is a runaway system accompanied by a bow shock (Kaper et al., 1997).

*GRO J1008-57* was discovered during the bright outburst in 1993 with *CGRO/BATSE* as an X-ray source pulsating with a period $93.587 \pm 0.005$ s (Stollberg et al., 1993). An optical counterpart was identified with a B1-B2 Ve star (Coe et al., 2007) at a distance of $\sim$ 5 kpc (Coe et al., 1994a). Orbital parameters were determined using data from different observatories as follows: $P_{orb} = 249.46 \pm 0.10$ d, $a_x \sin i = 530 \pm 60$ lt-s, $\omega = -26 \pm 8$ deg, $e = 0.68 \pm 0.02$ (Levine and Corbet, 2006a; Coe et al., 2007; Kuehnel et al., 2012). The combined spectrum from the *CGRO* and *ASCA* observations can be well approximated by a power law with the high energy cutoff and a 6.4-keV iron emission line (Shrader et al., 1999). An approximation of the *INTEGRAL* data results in the following spectral parameters: photon index $1.4 \pm 0.1$, cut-off energy $E_{cut} = 8.0 \pm 1.0$ keV, exponential folding $E_{fold} = 21 \pm 2$ keV (Coe et al., 2007). Based on the *Suzaku* data Yamamoto et al. (2013) discovered a cyclotron line in the source spectrum at $E_{cyc} = 75.5^{+2.5}_{-1.5}$ keV. This detection reconfirms the previously suggested spectral feature around $\sim$ 80 keV (Shrader et al., 1999).

*IGR J10101-5654* was detected by the *INTEGRAL* observatory at high energies (> 20 keV) in 2004 during observations of the Carina region (Kuiper et al., 2006). The NIR spectrum is very rich with many strong emission lines, originating from different media, that suggests the presence of a stratified circumstellar environment. This allowed Coleiro et al. (2013) to suggest the companion star to be a sgB[e]. *Chandra* observations revealed a significant change in the mass accretion rate onto the compact object and determined spectral parameters as: photon index $1.0^{+0.5}_{-0.4}$, photoabsorption $N_H = 3.2^{+1.2}_{-1.0} \times 10^{22}$ cm$^{-2}$ (Tomsick et al., 2008).

*3U 1022-55* (also known as 4U 1036-56 and RX 1037.5-5647) appeared in the *Uhuru* catalog (Giacconi et al., 1972). The optical counterpart of the system is a B0 V-IIIe star LS 1698 at the distance of $\sim 5$ kpc (Motch et al., 1997). Timing analysis of the *RXTE* data revealed pulsations in the source flux with a period of $P \simeq 860$ s and it was suggested that the system belongs to the subclass of persistent Be/X-ray systems with slowly rotating neutron stars (Reig, 2011). A possible association of 4U 1036-56 with the unidentified transient gamma-ray source AGL J1037-5708 was discussed by Li et al. (2012a). It is interesting to note that the black body component with $kT_{BB} = 1.26^{+0.16}_{-0.09}$ keV and $R_{BB} = 128^{+13}_{-21}$ m (La Palombara et al., 2009) is present in the source spectrum in addition to the typical pulsars components – power law and the high energy cutoff (White et al., 1983). This thermal emission suggests its polar-cap origin and can be characteristic of all low-luminosity Be systems (see, e.g., La Palombara and Mereghetti, 2006; La Palombara et al., 2009; Tsygankov et al., 2012).

*Cen X-3* is the first X-ray pulsar discovered with a spin period of 4.8 sec (Giacconi et al., 1971). It orbits an O6.5 II-III supergiant, located at 5-8 kpc (Day and Tennant, 1991; Krzeminski, 1974), in 2.1 days with a small eccentricity, if any (van der Meer et al., 2007; Falanga et al., 2015). Mass transfer probably occurs through a combination of wind and disk accretion (Petterson, 1978; Tsunemi et al., 1996; Tjemkes et al., 1986; Kohmura et al., 2001; Suchy et al., 2008). A cyclotron absorption feature is detected (Nagase et al., 1992; Santangelo et al., 1998; Heindl and Chakrabarty, 1999). Iron line variability indicate fluorescence on several components (Devasia et al., 2010).

*1A 1118-615* is a peculiar Be system with a long spin period (406 s, Staubert et al., 2011) for a short orbital period (24 days, Staubert et al., 2011). Three type II outbursts have been detected in 38 years featuring correlated X-rays and H$\alpha$ fluxes (Coe et al., 1994b). A cyclotron absorption feature at 55 keV (Doroshenko et al., 2010b) and QPOs (Nespoli and Reig, 2011) have been detected leading to the magnetic field estimations of $(7-8) \times 10^{12}$ G. The companion is a O9.5IV-Ve star located at 3-7 kpc (Janot-Pacheco et al., 1981).

*IGR J11215-5952* is an SFXT displaying a regular outbursting activity during the periastron passage (Sidoli et al., 2007). The system geometry could be well understood through the long term monitoring performed with *Swift*/XRT and the orbital period has been measured at ~165 days (Romano et al., 2009c). This source has been observed in outburst many times with *INTEGRAL* and *Swift*, and it is known to host a ~186 s spinning NS (Swank et al., 2007). Due to the peculiar regularity in the occurrence of its outburst, IGR J11215-5952 is suggested to be an evolutionary link between SFXTs and BeXRBs (Liu et al., 2011). A detailed study of the supergiant star hosted in this system was presented by Lorenzo et al. (2014). This study did not reveal any particularly relevant peculiarity from the star that is classified as a normal B0.5 Ia supergiant.

*IGR J11305-6256* was discovered by *INTEGRAL* in 2004 (Produit et al., 2004). The companion star was identified as the B0IIIe star HD100199 located at about 3 kpc (Masetti et al., 2006a). The broad-band X-ray spectrum, moderate absorption, and transient X-ray activity led La Parola et al. (2013) to classify the source as a Be X-ray binary. These authors also reported on the discovery of the source orbital period at 120.83 d, and noticed that the average orbital modulation of the X-ray emission from IGR J11305-6256 is relatively low compared to other sources in the same class. No X-ray pulsations have been detected so far.

*IGR J11435-6109* was discovered by *INTEGRAL* in 2004 (Grebenev et al., 2004a). Pulsations at a period of ∼166 s were first reported by Swank and Markwardt (2004) and later confirmed by Revnivtsev et al. (2005). The orbital period of the source is 52.5 days (Corbet and Remillard, 2005). The precise X-ray localization of IGR J11435-6109 obtained through a *Chandra* observation (Tomsick et al., 2007) permitted to identify the companion star in this object as a B0Ve/B2Ve located at $\gtrsim$6-10 kpc (Negueruela et al., 2007b). The source is thus a distant Be X-ray binary (see also Coleiro et al., 2013).

*4U 1145-619* was firstly mentioned in the second *UHURU* catalog (Giacconi et al., 1972). Subsequent examination of its localization error box in optics revealed inside a relatively bright star ($V \simeq 9$) with a spectral type B1Vne (Jones et al., 1974). These optical identification and spectral classification were later confirmed by Dower et al. (1978) and Hutchings et al. (1981). In the meantime, two close X-ray periodicities with periods of $\sim 292$ and $\sim 297$ sec were discovered from the vicinity of 4U 1145-619 (White et al., 1978). The situation was clarified with the discovery of another X-ray source 1E 1145.1-614 located only 15' away from 4U 1145-619 and demonstrated pulsations with the period of $\sim 298$ s, while pulsations with the period $\simeq 290$ s were attributed to 4U 1145-619 (Lamb et al., 1980). Long-term observations of 4U 1145-619 with the *Ariel V* observatory revealed outbursts from the source, which occurred at regular intervals of $\simeq 187.5$ days and which were interpreted as a motion of a neutron star in a highly eccentric $e > 0.6$ orbit with the corresponding period (Watson et al., 1981). Further observations of the source performed by different observatories allowed to trace its pulse period history (see, e.g., Lutovinov et al., 1994; Bildsten et al., 1997, and references therein), to measure for the first time the source spectrum up to 100 keV (Filippova et al., 2005), etc. It is necessary to note, that distance estimations from spectroscopic observations $d = 3.1 \pm 0.5$ kpc (Stevens et al., 1997) is several times larger than that from the parallax measurements $d = 0.51 \pm 0.24$ kpc (Clark and Dolan, 1999). But for the latter measurement authors indicate that the 90% confidence interval on the Hipparcos parallax measurement of 4U 1145-619 extends to a distance of 2.3 kpc and thus spectroscopic and astrometric parallaxes practically overlap.

*1E 1145.1-6141* is a persistent sgHMXB with a pulsar (297 sec spin) orbiting a companion in 14.365 days with an eccentricity of 0.2. The spectrum features constant intrinsic absorption ($10^{23}$ cm$^{-2}$). Both spin-up and spin-down have been observed (Ray and Chakrabarty, 2002; Ferrigno et al., 2008).

*1ES 1210-646* is a poorly studied X-ray source, which was found during the *Einstein* Slew Survey (Elvis et al., 1992). Based on optical spectroscopy the system was classified as a HMXB (Masetti et al., 2009). An orbital modulation with a period of about 6.7 days was found by Corbet and Mukai (2008) in the *RXTE/ASM* data. The source spectrum near its maximum flux can be well approximated by a power law continuum with a photon index $\simeq 1.41$, high energy cutoff ($E_{cut} = 6.0$ keV, $E_{fold} = 5.7$ keV) and an Fe K line at 6.56 keV with an equivalent width $\simeq 300$ eV (Corbet and Mukai, 2008). Later Masetti et al. (2010a) using data from the *SWIFT/XRT* telescope showed that the iron line has a transient nature and is tied to the orbital motion of the neutron star.

*GX 301-2* is among the brightest HMXB ($L_X \sim 10^{37}$ erg/s), thanks to the slow (3-400 km/s) and very dense stellar wind of its hyper giant companion (Kaper et al., 1995, 2006). The highly eccentric (e$\sim$ 0.5) pulsar orbit (Sato et al., 1986; Koh et al., 1997) generates a strong orbital modulation of the accretion rate with a broad maximum at phase 0.95 linked with a circumstellar disk or an accretion stream (Leahy and Kostka, 2008). The absence of eclipse constrains the inclination angle in the range ($44° - 78°$). A deep and variable cyclotron resonance feature is observed at hard X-rays (Kreykenbohm et al., 2004; Filippova et al., 2005; Fürst et al., 2011b). Spin up episodes have been observed and explained by the formation of transient accretion disks (Koh et al., 1997). Fast spin-down episodes have been interpreted as evidence for a $10^{14}$ G surface magnetic field (Doroshenko et al., 2010a) or as accretion of magnetised material (Ikhsanov and Finger, 2012). Off-states were detected (Göğüş et al., 2011; Suchy et al., 2012), similar to the ones observed in Vela X-1. The soft X-ray spectrum is affected by variable partial coverage of two different absorbers (Watanabe et al., 2003; Suchy et al., 2012).

*GX 304-1* was discovered during a high-energy X-ray balloon observations in 1967. A pulsar nature of the source was established with the detection of $\sim$ 272 s pulsations (Huckle et al., 1977; McClintock et al., 1977). Later a long-term study with the *Vela 5B* satellite revealed a 132.5-day periodicity of flaring events (Priedhorsky and Terrell, 1983), attributable to the binary period. An optical companion in the system is a Be star (Mason et al., 1978) at a distance of $2.4 \pm 0.5$ kpc (Parkes et al., 1980). Recently it was shown that additionally to the standard X-ray pulsars spectrum model an inclusion of the cyclotron absorption line with energy $E_{cyc} = 50.8 \pm 0.5$ keV, width $\sigma = 8.2 \pm 1.4$ keV and depth $\tau = 0.76 \pm 0.05$ is required for the correct approximation of the source spectrum (Mihara et al., 2010). Later, Yamamoto et al. (2011) and Klochkov et al. (2012) using the data from different observatories (including *INTEGRAL*) revealed a positive correlation between the cyclotron line energy and the source flux (see Fig.11). Observations with the *Fermi/GBM* instruments showed, that a strong outburst activity of the source is accompanied by significant changes in the source pulse period. The latter can be explained in the frame of the quasi-spherical settling accretion onto the neutron star (Postnov et al., 2015).

*2RXP J130159.6-635806* is a faint X-ray source, discovered by the *ROSAT* observatory during all sky survey (sometimes this source is named IGR 13020-6359 as well, due to its first detection in hard X-ray with *INTEGRAL*). The sky field around the source was observed several times in different epochs by different observatories (*ASCA, BeppoSAX, XMM-Newton*), but only after the source detection with the *INTEGRAL* observatory (Chernyakova et al., 2004) the detailed analysis of the archival and follow-up data was done. This analysis allowed Chernyakova et al. (2005) to discover pulsations from the source with the period $P_{spin} \simeq 700$ s and trace its evolution up to $\sim$ 10 years before. The study of a set of observations has shown that the pulse period changed from $\sim$ 735 sec in 1994 to $\sim$ 704 sec in 2004. (Chernyakova et al., 2005) proposed also a possible optical counterpart of the source as a Be-star and obtained a tentative estimate of the distance to the binary system as 4-7 kpc. Subsequent infrared spectral observations confirmed suggestions about the source nature and allowed Coleiro et al. (2013) to constrain its spectral type to B0.5Ve. Recent observations with the $N$uSTAR observatory revealed an unusually steady long-term spin-up in this system, when the pulse period was dramatically changed of about 100 sec during $\sim$ 20 years (Krivonos et al. 2015).

*4U 1416-62/2S 1417-624* is a well-known transient X-ray pulsar in a binary system with a Be-companion star, which was discovered with the *SAS-3* observatory. Using these data Apparao et al. (1980) found pulsations from the source with the period $P_{spin} \simeq 17.64$ sec. Based on the accurate measurements of the source position with the *Einstein* observatory and following optical observations it was shown that an optical counterpart in the system is a Be star (Grindlay et al., 1984) with a spectral type B1Ve (Reig, 2011). Long observations performed with the *BATSE* instrument on board the *Comptom-GRO* observatory in 1994 allowed to determine orbital parameters of the system and showed that a neutron star is orbiting in a highly eccentric orbit (eccentricity $e = 0.446$) with the period $P_{orb} = 42.12$ days (Finger et al., 1996a). Estimates of the distance to the system have still a large uncertainty, 1.4-11.1 kpc (Grindlay et al., 1984).

*IGR J14331-6112* was discovered by *INTEGRAL* in 2003 (Keek et al., 2006). The soft X-ray counterpart was detected first with *Swift*/XRT and later confirmed by *Chandra* (Tomsick et al., 2009b). Masetti et al. (2008b) suggested that the spectral type of the companion star is BIII/BV, but this classification is still a matter of debate (Coleiro et al., 2013).

*IGR J14488-5942* was presented for the first time in the 4th *IBIS* catalog (Bird et al., 2010b) as a transient source. Inside the *INTEGRAL/IBIS* error circle two X-ray sources were detected with the *Swift* observatory (Landi et al., 2009; Rodriguez et al., 2010). One of them, Swift J144843.3-594216, was suggested to be a true counterpart of IGR J14488-5942. A modulation of the hard X-ray flux (15−100 keV) with period around 49 days has been discovered using *Swift/BAT* data (Corbet et al., 2010b). Based on the NIR spectroscopy Coleiro et al. (2013) concluded that this HMXB is more likely an Oe/Be HMXB than a supergiant one.

*4U 1538-522* is an eclipsing persistent sgHMXB (spin 530.4 sec) with a short orbital period of 3.728 days and an eccentricity $> 0.08$ (Davison et al., 1977; Becker et al., 1977; Makishima et al., 1987; Corbet et al., 1993; Clark et al., 1994; Clark, 2000). The companion is a B0I star located at 5.5 kpc (Becker et al., 1977; Reynolds et al., 1992). Variability of the absorption at eclipse egress allows to measure the stellar wind parameters (Clark et al., 1994). The X-ray spectrum (Robba et al., 1992) displays two cyclotron absorption features (Clark et al., 1990; Robba et al., 2001; Rodes-Roca et al., 2009). Emission lines from an extended ionised region have been detected during eclipses (Rodes-Roca et al., 2011). Small spin-up and down have been detected (Rubin et al., 1997).

*XTE J1543-568* was discovered as a transient X-ray pulsar with the pulse period $P_{spin} =$ 27.12 s with the *PCA/RXTE* spectrometer (Marshall et al., 2000). A long term observational program during about a year allowed in't Zand et al. (2001) to determine the orbital parameters of the system, in particular, its orbital period $P_{orb} = 75.56$ d. Taking into account the source position on the pulse period – orbital period diagram and its temporal behavior these authors suggested that XTE J1543-568 is a Be system with an unusually low eccentricity ($e = 0.03$). No optical counterpart has been reported to the date.

*IGR J16195-4945* was discovered by *INTEGRAL* in 2003 (Walter et al., 2004) and associated with the ASCA source AX J161929-4945 (Sugizaki et al., 2001; Sidoli et al., 2005b). The fast flaring activity detected from this source with *INTEGRAL* led Sguera et al. (2006b) to associate this source with the SFXT class (see also Morris et al., 2009a). A *Chandra* observation performed in the direction of the source permitted to identify the supergiant companion and provide further support to this association (Tomsick et al., 2006a; Rahoui et al., 2008a).

*IGR J16207-5129* was discovered by *INTEGRAL* in 2003 (Walter et al., 2004). The companion star (Masetti et al., 2006a; Negueruela et al., 2007a) was classified as a B1 Ia star at ∼6.1 kpc by Nespoli et al. (2008b). Due to its relatively high persistent flux a and the lack of prominent outbursts, Walter and Zurita Heras (2007b); Tomsick et al. (2009c) suggested that IGR J16207-5121 belong to the class of the highly absorbed HMXBs, rather than to the SFXT class. This is supported by the results of *XMM-Newton* and *Chandra* observations, which measured an absorption column density of $\gtrsim 10^{23}$ cm$^{-2}$ (Tomsick et al., 2009c; Bodaghee et al., 2010). The classification of this source is, however, still a matter of debate.

*IGR J16318-4848* is the first source discovered with INTEGRAL (Courvoisier et al., 2003; Walter et al., 2003). XMM-Newton observation indicated that the source is Compton thick with $N_H \approx 2 \times 10^{24}$ cm$^{-2}$ (Matt and Guainazzi, 2003; Walter et al., 2003). Archive and further X-ay observations indicated a persistently bright and Compton thick source (Revnivtsev et al., 2003; Revnivtsev, 2003; Ibarra et al., 2007). The hard X-ray flux detected by INTEGRAL varies by a factor of up to 10 with doubling timescale of the order of 1 hour. The absorbing column density varies significantly by a factor of two Ibarra et al. (2007). The weakness of the Iron 6.4 keV fluorescence line Compton shoulder suggests that the absorption column density is larger on the line of sight than on average (Matt and Guainazzi, 2003; Barragán et al., 2009), pointing towards a disk like geometry. The source was associated with an infrared counterpart (Foschini et al., 2003) of spectral type sgB[e] (Filliatre and Chaty, 2004), indicating a very rare system surrounded by dense circumstellar gas and dust (Kaplan et al., 2006) that could be the signature of an equatorial disk (Rahoui et al., 2008b; Chaty and Rahoui, 2012) or of a close to LBV phase (Moon et al., 2007). No period has been detected in the system.

*IGR J16320-4751* is a persistent source (in 't Zand et al., 2003) serendipitously discovered with INTEGRAL (Tomsick et al., 2003). The source is highly absorbed with $N_H \approx (1-2) \times 10^{23}$ cm$^{-2}$ (Rodriguez et al., 2003). X-ray pulsations with a period of $(1309 \pm 40)$s (Lutovinov et al., 2005c) and an orbital period of 8.986d (Corbet et al., 2005a; Manousakis and Walter, 2012) (but no eclipse) have been detected. The hard X-ray flux detected by *INTEGRAL* varies by a factor larger than 10 and can do so in a few hours. The most likely companion star is an highly reddened O8I supergiant located at $\sim$ 3.5kpc (Rahoui et al., 2008b). We note that IGR J16320-4751 is not related to the pulsar wind nebula HESS J1632-478 (Balbo et al., 2010).

*IGR J16328-4726* was discovered with *INTEGRAL* by Bird et al. (2007a). The source is also classified as an hard X-ray transient in the *INTEGRAL*/ISGRI and *Swift*/BAT catalogues (Bird et al., 2010b; Cusumano et al., 2010). The first study of the source in the soft X-ray domain was performed as a follow-up to the bright outburst from the source caught with the *Swift*/BAT in 2009 (Grupe et al., 2009). In this occasion, the *Swift*/XRT could follow the evolution of the X-ray flux from the source up to 4 days after the onset of the outburst, and revealed a typical behavior of the SFXT sources (Fiocchi et al., 2010). Corbet et al. (2010a) reported on the discovery of the source orbital period at $\sim$10 d by using archival *Swift*/BAT data. A devoted *XMM-Newton* observation also evidenced a pronounced flaring activity during faint X-ray states (Bozzo et al., 2012b), a behavior already observed in a number of SFXTs. A similar flaring activity was also found in archival *Beppo-SAX* data (Fiocchi et al., 2013). The companion star hosted in this system is classified as a O8Iafpe supergiant (Coleiro et al., 2013).

*IGR J16393-4643* is a likely persistent sgHMXB. The source is a highly absorbed ($N_H \approx 2.5 \times 10^{23}$ cm$^{-2}$ Bodaghee et al., 2006) pulsar with a spin period of 911s (Bodaghee et al., 2006). The orbital period is under debate with a most likely value of 4.24d (Pearlman et al., 2011; Thompson et al., 2006; Islam et al., 2015). The companion is not yet identified (Bodaghee et al., 2012a) but its dynamical mass is estimated as $> 7.5 M_\odot$ (Pearlman et al., 2011; Nespoli et al., 2010a; Chaty et al., 2008).

*IGR J16418-4532* was discovered by *INTEGRAL* in 2003 (Tomsick et al., 2004) and later classified as an SFXT on the basis of its fast flaring activity (Sguera et al., 2006b). The discovery of the orbital period of the source at 3.75 days, together with some hint for the presence of an X-ray eclipse, was reported by (Corbet et al., 2006). The presence of an X-ray eclipse was later confirmed and analyzed in details by Drave et al. (2013). The source was detected in outburst few times with *Swift* (Romano et al., 2011c, 2012d), and monitored along its orbit with both the *Swift*/XRT (Romano et al., 2012e) and *XMM-Newton* (Sidoli et al., 2012). The observations confirmed the presence of prominent flaring activity in different X-ray luminosity states and led to the discovery of pulsations at a period of ∼1212 s (see also Walter et al., 2006a). Drave et al. (2013) showed that the apparent transient behavior of the source is most likely due to its large distance (and the consequently low intrinsic X-ray flux). When the latter is taken into account, the source behavior in X-rays is similar to that of classical sgHMXBs (see also Bozzo et al., 2015). IGR J16418-4532 is one of the few sources for which a superorbital modulation has been detected (the period of the modulation is 14.7 days; Corbet and Krimm, 2013).

*IGR J16465-4507* , discovered with *INTEGRAL* (Lutovinov et al., 2004), is a transient X-ray pulsar (spin period ∼228 s, Lutovinov et al. 2005a) which displays on average properties very similar to those of the highly absorbed HMXBs (Walter et al., 2006b) but was tentatively associated to the SFXT class due to the detection of fast flaring activity with *INTEGRAL* (Walter and Zurita Heras, 2007b). The supergiant companion was first identified by Smith (2004) and then confirmed by Negueruela et al. (2005). The measured orbital period of the source is 30.3 days (Clark et al., 2010; La Parola et al., 2010a). Despite the initial association with the SFXT class, the long term monitoring of the source carried out with *Swift* showed that its X-ray flux variability is fairly limited and the X-ray behavior is close to that of classical sgHMXBs (Romano et al., 2014a; Bozzo et al., 2015).

*IGR J16479-4514* is a confirmed SFXT source. It was discovered with *INTEGRAL* (Molkov et al., 2003b) and observed in outburst several times with *Swift* and *INTEGRAL* (Romano et al., 2008c,b; Sguera et al., 2008). This object is known to have at present the shortest orbital period among the other sources of the same class (3.3 days Romano et al., 2009b), and is the only one displaying X-ray eclipses (Bozzo et al., 2008d). The source undergoes regularly a peculiar flaring activity close to the periastron passage, which has been reported first by (Bozzo et al., 2009) and then studied in detail through a nearly complete orbital monitoring performed with *Suzaku* (Sidoli et al., 2013a). The latter observation also did not reveal strong variation in the spectral parameters in different orbital phases, at odds with the behavior displayed by other SFXT sources. IGR J16479-4514 is one of the few sources for which a superorbital modulation has been detected (the period of the modulation is 11.88 days; Corbet and Krimm, 2013).

*IGR J16493-4348* is an eclipsing sgHMXB system with a 6.78 day orbital period and a 1093 sec spin period (Pearlman et al., 2013). The X-ray spectrum shows signatures for intrinsic absorption ($5 - 9 \times 10^{22}$cm$^{-}$2) and for a cyclotron absorption feature (Morris et al., 2009b; D'Aì et al., 2011a). The companion star is a B0.5 Ib supergiant (Nespoli et al., 2010b). IGR J16493-4348 is one of the few sources for which a superorbital modulation has been detected (the period of the modulation is 20.07 days; Corbet and Krimm, 2013).

*OAO 1657-415* is a persistent eclipsing sgHMXB with a pulsar (spin 37 sec) orbiting a O or WR companion in 10.448 days on an eccentric (e=0.11) orbit (Mason et al., 2012). The absorbing column density is $\geq 2 \times 10^{22}$ cm$^{-2}$. The accretion mode alternates between disk and wind accretion (Jenke et al., 2012). The wind density profile could be constrained by the hard X-ray eclipse profile (Denis et al., 2010).

*4U 1700-37* is an eclipsing X-ray source associated with a very massive companion of type O6.5Iaf+ (Jones et al., 1973). The orbital parameters have been reconstructed and the most likely mass of the compact object is 2.4M$_\odot$ (Corbet et al., 2010c). The detection of QPOs and the absence of pulsation (Dolan, 2011) favour a black hole compact object while the hard X-ray spectral shape is typical for an accreting pulsar. The binary may have escaped the Sco OB1 association 2 millions years ago (Ankay et al., 2001). High-ionisation lines have been observed also during eclipses, indicating that the stellar wind is very inhomogeneous (Boroson et al., 2003). The hard X-ray flux varies by a factor of several hundreds. The absorbing column density increases around eclipses as expected for a spherical wind plus a stream trailing the neutron star (Haberl et al., 1989).

*AX J1700.2-4220* was discovered as a faint ASCA source. RXTE and Swift monitoring of the source allowed to characterize it as a Be system ($P_S$=54s; $P_{orb}$=44d). The optical counterpart is not yet identified.

*IGR J17200-3116* was discovered during a deep observations of the Galactic Center with the *INTEGRAL* observatory in 2003 (Revnivtsev et al., 2004; Walter et al., 2004). The exact class of the optical counterpart and distance to the source are still unknown. Based on the *XRT/Swift* data Nichelli et al. (2011) discovered pulsations from the source with the period $P_{spin} \simeq 328$ s, that allowed to suggest this source as a X-ray pulsar in the high mass X-ray binary system. More observations are required to determine the spectral type of this HMXB.

*EXO 1722-363* was discovered with EXOSAT (Warwick et al., 1988) and identified as an highly obscured X-ray pulsar with Ginga (Makino, 1988; Tawara et al., 1989; Takeuchi et al., 1990). The source position was refined with INTEGRAL (Lutovinov et al., 2003a; Walter et al., 2004) and further with XMM-Newton, which allowed an association with an infrared counterpart (Zurita Heras et al., 2006). Its infrared spectrum was identified with that of a supergiant B0-B1Ia star, located at a distance of $7.1 - 7.9$ kpc (Chaty et al., 2008; Mason et al., 2009, 2010). The orbital period of 9.742d, determined with RXTE (Markwardt and Swank, 2003; Corbet et al., 2005b) and refined with INTEGRAL (Manousakis and Walter, 2011) thanks to the presence of X-ray eclipses, established the system as a sgHMXB. The orbital eccentricity is smaller than 0.15. Outside of the X-ray eclipses, the X-ray (2-10 keV) luminosity varies in the range $(0.25 - 2) \times 10^{36}$ erg/s and a soft component is detected at a level of $3 \times 10^{33}$ erg/s. The spectrum is typical for an accreting pulsar with $\Gamma \sim 0$ and a cutoff energy of $E_C \sim 8.2$ keV. An Iron line is detected with an equivalent width of $\sim 100$ eV, generated by material very close to the neutron star. The X-ray pulsar features a spin period of 413.89 s (with short time scale variability as large as 1$\mu$s/s) and a persistently high obscuration, with an absorbing column density varying along the orbit and averaging to $2 \times 10^{22}$ cm$^{-2}$ (Walter et al., 2006a; Manousakis and Walter, 2011). Detailed hydrodynamic simulations of EXO 1722-363 indicated that its high obscuration is linked with the low velocity ($\sim 500$ km/s) of the companion stellar wind and constrained the neutron star mass to 1.75-2.15 (Manousakis et al., 2012b), a value slightly larger but compatible with the kinematic value of $1.5 \pm 0.4$ (Mason et al., 2010).

*IGR J17354-3255* was discovered with *INTEGRAL* in 2006 (Kuulkers et al., 2006). The source only sporadically displays relatively short flares with duration from few hours to $\sim$1 day (Vercellone et al., 2009a; Tomsick, 2009) and has an orbital period of 8.4 days. For these reasons it was associated to the SFXT class (D'Aì et al., 2011b; Sguera et al., 2011). The source is also positionally coincident with the high energy AGILE transient AGL J1734−3310, even though the localization uncertainties are still too large to claim a firm association (Vercellone et al., 2009b). An *XMM-Newton* observation aimed at the source failed to detect it (Bozzo et al., 2012b) and set a lower limit to the dynamic range of its X-ray luminosity of $\gtrsim 10^4$. An orbital monitoring of the source with *Swift* suggested the presence of a possible X-ray eclipse (Ducci et al., 2013b).

*XTE J1739-302* (other name IGR J17391-3021) was discovered with *RXTE* during a bright outburst in 1997 (Smith et al., 1998). Several outbursts from this source were detected with ASCA (Sakano et al., 2002), RXTE (Smith et al., 2006), *INTEGRAL* (Sunyaev et al., 2003a; Lutovinov et al., 2005b; Sguera et al., 2005, 2006b; Blay et al., 2008), and *Swift*/BAT (Sidoli et al., 2009a,c; Romano et al., 2009b, 2011b). The source was also observed during faint X-ray states by *Chandra* and *XMM-Newton*, revealing the typical variability of the SFXT sources (Smith et al., 2006; Bozzo et al., 2010; Bodaghee et al., 2011b). The discovery of the source orbital period at 51.47 days was reported by Drave et al. (2010). The identification of the supergiant companion of XTE J1739-302 was reported by (Rahoui et al., 2008a).

*AX J1749.1-2733 and AX J1749.2-2725* are two closely spaced (angular distance is about 7') faint X-ray sources discovered by the *ASCA* observatory in the direction to the Galactic Center (Sakano et al., 2002; Torii et al., 1998). The latter one was initially recognized as an X-ray pulsar with the period $P_{spin} \simeq 220$ s (Torii et al., 1998); pulsations with the period $P_{spin} \simeq 132$ s from AX J1749.1-2733 were detected later, based on the *XMM-Newton* and *INTEGRAL* data (Karasev et al., 2007, 2008). The *INTEGRAL* observatory detected these sources in hard X-rays: AX J1749.1-2733 during the outburst (Grebenev and Sunyaev, 2007b) and on the average map (Krivonos et al., 2012), AX J1749.2-2725 – on the average map (Krivonos et al., 2012). Spectra of both sources demonstrate a presence of the strong photoabsorption, which significantly exceeds the interstellar one and indicates the massive nature of their companions. An optical identification of both sources was problematic a long time. The infrared data from the *NTT/SOFI* telescope allowed Karasev et al. (2010a) to determine optical counterparts in both systems and estimate their spectral classes as B1-3 and B3 for AX J1749.1-2733 and AX J1749.2-2725, respectively. Moreover, based on the currently developed methods of distance estimation according to the position of red clump giants on the color-magnitude diagram (Karasev et al., 2010b), Karasev et al. (2010a) also estimated the distances to the sources as $13-16$ and $\sim 14$ kpc, respectively.

*GRO 1750-27* is a transient X-ray pulsar with a pulse period $P_{spin} = 4.45$ s. It was discovered by the *BATSE* instrument on board the *Comptom-GRO* observatory during a strong outburst in 1995 (Scott et al., 1997). Besides the pulse period a strong modulation of the source flux was found on a time scale of 29.8 d and interpreted as the orbital period in the binary system (Scott et al., 1997). A second outburst from the system was detected in 2008 by the *Swift* observatory (Krimm et al., 2008) and was monitored by several instruments. These observations allowed to measure the broadband X-ray spectrum of the source and trace the evolution of its hardness, which demonstrated a gradual softening during the outburst (Shaw et al., 2009). Moreover, the accuracy of the determination of the orbital period was improved to $P_{orb} = 29.806 \pm 0.001$ d (Shaw et al., 2009). Based on the source behaviour and on a relation between pulse and orbital periods (Scott et al., 1997) assumed a Be-nature of its optical counterpart and estimated a distance to the system as $\sim 18$ kpc, however this result still needs to be confirmed.

*IGR J17544-2619* , discovered with *INTEGRAL* (Sunyaev et al., 2003b), is one of the most extreme and well-studied SFXT sources (in't Zand, 2005b). The companion star was spectroscopically identified by Pellizza et al. (2006) (but see also Rahoui et al., 2008a) and the orbital period was measured at 4.92 days (Clark et al., 2009). A possible indication of pulsations from the direction of the source at 71 s was reported by Drave et al. (2012) by using the RXTE/PCA, but then retracted (Drave et al., 2014). The deepest observation available was performed with the XIS on-board Suzaku (Rampy et al., 2009). In these data, the authors found evidence for the presence of clumps using hardness ratio measurements, caused by variations of the local absorption. The source was also monitored with Swift/XRT for more than two years (Romano et al., 2011a), during which a number of typical SFXT outbursts were identified. Enhanced variability in the X-ray domain was also evidenced in two relatively short observations performed with *XMM-Newton* in 2003 (González-Riestra et al., 2004). An unprecedentedly bright outburst was detected by the source in 2014, leading to the suggestion that temporary accretion disks might form around the neutron star hosted in this system. A possible detection of pulsations at 11.6 s (Romano et al., 2015) and of a cyclotron line at 17 keV (Bhalerao et al., 2015) were also reported.

*IGR J17586-2129* was first reported by Bird et al. (2007a). Using follow-up observations with the Chandra observatory Tomsick et al. (2009b) improved an accuracy of the source coordinates and determined the infrared 2MASS counterpart. In addition, these authors found a significant absorption ($\simeq 10^{23}$ cm$^{-2}$) in the IGR J17586-2129 spectrum and stated that the source is a candidate to the absorbed HMXB. The infrared spectroscopic measurements revealed only the Br(7-4) emission line, that in combination with the measured spectral energy distribution points toward a supergiant companion star (Coleiro et al., 2013).

*IGR J18027-2016* is a persistent eclipsing X-ray pulsar detected by *INTEGRAL* (Revnivtsev et al., 2004; Lutovinov et al., 2005b) and BeppoSAX (Augello et al., 2003). With a spin period of 139.612 sec and an orbital period of 4.4696 days, its orbit could be reconstructed (Hill et al., 2005; Mason et al., 2011). Its X-ray continuum, typical of an accreting pulsar, is moderately absorbed with $N_H \approx 0.9 \times 10^{23}$ cm$^{-2}$ and the presence for an Iron line (Walter et al., 2006a). The companion star is likely a supergiant B1Ib located at a distance of $\sim$ 12.4kpc (Masetti et al., 2008a; Chaty et al., 2008; Torrejón et al., 2010).

*IGR J18151-1052* was discovered by the *INTEGRAL* observatory during the Galactic plane survey (Krivonos et al., 2009). Follow-up observations of the source, performed with the *XRT* telescope aboard the *Swift* observatory, revealed a significant photoabsorption in its spectrum – up to $3.4 \times 10^{22}$ cm$^{-2}$, that is much higher than that in the Galactic interstellar medium. A strong $H_\alpha$ emission line at zero redshift was detected in the spectrum of its optical counterpart. This suggests that the object is definitely an X-ray binary in our Galaxy, probably an absorbed OB-star (Burenin et al., 2009). The further detailed analysis showed that the identification of the system as a cataclysmic variable cannot be fully ruled out and might be preferable (Lutovinov et al., 2012b; Masetti et al., 2013).

*IGR J18179-1621* is a hard X-ray transient source discovered during the inner Galactic disk observations in February 2012 (Tuerler et al., 2012). X-ray pulsations with a period of $P_{spin} \simeq 11.82$ sec were discovered immediately in the source light curve during follow-up observations with the *XRT/Swift* telescope (Halpern, 2012b). The broadband spectrum of the source can be described by a powerlaw model modified by a high energy cutoff and strong photoabsorption ($N_H \simeq 12 \times 10^{22}$ cm$^{-2}$) at low energies (Li et al., 2012b). Thus, it can be concluded that IGR J18179-1621 is a new heavily absorbed X-ray pulsar in a HMXB. Finally note, that a type of its optical companion is still not determined.

*SAX J1818.6-1703* is one of the confirmed SFXT sources, and was discovered in 1998 by *Beppo-SAX* (in 't Zand et al., 1998a). Several outbursts from the source were detected with *INTEGRAL* and *Swift* (see, e.g. Sidoli et al., 2009b, and references therein). Bird et al. (2010a) and Zurita Heras and Chaty (2009a) determined the best orbital period of the source at 30±0.1 d. Zurita Heras and Chaty (2009a) also found that most of the discovered outbursts took place close to the periastron passage, and that the source usually remains relatively bright in X-rays for about ~6 d around this orbital phase. Outbursts in several periastron passages were missing. SAX J1818.6-1703 was also observed twice with *XMM-Newton* close to the apastron, but not detected Bozzo et al. (2008b, 2012b).

*AX J1820.5-1434* is a faint X-ray pulsar with the neutron star spin period $P_{spin} \simeq 152$ s, discovered with the *ASCA* observatory during the Galactic plane survey (Kinugasa et al., 1998). These observations revealed also a strong absorption in the X-ray spectrum ($N_H \sim 10^{23}$ cm$^{-2}$). It was interpreted as an indication that AX J1820.5-1434 is a high-mass X-ray binary system, but a clear optical identification and determination of the spectral class of the optical star are still problematic (Negueruela and Schurch, 2007). A detection of the hard X-ray emission from from AX J1820.5-1434 with the *INTEGRAL* observatory (Lutovinov et al., 2003b) allowed to reconstruct the source spectrum up to $\sim$ 70 keV and to show that it is typical for X-ray pulsars in HMXB (Filippova et al., 2005). This source is also tentatively associated with the SFXT class due to the detection of fast flaring activity with *INTEGRAL* (Walter and Zurita Heras, 2007b). A timing analysis of the long-term observations with the *S*wift observatory revealed the detection of a coherent signal at $P_{orb} = 54.0 \pm 0.4$ d, which was interpreted as the orbital period of the binary system (Segreto et al., 2013).

*IGR J18410-0535* (other name AX J1841.0-0536) was discovered with *ASCA* in 1994 (Bamba et al., 2001), while undergoing two bright flares lasting about 1 hr each. Similar SFXT-like flaring activity was also recorded several times with MAXI and *INTEGRAL* (Rodriguez et al., 2004; Sguera et al., 2006b; Walter and Zurita Heras, 2007b). Hours-long outbursts were also detected by *Swift*/BAT and followed-up a few times by *Swift*/XRT (de Pasquale et al., 2010; Romano et al., 2010a, 2011b, 2012a,c). This behavior led to the association of IGR J1841.0-0536 with the SFXT class. This association was strengthened by the identification of the supergiant companion through infrared observations (Nespoli et al., 2008b). A 45 ks-long *XMM-Newton* observation performed in 2011 in the direction of IGR J1841.0-0536 caught the source undergoing a bright X-ray flare, which could be interpreted in terms of sudden "ingestion" of accreting material from the dense wind environment. This observation could not confirm the presence of pulsations at $\sim$4.7 s, as suggested by the analysis of previous data (Bamba et al., 2001; Sidoli et al., 2008). A possible association between IGR J18410-0535 and the transient MeV EGRET source 3EG J1837-0423 was suggested by Sguera et al. (2009). The discovery of the source orbital period at 6.5 days was reported by González-Galán (2015).

*GS 1843+00* is a transient X-ray pulsar discovered in 1988 by the *Ginga* observatory during a galactic plane scan (Makino and GINGA Team, 1988b). A pulse period of 29.5 s was measured shortly (Koyama et al., 1990a). Spectroscopic and photometric data indicate a B0-B2 IV-Ve star located at a distance of $\geq$ 10 kpc as an optical counterpart (Israel et al., 2001).

*IGR J18450-0435* (other name AX J1845.0-0433) was discovered by Yamauchi et al. (1995b) in 1993 with the *ASCA* observatory and classified as a transient X-ray source. It exhibited a few hours-long flaring activity and spectral properties similar to those displayed by the SFXTs. The supergiant companion was identified by Coe et al. (1996). The source has been observed several times during periods of enhanced X-ray activity with *INTEGRAL* (Molkov et al., 2004; Halpern and Gotthelf, 2006) and *Swift* (Sguera et al., 2007; Romano et al., 2009a, 2012b). In all cases, the X-ray flares displayed similar properties with respect to those detected originally with *ASCA*. IGR J18450-0435 was also observed with *XMM-Newton* and caught during the transition from a flaring to a quiescent state (Zurita Heras and Walter, 2009b). The *XMM-Newton* observation also revealed the presence of a soft spectral component at energies $\lesssim$2 keV, similar to that already detected from a number of SFXTs and interpreted in terms of X-ray emission from the supergiant wind itself or reprocessing of the NS X-rays within the wind material. The discovery of the source orbital period was reported by Goossens et al. (2013).

*A 1845-024* was initially found by *Ariel-5* (Seward et al., 1976). Later *Ginga* discovered a pulsating source GS 1843-024 with the period of $94.8 \pm 0.1$ s (Makino and GINGA Team, 1988a) at the same position. Soffitta et al. (1998) identified these two sources with a hard

X-ray object GRO J1849-03, which was discovered by *CGRO/BATSE* and demonstrated recurrent hard X-ray outbursts with a period of $\sim$ 241 days (Zhang et al., 1996). Assuming this periodicity to be orbital one Soffitta et al. (1998) classified this source as Be/XRP system using the Corbet diagram. The source spectrum is typical for X-ray pulsars, but modified by a large absorption at low energies $N_H = (1.5 - 3) \times 10^{23}$ cm$^{-2}$ (Koyama et al., 1990b). According to the *INTEGRAL* data the source spectrum above 20 keV can be approximated by a simple power-law (Doroshenko et al., 2008).

*IGR J18462-0223* was discovered by *INTEGRAL* during a few hours-long outburst very reminiscent of the event usually recorded from the SFXTs (Grebenev and Sunyaev, 2010). The source was also observed later with *XMM-Newton* (Bodaghee et al., 2012b), which provided an improved X-ray position within a few arcsec accuracy. The infrared counterpart it is, however, not securely identified yet. The *XMM-Newton* observation also led to the measurement of a strong absorption in X-rays local to the source, which is reminiscent of what is usually observed in the highly absorbed HMXBs, and the identification of X-ray pulsations at a period of 997 s. This confirmed the presence of a neutron star accretor in IGR J18462-0223 as expected for an SFXT source. The NIR counterpart of IGR J18462-0223 was identified by Sguera et al. (2013) and suggested to be a supergiant star located at $\sim$11 kpc.

*IGR J18483-0311* was discovered in 2003 by Chernyakova et al. (2003). The 18.5 days orbital period of the system was first identified by Levine and Corbet (2006b) using *RXTE* archival data, and later confirmed with *INTEGRAL* (Sguera et al., 2007). *INTEGRAL* data also showed that IGR J18483-0311 sporadically displays a few days-long X-ray active states ($\sim$3 days), during which fast flares with typical timescales of a few hours can be observed (Krimm et al., 2011; Romano et al., 2010b; Ducci et al., 2013a). Pulsations with a period of $\sim$21 s were first reported by Sguera et al. (2007). Giunta et al. (2009) discussed the possible detection of pulsations during the low X-ray intensity states of the source. These detections of pulsations were later questioned by Ducci et al. (2013a). The supergiant companion of IGR J18483-0311 was identified by Rahoui et al. (2008a).

*XTE J1855-026* was discovered during *RXTE* scans along the Galactic plane (Corbet et al., 1999b). The source exhibited pulsations with a period of $P_{spin} \simeq 360.7$ s and also a flux modulation with a period of $P_{orb} \simeq 6.07$ days, which was interpreted as the orbital period in the binary system (Corbet and Mukai, 2002). In the same paper other orbital parameters were determined as well: $a_x \sin i = 80.5 \pm 1.4$ lt-s, $\omega = 226 \pm 15$ deg, $e = 0.04 \pm 0.02$. An optical counterpart of XTE J1855-026 was identified as a B0 Iaep luminous supergiant star (Verrecchia et al., 2002b; Negueruela et al., 2008a). The source spectrum has a typical form for X-ray pulsars (White et al., 1983) modified by a significant photoabsorption $N_H \simeq (4 - 15) \times 10^{22}$ cm$^{-2}$ (Corbet et al., 1999b; Romano et al., 2008a).

*XTE J1858+034* is a hard X-ray transient pulsar discovered by *RXTE/ASM* in February 1998 (Remillard et al., 1998). The pulse period was measured with using of *RXTE/PCA* data to be $221.0 \pm 0.5$ s (Takeshima et al., 1998). The transient behaviour, hard X-ray spectrum and pulsations suggest the Be/XRP nature of the source (Takeshima et al., 1998). An orbital period value was estimated to be $\sim$ 380 days (Doroshenko et al., 2008). The hard X-ray spectrum obtained by the *INTEGRAL* observatory can be well described by a powerlaw model with the high energy cutoff and photo-absorption at low energies: photon index $1.26 \pm 0.08$, $E_{cut} = 26.7 \pm 0.7$ keV, $E_{fold} = 6.6 \pm 0.3$ keV, $N_H = (9.0 \pm 1.3) \times 10^{22}$ cm$^{-2}$ (Doroshenko et al., 2008).

*4U 1901+03* was detected by the All Sky Monitor of the *RXTE* observatory in January 2003 (Galloway et al., 2003). It was only a second appearance of this source on the X-ray sky after its discovery with the *UHURU* observatory (Forman et al., 1976). The follow-up observations, performed with the *PCA/RXTE* spectrometer, allowed to discover a coherent signal with the period $P_{spin} \simeq 2.763$ s in the source light curve. This discovery was confirmed soon with the *INTEGRAL* observatory, which observed this region of the sky (Molkov et al., 2003a). Moreover, these observations allowed to obtain for the first time a source broadband spectrum and demonstrate that it can be well approximated by a powerlaw model with a photon index $\Gamma \sim 1.9$ and a high energy cutoff ($E_{cut} \simeq 12$ and $E_{fold} \simeq 13.5$ keV), that is typical for X-ray pulsars. Using data of the *RXTE* observatory (Galloway et al., 2005) determined orbital parameters of the system and showed that it has a very small eccentricity ($e \simeq 0.036$) and moderate orbital period $P_{orb} \simeq 22.58$ d. The outburst have been lasted about five months. There are not firmly established optical counterpart of the source and its distance measurements, there are only tentative suggestions that the neutron star in 4U 1901+03 probably accretes from the wind of a main-sequence OB star (Galloway et al. 2005, but see also an identification of 4U 1901+03 with an early type giant star B0III by Jones et al. 1974).

*4U 1907+097* is a persistent sgHMXB with a pulsar (437.5 sec spin) orbiting an O8-9 Ia supergiant (located at 2-6kpc) in 8.37d with an eccentricity of 0.28 (Makishima et al., 1984; in 't Zand et al., 1998b; Cox et al., 2005; Nespoli et al., 2008a). Its X-ray emission is highly variable and feature cyclotron absorption features (Mihara et al., 1995; Cusumano et al., 1998; Rivers et al., 2010; Fürst et al., 2011a; Hemphill et al., 2013). The source spends $\sim 60\%$ of the time in X-ray off-states that can last from minutes to hours (in 't Zand et al., 1998b; Roberts et al., 2001; Rivers et al., 2010; Şahiner et al., 2012). Pulsations are detected during the off-states (Roberts et al., 2001; Doroshenko et al., 2012a). Limited random-walk spin period variations have been observed (Şahiner et al., 2012). The X-ray absorption is modulated by the orbit (Şahiner et al., 2012, but remains $< 10^{23}$ cm$^{-2}$) and could be modelled with an accretion stream trailing the neutron star (Kostka and Leahy, 2010). 4U 1907+097 is a runaway system accompanied by a bow shock (Gvaramadze et al., 2011).

*4U 1909+07* is a persistent X-ray pulsar discovered by the *UHURU* observatory (Forman et al., 1978) (also known as X 1908+075). An orbital periodicity of 4.4 days has been found in the *RXTE/ASM* data (Wen et al., 2000). Morel and Grosdidier (2005) reported a near-infrared identification of the counterpart consistent with a late O-type supergiant star lying at a distance of $\sim 7$ kpc. Using *RXTE/PCA* data Levine et al. (2004) found the pulse period of the neutron star of $\sim 605$ s and determined the binary orbit parameters $P_{orb} = 4.4007 \pm 0.0009$ days, $e = 0.021 \pm 0.039$, $a_x \sin i = 47.83 \pm 0.94$ lt-s, $f(M) = 6.07 \pm 0.35$ $M_{sun}$. A very strong stellar wind in the system leads to the substantial photoabsorption in the energy spectrum of X 1908+075, which consists of a powerlaw continuum modified by a turnover at high energies. The orbital phase resolved spectroscopy reveals an increase of the photoabsorption by a factor of 3 or more reaching values of $N_H \sim few \times 10^{23}$ cm$^{-2}$ around orbital phase 0 (Levine et al., 2004). Possible detection of the cyclotron scattering feature at 44 keV was reported by Jaisawal et al. (2013) based on the *Suzaku* data. 4U 1909+07 is one of the few sources for which a superorbital modulation has been detected (the period of the modulation is 15.18 days; Corbet and Krimm, 2013).

*IGR J19140+0951* is a persistent sgHMXB featuring a 13.55 days orbital period. The counterpart is a B0.5 supergiant located at 2-5 kpc (Hannikainen et al., 2007). It's accreting pulsar X-ray spectrum features absorption ($10^{23-24}$ cm$^{-2}$) modulated by the orbital period and a variable soft X-ray excess (Prat et al., 2008).

*IGR J19173+0747* was discovered by the *INTEGRAL* observatory during deep observations of the Sagittarius arm region (Pavan et al., 2011). Follow-up observations with the *XRT/Swift* telescope allowed to refine the source position, which was coincident with that of the *ROSAT* source 1RXS J191720.6+074755, and determine its optical counterpart. Based on the overall optical spectral shape and characteristics of an early-type star Masetti et al. (2012b) classified the object IGR J19173+0747 as a candidate to the high mass X-ray binary.

*IGR J19294+1816* was discovered in 2009 with the *INTEGRAL* observatory (Turler et al., 2009). Indications of coherent pulsations with a period of about 12.4 s were found with *Swift/XRT* data (Rodriguez et al., 2009). Corbet and Krimm (2009) have found an orbital modulation of the hard X-ray flux with a period of 117 days. The relation between orbital and pulsation periods, as well as transient nature, lead to the identification of this source as a Be binary system with the X-ray pulsar. The source broadband spectrum could be well fitted by a cut-off power law with photoabsorption: photon index $0.4 \pm 0.3$, $E_{cut} = 8.0^{+1.2}_{1.0}$ keV, $N_H = (3.1 \pm 0.7) \times 10^{22}$ cm$^{-2}$ (Bozzo et al., 2011a).

*XTE J1946+274* is the X-ray pulsar ($P_{spin} = 15.83$ s) discovered simultaneously with the *RXTE* observatory and *BATSE* instrument in 1998 (Smith and Takeshima, 1998). A strong outburst activity of the source in 1998-2001 allowed Wilson et al. (2003) to measure the orbital period of the system $P_{orb} = 169.2$ d and its eccentricity $e = 0.33$. Such long orbital periods in a combination with a relatively high eccentricity are typical for Be/X-ray binaries. In the case of XTE J1946+274, its optical counterpart have a spectral class B0-1V-IVe and is located at a distance of 8-10 kpc (Verrecchia et al., 2002a). Based on the *RXTE* data Heindl et al. (2001) found a cyclotron resonance scattering feature in the source hard X-ray spectrum near 36 keV. Such an energy corresponds to a magnetic field strength of $\simeq 3.1 \times 10^{12}$ G.

*KS 1947+300* is a transient X-ray pulsar, which was discovered in June 1989 by the *TTM* telescope aboard the *KVANT* module of the *Mir* space station (Borozdin et al., 1990). Later the *BATSE* monitor of the *Compton-GRO* observatory revealed the X-ray pulsar GRO J1948+32 with a period of 18.7 s in the same region of the sky (Chakrabarty et al., 1995). Subsequently, KS 1947+300 and GRO J1948+32 were found to be the same object (Swank and Morgan, 2000). Based on the association of the optical counterpart with a B0Ve star the distance to the source was estimated as $\sim$ 10 kpc (Negueruela et al., 2003). Later, using *INTEGRAL* and *RXTE* data a similar value for the distance to the source ($\sim$ 9.5 kpc) was derived from the spin-up rate of the neutron star (Tsygankov and Lutovinov, 2005b). Measurements of the orbital Doppler shift of the pulse period allowed Galloway et al. (2004) to determine the orbital parameters of the binary system: the orbital period $P_{orb} = 40.415 \pm 0.010$ d, the projected semimajor axis of the relativistic object $a_x \sin i = 137 \pm 3$ light seconds, and the eccentricity $e = 0.033 \pm 0.013$. The spectrum of KS 1947+300 in the $3 - 100$ keV energy range can be described by a power law with a high energy cut-off. Spectral parameters are slightly dependent on the source luminosity and in average consistent with a photon index of $\Gamma \sim 1.1$, $E_{cut} \sim 10$ keV and $E_{fold} \sim 25$ keV (Tsygankov and Lutovinov, 2005b). According to the NuSTAR observations an emission continuum is modified by the pulse phase dependent cyclotron scattering feature at $\sim$ 12.5 keV Fürst et al. (2014a).

*SWIFT J2000.6+3210* was recently discovered by the *Swift/BAT* telescope (Tueller et al., 2005) and optically identified with an early BV or mid BIII star (Halpern, 2006; Burenin et al., 2006; Masetti et al., 2008a). During one of two *Suzaku* observations a period of 1056 s was found and interpreted as the spin period of the neutron star (Morris et al., 2009b). Spectral analysis of these data revealed a significant photoabsorption $N_H \simeq \times 10^{23}$ cm$^{-2}$ (Morris et al., 2009b).

*EXO 2030+375* is a transient accreting X-ray pulsar with a spin period of $\sim$ 42 s discovered with the *EXOSAT* observatory during a giant outburst in 1985 (Parmar et al., 1985).

The optical counterpart in the binary system is a B0Ve star (Motch and Janot-Pacheco, 1987; Coe et al., 1988), the distance to the system is estimated as $\sim 7.1$ kpc (Wilson et al., 2002). Orbital parameters of the binary system were derived using *BATSE* data (Stollberg et al., 1999): the orbital period $P_{orb} = 46.016 \pm 0.003$ days, $e = 0.36 \pm 0.02$, $a_x \sin i = 261 \pm 14$ lt-s. The energy spectrum of the source is typical for X-ray pulsars and can be fitted by a power-law model with the high-energy cutoff and iron line. Some authors reported about a tentative detection of the cyclotron absorption feature at $\sim 36$ keV (Reig and Coe, 1999) and $\sim 11$ keV (Wilson et al., 2008) using *RXTE* data, and at $\sim 63$ keV using *INTEGRAL* data (Klochkov et al., 2008). However the existence of this feature in the source spectrum is still not proven reliably.

*SAX J2103.5+4545* is a member of a high-mass Be binary system with a moderate eccentricity ($e \simeq 0.4$) and one of the shortest orbital period $P_{orb} \simeq 12.68$ d known to date among such binaries (Baykal et al., 2000). The source was discovered as a X-ray pulsar with the period $P_{spin} = 358.6$ s based on the data of the *BeppoSAX* observatory (Hulleman et al., 1998). A subsequent monitoring of the pulse period revealed its strong evolution and periods of a drastic acceleration of the neutron star rotation (Sidoli et al., 2005a; Baykal et al., 2007). The accurate X-ray coordinates of the source, obtained with the *XMM-Newton* observatory allowed to determine unambiguously its optical counterpart, which turned out an O-B star with strong emission lines (Filippova et al., 2004). The spectral class of the optical star was determined as B0Ve (Reig, 2011). A distance to the source is estimated as $\simeq 4.5 - 6.9$ kpc (Baykal et al., 2007). The broadband spectrum of SAX J2103.5+4545, obtained with data of *RXTE* and *INTEGRAL* observatories is typical for X-ray pulsars and can be described by a powerlaw with high energy cutoff (see, e.g., Baykal et al., 2002; Lutovinov et al., 2003c; Filippova et al., 2004; Ducci et al., 2008).

*IGR J21343+4738* was discovered during deep observations with the *INTEGRAL* observatory (Krivonos et al., 2007; Bird et al., 2007b). An optical companion is a $V = 14.1$ B1IVe shell star located at a distance of $\sim$8.5 kpc (Reig and Zezas, 2014a). X-ray pulsations with the period of $\simeq 320$ seconds were discovered with the *XMM-Newton* observatory (Reig and Zezas, 2014b).

*4U 2206+543* appeared for the first time in the *UHURU* catalog (Giacconi et al., 1972). The optical counterpart in the system is a peculiar O9.5V star with a high He abundance at a distance of $\sim 2.6$ kpc (Blay et al., 2006). A possible orbital modulation with a period of $\simeq 9.57$ days was found in the *RXTE/ASM* data (Corbet and Peele, 2001). Pulsations at a period of $5559 \pm 3$ s were discovered using observations of *RXTE/PCA* (Reig et al., 2009). Some models predict the system 4U 2206+543 to harbour a magnetar (see, e.g. Finger et al., 2010). In addition to the spectral model typical for X-ray pulsars (power law with high energy cut-off), some evidence of a cyclotron resonance scattering feature at energies $\sim 30$ and $\sim 60$ keV were presented using *RXTE, BeppoSAX* (Torrejón et al., 2004), and *INTEGRAL* (Blay et al., 2005; Wang, 2009) data. However, later an existence of such features in the source spectrum was not confirmed (Wang, 2013).

*IGR J22534+6243* was discovered as a faint hard X-ray source on the Galactic plane map averaging about 9 years of *INTEGRAL* observations (Krivonos et al., 2012). Based on *Chandra* and *Swift* archival data, Halpern (2012a) found pulsations of the X-ray emission with a period $P_{spin} \simeq 46.67$ s. The broad band spectrum of IGR J22534+6243 obtained with *Chandra, Swift* and *INTEGRAL* observatories can be well described by a powerlaw model with a cutoff energy of $25 - 30$ keV, slightly higher than usually observed for X-ray pulsars (Lutovinov et al., 2013a). The proposed optical counterpart 2MASS J22535512+6243368 was observed later by Masetti et al. (2012a) and Lutovinov et al. (2013a), who revealed an optical spectrum typical for an early type star with superimposed $H\alpha$, $H\beta$ and $HeI$ emissions at redshift zero. Based on these measurements it was concluded that IGR J22534+6243 is a X-ray pulsar in a Be high mass X-ray binary system.

**Acknowledgements** Based on observations with INTEGRAL, an ESA project with instruments and science data centre funded by ESA member states (especially the PI countries: Denmark, France, Germany, Italy, Switzerland, Spain) and with the participation of Russia and the USA. This work was supported by a grant from the Scientific & Technological Cooperation Programme Switzerland-Russia. AAL acknowledges support by the grant of the Russian Science Foundation 14-22-00271. We thank Ed. van den Heuvel for carefully reading our manuscript and providing detailed comments.